# A Novel Approach for the Process Planning and Scheduling Problem Using the Concept of Maximum Weighted Independent Set

Kai Sun*

**Abstract**

Process Planning and Scheduling (PPS) is an essential and practical topic but a very intractable problem in manufacturing systems. Many research use iterative methods to solve such problems; however, they cannot achieve satisfactory results in both quality and computational speed. Other studies formulate scheduling problems as a graph coloring problem (GCP) or its extensions, but these formulations are limited to certain types of scheduling problems. In this paper, we propose a novel approach to formulate a general type of the PPS problem with resource allocation and process planning integrated towards a typical objective, minimizing the makespan. The PPS problem is formulated into an undirected weighted conflicting graph, where nodes represent operations and their resources; edges represent constraints, and weight factors are guidelines for the node selection at each time slot. Then, the Maximum Weighted Independent Set (MWIS) problem can be solved to find the best set of operations with their desired resources for each discrete time slot. This proposed approach solves the PPS problem directly with minimum iterations. We establish that the proposed approach always returns a feasible optimum or near-optimum solution to the PPS problem. The different weight configurations of the proposed approach for solving the PPS problem are tested on a real-world PPS example and further designated test instances to evaluate the scalability, accuracy, and robustness.

## 1 Introduction

Process Planning and Scheduling (PPS) is to process a set of prismatic parts into completed products effectively and economically in a manufacturing system. A prismatic part to be produced is generally described by features. For each feature, one or more corresponding operations are determined according to its feature geometry and available machining resources. Each operation requires a selection of critical resources; some examples of these vital resources include machines, tools, fixtures, or specially qualified technicians. The resource constraints are that one critical resource cannot be occupied by more than one operation at the same time. There are precedence relationship constraints among operations, according to the geometrical and technological considerations. Process planning in PPS is the determination of an optimum process plan, i.e., operations and their sequences, within the precedence relationship constraints and resource constraints. The scheduling is the allocation of the resources in the machine shop over time to manufacture the various parts (Zhang et al., 2003). One of the common objectives is to find the feasible schedule with the earliest finishing time of all parts, or formally, minimizing the makespan.

*Department of Mechanical & Aerospace Engineering, College of Engineering & Computer Science, Syracuse University, Syracuse, NY 13244, USA. (kasun@syr.edu)

PPS as one of the main functions of Computer-Aided Process Planning (CAPP) system, it becomes more critical for the effective allocation and utilization of resources in modern flexible manufacturing systems. The challenges are, firstly, a closer integration of process planning and scheduling is required. More specifically, the determination of the operation processing order in a machine shop and the allocation of resources for each operation need to be considered interactively. Secondly, non-iteration or light-iteration methodologies with satisfactory accuracy are desired for the PPS problem. More specifically, the PPS problem is usually solved in a trial and error fashion using methods such as generic algorithms and metaheuristics. However, such methodologies do not guarantee an optimal solution is ever found, and they usually do not scale well with complexity. Also, these methods operating on dynamic data sets is difficult, as genomes begin to converge early on towards solutions which may no longer be valid for later data.

In this paper, we propose a novel approach for formulating and solving the resource-constrained PPS optimization problem. The two procedures, the resource selection and process scheduling, are integrated and formulated into an undirected weighted conflicting graph due to the nature of sequencing and resource constraints. A node in the conflicting graph represents one operation with one possible combination of its required resources during one time slot, and an edge indicates that there is a conflict between the two nodes at both ends of the edge. Each node in the graph is assigned with a weight factor as the guidance for the node selection process to fulfill the optimization objective. The node with a higher possibility leading to the objective, is given a larger weight, so that they are more likely to be selected when generating the schedule. We utilize algorithms proposed in our work (Sun et al., preprint) to solve the Maximum Weighted Independent Set (MWIS) problem to realize this node selection process to get the optimum or a near-optimum solution. Note that this is a preprint of the paper that is scheduled to submit to Applied Discrete Mathematics.

The remainder of the paper is organized as follows. Section 2 provides a comprehensive literature review on the PPS problem. Section 3 provides the mathematical modeling and implements the Integer Programming (IP) model. Section 4 discusses how the conflicting graph is generated for the resource-constrained PPS problem. Section 5 explains how we configure the weight factors in the conflicting graph with the proposed MWIS algorithms (Sun et al., preprint) to achieve the optimization objective. In Section 6, a simplified PPS example problem from the literature (Zhang et al., 2014) is employed to illustrate the proposed approach. Section 7 describes our approach with a real-world problem (Zhang et al., 2014; Chu & Gadh, 1996; Zhang et al., 2003; Li et al., 2005; Li & McMahon, 2007) to verify the practicability. Section 8 discusses and analyzes the results of test instances. Finally, a few concluding remarks are given in Section 9.

## 2 Literature Review

Process planning and scheduling are usually complementary procedures. The former, process planning, can be used to plan manufacturing resources and operations for a part to ensure the application of good manufacturing practice and maintain the consistency of the desired functional specifications of the part during its production processes. Process planning activities include interpretation of design data, selection and sequencing of operations to manufacture the part, selection of machines and cutting tools, determination of cutting parameters, choice of jigs and fixtures, allocation of other resources required by the processes, and calculation of machining times and costs. To clarify process planning, parts are represented by manufacturing features. Figure 1 shows a part composed of $m$ features in which each feature can be manufactured by one or more machining operations ($n$ operations in total for the part). Each operation can be executed by several alternative plans if different machines, cutting tools, or set-up plans are chosen for this operation (Case & Harun Wan, 2000; Maropoulos & Baker, 2000). The latter, scheduling, specifies the schedule of manufacturing resources on each operation of the parts according to the importance of parts (jobs), availability of resources and time constraints, and in the meantime, achieves the optimization objectives (Zhang et al., 2003).

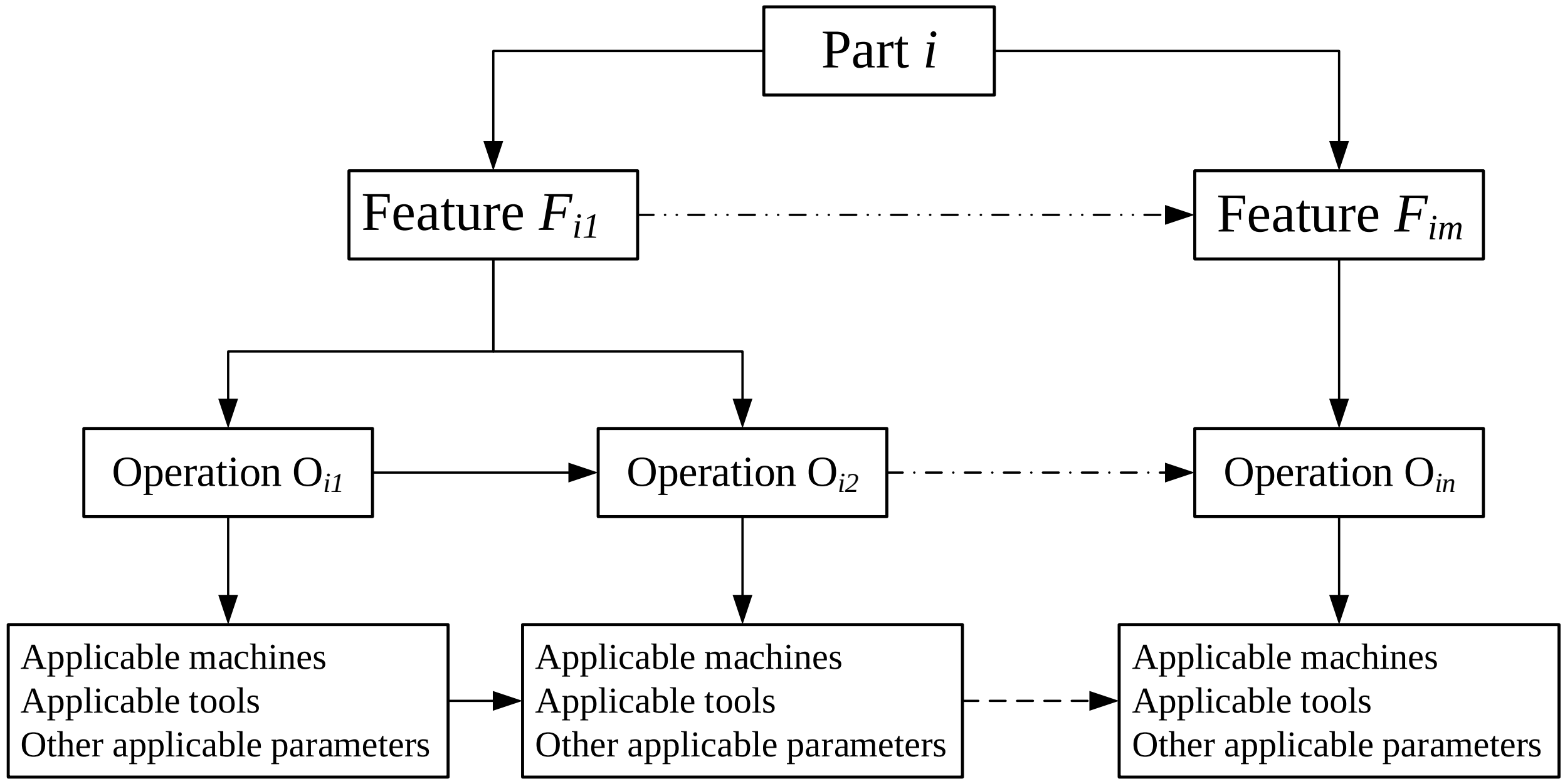

Figure 1. Representation of the Process Plan (adapted from Salehi & Bahreininejad, 2011)

PPS problems vary in complexity. However, seeking an optimum solution rapidly and effectively from all of the permutations, combinations of all of the tasks, manufacturing resources according to specified criteria is very difficult for decision-makers. Lenstra et al. (Lenstra et al., 1977) show that while some classical machine scheduling problems are efficiently solvable, others are NP-hard.

Due to its importance, practicality, and difficulty, in the past decade, many research studies have addressed the PPS problem. Traditionally, such a problem is usually solved in a trial and error fashion adopting methods such as generic algorithms and metaheuristics (Alander, 2014; Milosevic et al., 2016). These approaches include simulated annealing algorithm (Zhang et al., 2003; Tiwari et al., 2006; Li & McMahon, 2007; Chan et al., 2009), tabu search algorithm (Yan et al., 2003), agent-based approach (Shen et al., 2006; Wong et al., 2006), particle swarm optimization algorithm (Guo et al., 2006) and genetic algorithm (Zhang et al., 1997; Morad & Zalzala, 1999; Jia et al., 2002, 2003, 2007; Kim et al., 2003; Chan et al., 2005, 2006, 2008; Moon & Seo, 2005; Li et al., 2005; Zhang & Yan, 2005; Chan et al., 2006; Zhang & Gen, 2010; Salehi & Bahreininejad, 2011; Chaube et al., 2012; Qiao & Lv, 2012; Zhang et al., 2014). Researchers also solved the PPS problem partially as an operation sequencing problem with individual parts (Salehi & Bahreininejad, 2011; Su et al., 2018).

According to the discussions above, the integration and interactions of PPS are through an iterative and empirical fashion. The process planning system first generates a reasonable process plan for each part. Crucial processes in the system include determining suitable manufacturing resources (such as machines and tools), selecting set-up plans, and sequencing machining operations of the part. The scheduling system then specifies the schedule of manufacturing resources on each operation (task) of the parts according to the importance of operations, availability of resources, and time constraints. It is usually difficult to produce a satisfactory result in a single iteration of the execution of the two systems. For the process planning system, the decision of selecting machines and tools is usually made based on objectives to achieve the minimal manufacturing cost and ensure the good manufacturability of a part. Not all the generated process plans for a group of parts could be schedulable according to the time and resource feasibility in a job shop. To overcome this issue, it is necessary iteratively to re-invoke the process planning system to produce alternative plans for further trials until an acceptable scheduling solution is obtained. However, the above iterative process brings forth two severe problems in practical applications. First, it is quite tedious and time-consuming to search for a feasible solution to meet the requirements of process planning and scheduling simultaneously, and an overall optimized target is even more difficult to achieve. Meanwhile, the value of a process plan can be severely discounted since the assumption that all resources are available during the process planning stage might not be entirely valid in the scheduling stage. For instance, the generated process plans sometimes cause some machines to be overloaded, further, to create bottlenecks and restrict the capabilities of machines. Second, the PPS problem has vast solution spaces due to its combinatorial nature. Each time period can schedule one of the feasible operation sets, a feasible operation set can be any non-empty combination of feasible operations, and each operation can be one instance among all the feasible combinations of the available resources. The iteration-based approach needs to be carried out again and again in this vast discrete solution space. Furthermore, the outputs of such methodologies are easily trapped at local optimum, and the local optimum is hard to detect due to the combinatorial nature of such a problem.

Modeling a PPS problem as a Graph Coloring Problem (GCP) is particularly relevant in the presence of incompatible jobs. Extensions of the GCP have been proposed to cope with these scheduling environments (Epstein et al., 2009; Fukunaga et al., 2007; De Werra et al., 2005; Giaro et al., 2009; Halldórsson et al., 2004; Meuwly et al., 2010; Thevenin et al., 2018). As we identify in the previous section, the structural nature of some scheduling problems makes graph coloring an attractive formulation. Gamache et al. (Gamache et al., 2007) use graph coloring methods to determine a feasible schedule for crew scheduling problems within the airline industry. Moreover, they propose a new methodology to determine the existence of a feasible solution based on a graph coloring model and a tabu search algorithm. However, these methodologies often require a specific application environment. For example, Blöchliger and Zufferey (Blöchliger & Zufferey, 2013), Thevenin et al. (Thevenin et al., 2018) formulate the PPS problem as a graph multi-coloring problem. They require that the production system uses continuous flow production, and each job is leading to the end product with no resource change. And still, unlike the particular case of the scheduling problem they are attempting, a typical PPS problem often requires multiple operations to be performed with different resource selections for each job following sequencing constraints. For those reasons, the graph multi-coloring formulations of the PPS problem could be limited in terms of universality.

As a consequence of the review, the requirements are, firstly, a closer integration of process planning and scheduling is required. More specifically, determining the operation processing order in a machine shop and allocation of resources for each operation needs to be considered interactively. Secondly, a direct method or a method with fewer iterations is desired to solve the PPS problem.

Starting with the nature of the PPS problem, we proposed a novel approach to formulate a general type of the PPS problem with resource allocation and process planning integrated towards a typical objective, minimizing the makespan. The PPS problem is formulated into an undirected weighted conflicting graph. In this conflicting graph, nodes stand for operations and their resources; edges stand for constraints; weight factors are the guidelines for the node selection at each time slot. A variation of GCP, the MWIS problem, can be solved to find the best set of operations with their desired resources for each discrete time slot. This proposed approach can solve the problem directly, or it can be applied with few iterations for improving the quality of results.

The performance of the proposed approach depends on the accuracy and computational speed of the MWIS algorithms. We utilize algorithms reported in our work (Sun et al., preprint) to solve the MWIS problem to realize this node selection process to get the optimum or a near-optimum solution.

# 3 Process Planning and Scheduling Problem

## *3.1 Problem Description*

As an example of the PPS problem in a manufacturing system, there are four parts to be processed by four machines with a number of tools. Each part requires several operations (four parts have 4, 3, 3, and 4 operations, respectively), and each operation can be performed on at least one available machine with different processing times. Table 1 shows the operation information of the four parts. Each column describes the part ID, operation ID, successors, operation name, machine candidates, tool candidates, and machining time, respectively. The illustration of one feasible solution to this example problem is shown in Figure 2.

**Table 1. Operation Information of Part 1-4**

| **Part-ID** | **Op-ID** | **Successor** | **Operations** | **Machine Candidates** | **Tool Candidates** | **Machining time (time unit)** |
|---|---|---|---|---|---|---|
| **Part 1** | $O_{1,1}$ | $O_{1,2}$, $O_{1,3}$ | Milling | $M_2$, $M_3$, $M_4$ | $T_6$, $T_7$ | 40, 40, 30 |
| | $O_{1,2}$ | $O_{1,4}$ | Milling | $M_2$, $M_3$, $M_4$ | $T_6$, $T_7$ | 40, 40, 30 |
| | $O_{1,3}$ | $O_{1,4}$ | Milling | $M_2$, $M_3$, $M_4$ | $T_6$, $T_7$ | 20, 20, 15 |
| | $O_{1,4}$ | - | Drilling | $M_1$, $M_2$, $M_3$, $M_4$ | $T_2$ | 12, 10, 10, 7.5 |
| **Part 2** | $O_{2,1}$ | $O_{2,2}$, $O_{2,3}$ | Drilling | $M_1$, $M_2$, $M_3$, $M_4$ | $T_1$ | 12, 10, 10, 7.5 |
| | $O_{2,2}$ | - | Milling | $M_2$, $M_3$, $M_4$ | $T_{12}$ | 20, 20, 15 |
| | $O_{2,3}$ | - | Milling | $M_2$, $M_3$, $M_4$ | $T_6$, $T_7$, $T_{11}$ | 18, 18, 13.5 |
| **Part 3** | $O_{3,1}$ | $O_{3,2}$ | Milling | $M_2$, $M_3$, $M_4$ | $T_7$, $T_8$ | 20, 20, 15 |
| | $O_{3,2}$ | - | Milling | $M_2$, $M_3$, $M_4$ | $T_7$, $T_8$ | 20, 20, 15 |
| | $O_{3,3}$ | $O_{3,2}$ | Milling | $M_2$, $M_3$, $M_4$ | $T_7$, $T_8$ | 15, 15, 11.25 |
| **Part 4** | $O_{4,1}$ | $O_{4,3}$ | Milling | $M_2$, $M_3$ | $T_6$, $T_9$ | 12, 15 |
| | $O_{4,2}$ | $O_{4,4}$ | Milling | $M_2$, $M_3$ | $T_9$, $T_{10}$ | 21, 18 |
| | $O_{4,3}$ | - | Milling | $M_2$, $M_3$ | $T_3$ | 18, 25 |
| | $O_{4,4}$ | - | Milling | $M_2$, $M_3$ | $T_1$, $T_3$ | 27, 25 |

The PPS problem herein is to determine a process plan and schedule (Gantt chart is shown in the lower part of Figure 2), which provides the information for decision-makers on how, when, and in which sequence to allocate these operations of parts to suitable manufacturing resources effectively. When determining the process plan, the best practice operation sequence should be decided first. Then, manufacturing resources such as a machine and one tool should be assigned to every operation. All the manufacturing resources are assumed available in this phase. The determination of schedule is to decide the most appropriate moment to execute each operation with competitive resources like machines, tools, and other possible critical resources. Precedence constraints and resource constraints should be satisfied while determining the process plan and schedule. Moreover, this process plan and schedule should also satisfy the optimization objectives (in this case, minimizing the makespan) concurrently while maintaining the feasibility.

**The problem can be defined as follows:**

(i) Part scheduling: determining how and when to allocate the manufacturing resources to the parts and satisfying the best practice operation sequencing for all the parts.

(ii) Machine and tool selecting: determining the resource selection according to the feature geometry and available machining resources.

**The PPS problem subjects to the following assumptions:**

**A1.** Each resource set (a set of resources needed for processing an operation) can only handle one operation at each time;

**A2.** Each operation is completed before another operation is loaded;

**A3.** The sequence of the operations of each part complies with manufacturing constraints;

**A4.** All parts, machines, tools and other possible resources are available at time zero simultaneously;

**A5.** Each operation is performed on a single resource set, and each resource can only be occupied by one operation at a time;

**A6.** The time for setup change is considered as part of the operation. The time for a machine change or a tool change follows the same assumption;

**A7.** Machines are continuously available for production.

As for the constraints, there are precedence constraints among the operations of each part. These precedence relationships must not be violated in the manufacturing process. For example, a best practice operation sequence of 14 operations from example PPS problem is shown as in the top part of Figure 2. According to this operation sequence, the manufacturing resources can be specified (machines, tools, and other possible critical resources), and then, the schedule can be determined.

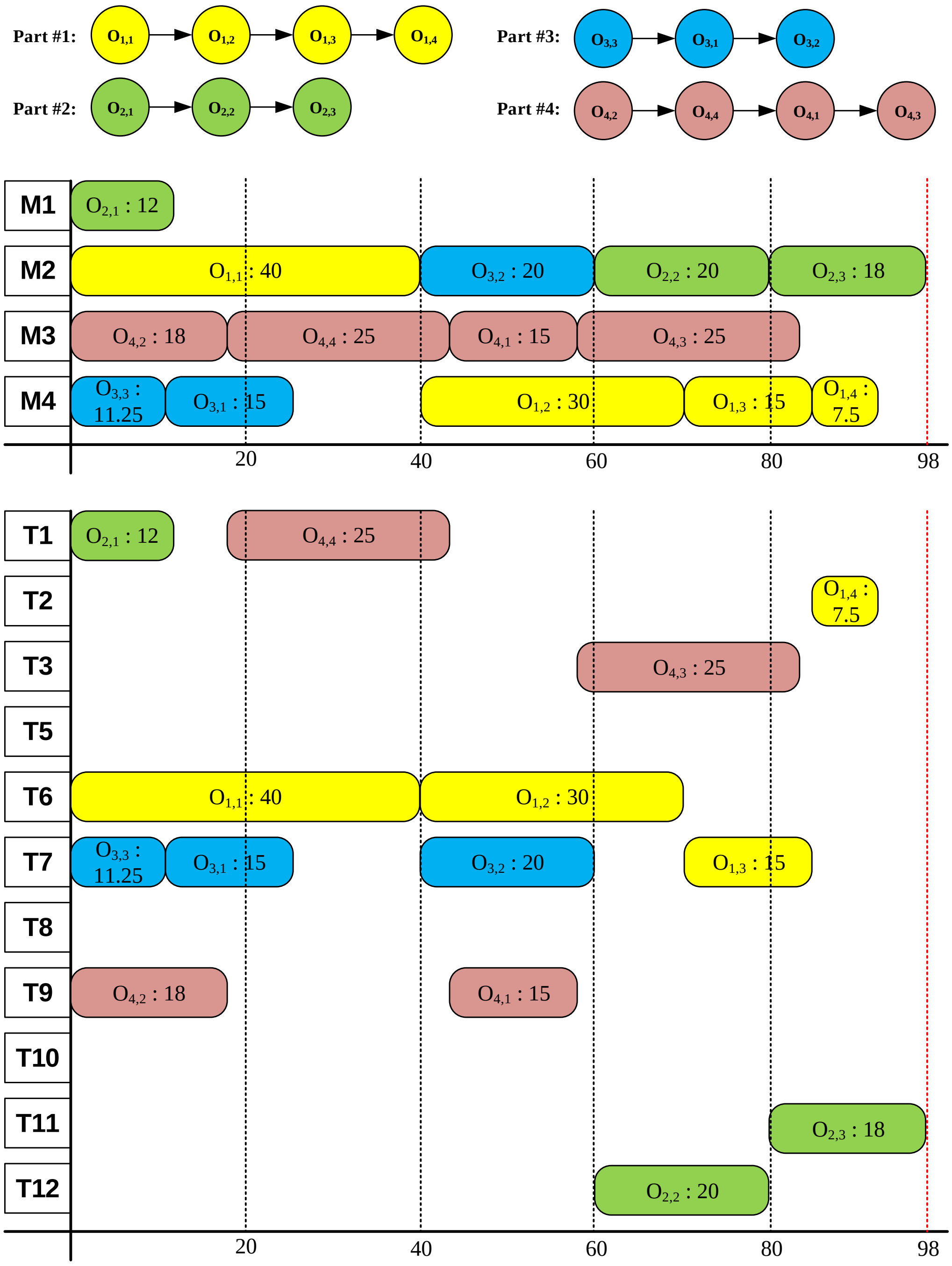

Figure 2. Illustration of the PPS Example Problem

### 3.2 *Mathematical Formulation of the PPS Problem*

Many important and frequently-used objectives in both literature and real-life are applied in the PPS problem. To name a few, there are minimizing the makespan, variation of workload for each machine, minimizing cost, maximizing capacity utilization, delivery dates, or profit optimizations. In this work, we are focusing on minimizing the makespan as the main objective for our solution to the PPS problem. Minimizing the makespan means that the manufacturing system can get high production in a limited period. Or, in other words, the earliest time for finishing all the planned parts. The mathematical model of the problem is expressed in the following notations:

**Indices**

$i, k$: indices of part, $(i, k = 1, 2, \ldots, I)$.

$j, h$: indices of operation for part $i$, $(j, h = 1, 2, \ldots, J_i)$.

$m$: index of machine, $(m = 1, 2, \ldots, M)$.

$l$: index of tool, $(l = 1, 2, \ldots, L)$.

**Parameters**

$I$: number of parts.

$J_i$: number of operations for part $i$.

$M$: number of machines.

$L$: number of tools.

$O_i$: set of operations for part $i$, $O_i = \{o_{i,j} \mid j = 1, 2, \ldots, J_i\}$.

$o_{i,j}$: the $j$th operation of part $i$.

$m_m$: the $m$th machine.

$t_l$: the $l$th tool.

$M_{i,j}$: a set of machines that can process $o_{i,j}$.

$L_{i,j}$: a set of tools that can process $o_{i,j}$.

$A_m$: a set of operations that can be processed on machine $m$.

$A_l$: a set of operations that can be processed with tool $l$.

$r_{i,j,h}$: precedence constraints. if $o_{i,j}$ is predecessor of $o_{i,h}$, $r_{i,j,h} = 1$; otherwise, 0.

$t^P_{m,i,j}$: processing time of $o_{i,j}$ by machine $m$. All the process related time such as setup time, tool and machine change time are integrated with $t^P_{m,i,j}$.

$t^C_{m,i,j}$: completion time of $o_{i,j}$ by machine $m$, it should satisfy the inequality $t^C_{m',i,(j-1)} + t^P_{m,i,j} \leq t^C_{m,i,j}$ that means for every operation, its direct predecessor's completion time plus its processing time might be shorter than its completion time.

**Decision variables**

$$x^M_{m,i,j} = \begin{cases} 1, \ if \ o_{i,j} \ is \ performed \ by \ machine \ m, \\ 0, \ otherwise \end{cases} \tag{1}$$

$$x^L_{l,i,j} = \begin{cases} 1, \ if \ o_{i,j} \ is \ performed \ by \ tool \ l, \\ 0, \ otherwise \end{cases} \tag{2}$$

$$y_{i,j,k,h} = \begin{cases} 1, \ if \ o_{i,j} \ is \ performed \ on \ the \ same \ machine \ right \ before \ o_{k,h}, \\ 0, \ otherwise \end{cases} \tag{3}$$

$$\Omega(X,Y) = \begin{cases} 1, \ if \ X \neq Y, \\ 0, \ otherwise \end{cases} \tag{4}$$

The mathematical model for minimization of makespan can be formulated as the following the mixed-integer programming model:

$$\min t_M = \max_{m,i,j}\{t^C_{m,i,j}\} \tag{5}$$

$$s.t. \left(t^C_{m,k,h} - t^P_{m,k,h} - t^C_{m,i,j}\right) * x^M_{m,i,j} * x^M_{m,k,h} * y_{i,j,k,h} = 0, \forall(i,j),(k,h),m \tag{6}$$

$$s.t. \left(t^C_{m,k,h} - t^P_{m,k,h} - t^C_{m,i,j}\right) * x^L_{l,i,j} * x^L_{l,k,h} * y_{i,j,k,h} = 0, \forall(i,j),(k,h),l \tag{7}$$

$$r_{i,j,h} * y_{i,h,i,j} = 0, \forall(i,j),h \tag{8}$$

$$y_{i,j,i,j} = 0, \forall(i,j) \tag{9}$$

$$\textstyle\sum_{m-1}^{M} x^M_{m,i,j} = 1, \forall(i,j) \tag{10}$$

$$\textstyle\sum_{l-1}^{L} x^L_{l,i,j} = 1, \forall(i,j) \tag{11}$$

$$x^M_{m,i,j} = 0, \forall(i,j) \notin A_m, \forall m \tag{12}$$

$$x^L_{l,i,j} = 0, \forall(i,j) \notin A_l, \forall l \tag{13}$$

$$y_{i,j,k,h} \in \{0,1\}, \forall(i,j),(k,h) \tag{14}$$

$$x^M_{m,i,j} \in \{0,1\}, \forall m,(i,j) \tag{15}$$

$$x^L_{l,i,j} \in \{0,1\}, \forall l,(i,j) \tag{16}$$

$$t^C_{m,i,j} \geq 0, \forall m,(i,j) \tag{17}$$

$$t^P_{m,i,j} \geq 0, \forall m,(i,j) \tag{18}$$

Firstly, the objective function for the PPS problem. Equation (5) illustrates the objective function, which is the minimization of makespan $t_M$. Makespan $t_M$ is the last operation's finishing time, i.e., the maximization of completion time among all the operations. Secondly, the sequencing constraints. Equations (6) and (7) imposes that any machine or tool cannot be selected for one operation until the predecessor is completed. The precedence constraint is defined as Equation (8). Equation (9) ensures the feasible operation sequence. Thirdly, the incompatible resource constraints. The feasible resource selection is defined by Equations (10) and (11). Equation (10) ensures that one operation is only performed on a single machine, and Equation (11) ensures that one operation requires only one tool. Equation (12) and (13) denotes that the assignment of machine and tool for each operation should be selected from the available machine candidates and tool candidates. Lastly, Equations (14), (15), (16), (17) and (18) impose nonnegative condition.

## 3.3 *Implementation of Integer Programming (IP) Model*

In order to get the optimum solution to the PPS problem, the IP model is implemented and tested based on the mathematical modeling discussed in Section 3.2. The IP model is implemented with python package "pyomo" in Python 3.7.5. The solver utilized in this implementation is "glpk (GNU Linear Programming Kit)."

Part #1: $\frac{T_{1,1}[(R_1,R_2,R_3)_2]}{2} \rightarrow \frac{T_{1,2}[(R_1,R_2,R_3)_2]}{1} \rightarrow \frac{T_{1,3}[(R_1,R_2,R_3)_1]}{2} \rightarrow \frac{T_{1,4}[(R_4)_1]}{1}$

Part #2: $\frac{T_{2,1}[(R_1,R_2,R_3)_1]}{1} \rightarrow \frac{T_{2,2}[(R_1)_1 and (R_2,R_3)_1]}{2} \rightarrow \frac{T_{2,3}[(R_4)_1]}{1}$

Part #3: $\frac{T_{3,1}[(R_1,R_2)_1]}{1} \rightarrow \begin{pmatrix} \frac{T_{3,2a}[(R_1,R_2)_2]}{1} \\ \frac{T_{3,2b}[(R_3)_1]}{2} \end{pmatrix}^1 \rightarrow \frac{T_{3,3}[(R_3,R_4)_1]}{3}$

Part #4: $\frac{T_{4,1}[(R_1,R_2,R_3)_3]}{1} \rightarrow \frac{T_{4,2}[(R_1,R_2,R_3)_2]}{2} \rightarrow \frac{T_{4,3}[(R_1,R_2,R_3)_3]}{1} \rightarrow \frac{T_{4,4}[(R_1,R_2,R_3)_2]}{1} \rightarrow \frac{T_{4,5}[(R_4)_1]}{1}$

Where, $R_1, R_2, R_3, R_4$ are the four different resources. For $\frac{T_{1,1}[(R_1,R_2,R_3)_2]}{2}$, it means that the first operation (task $T_{1,1}$) of Part #1 requires any combination of two resources among $(R_1, R_2, R_3)$ and the duration is 2 time slots. $\begin{pmatrix} \frac{T_{3,2a}[(R_1,R_2)_2]}{1} \\ \frac{T_{3,2b}[(R_3)_1]}{2} \end{pmatrix}^1$ means that the task $T_{3,2}$ can be processed with two task options $T_{3,2a}$ and $T_{3,2b}$.

Figure 3. Parts Information with Simplified Duration Information

```
['P1',['T11','T12','T13','T14'],[2,1,2,1],
        [
            [['2'],['R1','R2','R3'],1],
            [['2'],['R1','R2','R3'],1],
            [['1'],['R1','R2','R3'],1],
            [['1'],['R4'],1]
        ]
    ],
```

Figure 4. Input of Part #1 Operations

Assume that the best practice sequence of operations of four parts is given in Figure 3. Note that, we generalize the machines, tools and all other possible resources as $r$ number of resources, $(R_1, R_2, \ldots, R_r)$. The input format, taking Part #1 operations as an example, is shown in Figure 4.

The inputs are then transformed into the dictionary shown in Figure 5 to fulfill the solver's requirements. Each job is broken down into task-resource pairs associated with its duration and sequencing information. For tasks that require more than one resource, each required resource is generated as one task-resource pair instance.

```
Results_Transformed = {
        ('T11', 'R1'): {'dur': 2},
        ('T11', 'R2'): {'dur': 2},
        ('T12', 'R1'): {'dur': 1, 'prec': ('T11', 'R2')},
        ('T12', 'R2'): {'dur': 1, 'prec': ('T11', 'R2')},
        ('T13', 'R1'): {'dur': 2, 'prec': ('T12', 'R2')},
        ('T14', 'R4'): {'dur': 1, 'prec': ('T13', 'R1')},
        ('T21', 'R1'): {'dur': 1},
        ('T22', 'R1'): {'dur': 1, 'prec': ('T21', 'R1')},
        ('T22', 'R2'): {'dur': 1, 'prec': ('T21', 'R1')},
        ('T23', 'R4'): {'dur': 1, 'prec': ('T22', 'R2')},
        ('T31', 'R1'): {'dur': 1},
        ('T31', 'R2'): {'dur': 1},
        ('T32a', 'R1'): {'dur': 1, 'prec': ('T31', 'R2')},
        ('T33', 'R3'): {'dur': 3, 'prec': ('T32a', 'R1')},
        ('T41', 'R1'): {'dur': 1},
        ('T41', 'R2'): {'dur': 1},
        ('T41', 'R3'): {'dur': 1},
        ('T42', 'R1'): {'dur': 2, 'prec': ('T41', 'R3')},
        ('T42', 'R2'): {'dur': 2, 'prec': ('T41', 'R3')},
        ('T43', 'R1'): {'dur': 1, 'prec': ('T42', 'R2')},
        ('T43', 'R2'): {'dur': 1, 'prec': ('T42', 'R2')},
        ('T43', 'R3'): {'dur': 1, 'prec': ('T42', 'R2')},
        ('T44', 'R1'): {'dur': 1, 'prec': ('T43', 'R3')},
        ('T44', 'R2'): {'dur': 1, 'prec': ('T43', 'R3')},
        ('T45', 'R4'): {'dur': 1, 'prec': ('T44', 'R2')}
        }
```

Figure 5. Inputs Dictionary Format for Package "pyomo" in Python

The mathematical modeling of PPS problem from Section 3.2 is transformed into the format for the python package "pyomo" as well as the solver "glpk". The new formulation is as below:

(1) The variables

- model.start = pyo.Var(PARTS, RESOURCES, domain = pyo.NonNegativeReals)
- model.makespan = pyo.Var(domain=pyo.NonNegativeReals)
- model.y = pyo.Var(PARTS,PARTS,RESOURCES, domain = pyo.Boolean)

(2) The objective

- model.Obj = pyo.Objective(expr = model.makespan, sense = pyo.minimize)

(3) The constraints

For the instances of the same tasks but different resources, these instances must have the same start time.

- model.cons.add(model.start[j,r] <= model.start[m,n])

- model.cons.add(model.start[m,n] <= model.start[j,r])

The makespan is the finishing time for all tasks.

- model.cons.add(model.start[j,m] + TASKS[(j,m)]['dur'] <= model.makespan)

For a task which requires a predecessor, it can only be scheduled after the predecessor is finished.

- model.cons.add(model.start[j,m] >= model.start[k,n] + TASKS[(k,n)]['dur'])

For the tasks who shares resources, they cannot be scheduled in the same time.

- model.cons.add(model.start[j,m] + TASKS[(j,m)]['dur'] <= model.start[k,m]

or

model.cons.add(model.start[k,m] + TASKS[(k,m)]['dur'] <= model.start[j,m]

### 3.4 *Numerical Results of IP Model*

We introduce the **I**nput **C**omplexity **I**ndex (ICI) to measure the complexity of inputs. It is essentially a reference value describing the relative size of the possible number of combinations of results of the PPS problem. The PPS problem can be understood as a conflicting graph so that we can utilize some parameters of this graph to calculate the ICI. The ICI can be defined as,

$$ICI = |P|^{(\frac{|T|}{|P|}-1)} * |N| * |O| * |D|$$

Where, $|P|$ is the number of parts; $|T|$ is the number of tasks; $|O|$ is the total number of the options of the tasks. $|D|$ is the density of the conflicting graph, and $|N|$ is the number of nodes of the conflicting graph.

The graph density is defined as follows (Diestel, 2006),

$$D = \frac{2|E|}{|N|(|N| - 1)}$$

Where, $|E|$ is the number of edges in the conflicting graph.

IP is NP-complete on discrete problems, which means that its computation time should increase exponentially with the size of the inputs. To verify this hypothesis, we simulated 10 PPS problems considering a different number of parts and operations, as well as diverse information for operations. The results are shown in Table 2. In Table 2, each column describes the test ID, number of parts, number of tasks, number of edges, number of nods, graph density, number of options, ICI, 3-run average clock time in seconds, minimum makespan, respectively. Note that all computational experiments in this work are performed on a virtual server at Syracuse University. The CPU is Intel Xeon E5-2699 with a fixed maximum speed at 2.3 GHz, and the memory is 32 GB. All the implementations mentioned in this paper are in single threading.

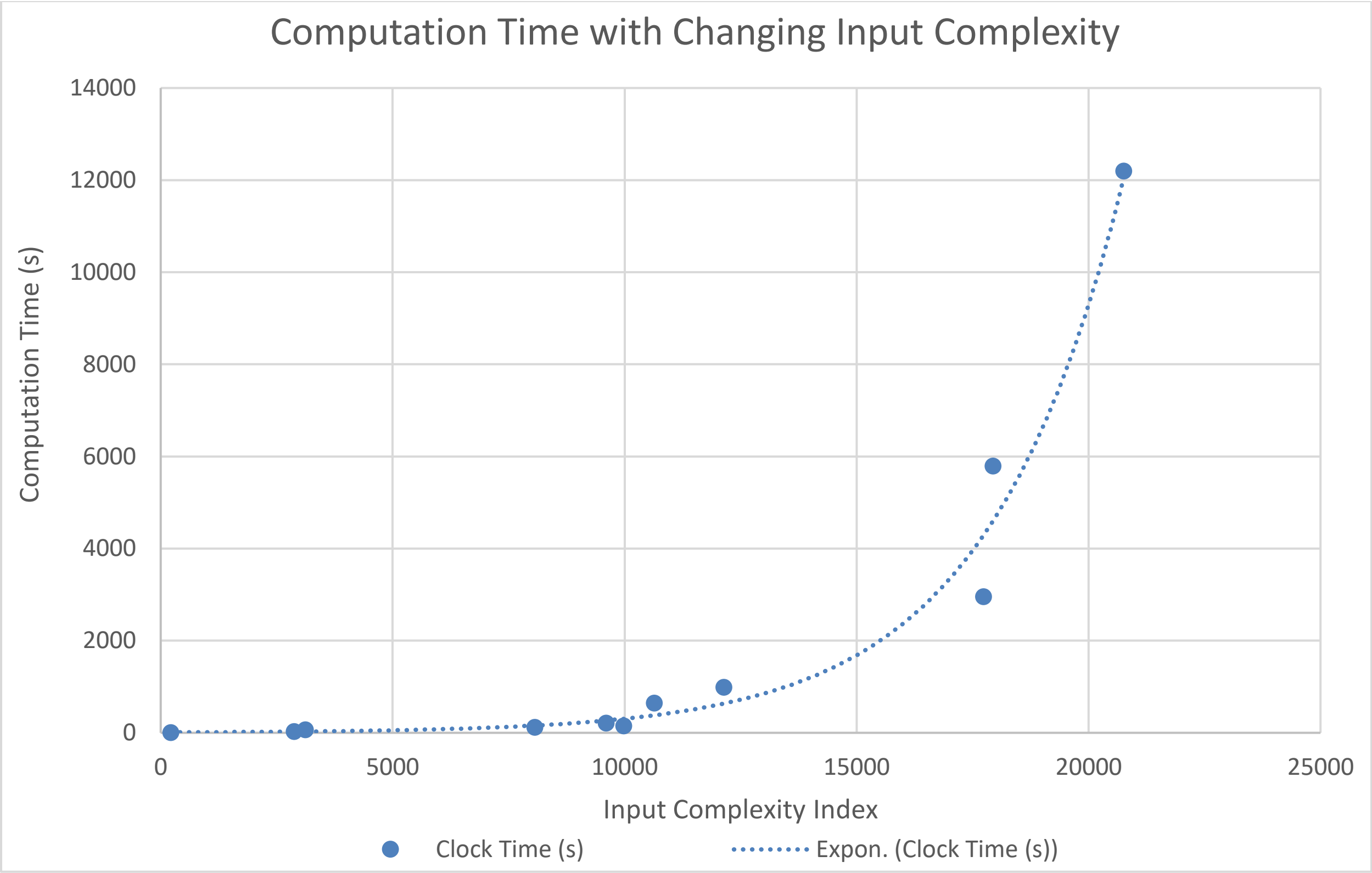

Figure 6. Computation Time with Changing ICI

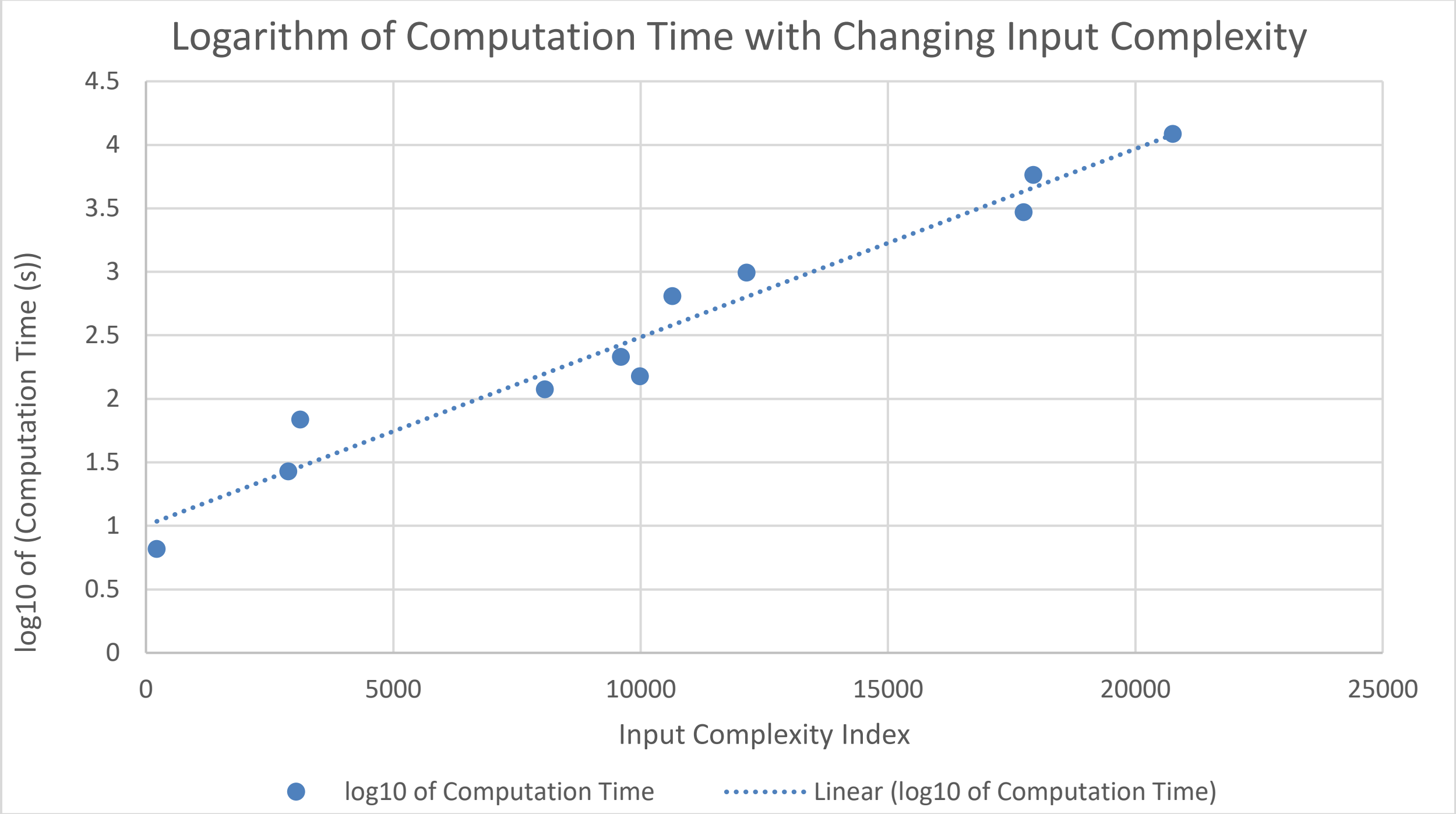

Figure 7. Logarithmic Computation Time with Changing ICI

**Table 2. Integer Programming Model Numerical Results**

| Test -ID | # of Parts | # of Tasks | # of Edges | # of Nodes | Graph Density | # of options | Input Complexity Index | Clock Time (s) | Minimum Makespan |
|---|---|---|---|---|---|---|---|---|---|
| **1** | 2 | 7 | 111 | 24 | 0.40 | 1 | 218.40 | 6.59 | 7 |
| **2** | 3 | 10 | 227 | 35 | 0.38 | 2 | 3119.84 | 68.67 | 8 |
| **3** | 3 | 11 | 307 | 37 | 0.46 | 1 | 2873.64 | 26.97 | 16 |
| **4** | 4 | 12 | 315 | 41 | 0.38 | 2 | 8064 | 118.94 | 11 |
| **5** | 4 | 12 | 390 | 41 | 0.48 | 2 | 9984 | 150.77 | 10 |
| **6** | 5 | 13 | 316 | 40 | 0.41 | 2 | 10640.80 | 643.77 | 7 |
| **7** | 5 | 14 | 396 | 41 | 0.48 | 2 | 17938.30 | 5794.50 | 10 |
| **8** | 5 | 14 | 504 | 45 | 0.51 | 2 | 20755.05 | 12196.93 | 10 |
| **9** | 4 | 14 | 375 | 41 | 0.46 | 1 | 9600 | 213.92 | 18 |
| **10** | 4 | 14 | 1074 | 63 | 0.55 | 1 | 17738.32 | 2959.31 | 18 |

The computation time follows an exponential trendline with increasing input ICI in Figure 6, and the logarithmic computation time follows a straight trendline with increasing input ICI in Figure 7. It can be seen that the IP model follows an exponential complexity of the PPS problem. Although the solution of the IP model can provide the optimum solution to the PPS problem, the computational speed is unacceptable. But we can manipulate inputs based on the outputs of our approach, so that the outputs of our approach can be verified in terms of accuracy.

### 3.5 *Discussions on Formulating and Solving the PPS Problem via Conflicting Graph*

The PPS problem considered in this work, mainly have two types of constraints, the sequencing constraints and the incompatible recourse constraints. The former ensures the best practice operation sequence for each part, and the latter ensures no resource conflict for operations scheduled in parallel. Since the operation sequence of the parts is usually predefined, the PPS problem can be considered as selecting the best set of feasible operations that can be processed in parallel during every discrete time period (a time slot). The feasible operations refer to operations that can be scheduled for the current time period without resource and precedence conflicts. Usually, there is more than one set of feasible operations can be selected for the current time period. The best set of feasible operations is that by scheduling the set of feasible operations for the current time period, the global optimization objective, minimizing the makespan, is most likely to be achieved. If we consider each operation-resource pair (the operation along with one combination of the required resources during a unit discrete time period) as a node, and apply the edges to represent the constraints, a conflicting graph can be generated for the PPS problem. Furthermore, with a weight factor assigned to each operation-resource node as the guidance for selecting the best set of feasible operations, solving the PPS problem becomes solving the MWIS problem for each unit discrete time period. The output of the PPS problem is a combination of the best sets of feasible operations of each unit discrete time period. In the following Sections 4, 5, and 6, we discuss how the conflicting graph is generated, how the weight factor is calculated and assigned, and how we get the optimal or near-optimal solution. The discussions are based on the example problem, as shown in Figure 2 and Table 1.

## 4 Generating the Conflicting Graph

Based on previous discussions, the PPS problem can be naturally represented as a conflicting graph. Then, the optimization is to find and schedule the best qualified **M**aximal **I**ndependent **S**et (MIS) for each time period, so that an optimal processing schedule can be constructed. In this section, we discuss how to construct the conflicting graph. There are two steps to construct the conflicting graph, Step 1, Operation Data Preparation, and Step 2, Generating the Conflicting Graph.

Step 1. Operation Data Preparation

Before we start to generate the conflicting graph, let us reformulate all operations of the parts that need to be produced. In this step, we need three types of information on the operations of the parts, they are (1) the best practice operation sequence, (2) the resource options of each operation, and (3) the processing time of each operation with each of its resource combinations. The top part of Figure 2 illustrates the best practice operation sequence of each part. And from Table 1, we understand machine candidates, tool candidates, and machining time associated with the machines, respectively. With this information, we can reformulate the operation information of the parts as Figure 8.

As described in Figure 8, it can be interpreted as the four operations for Part #1 need to be processed in the sequence of $O_{1,1} \rightarrow O_{1,2} \rightarrow O_{1,3} \rightarrow O_{1,4}$. Each operation of each part is corresponding to a detailed task unit. For instance, the first operation $O_{1,1}$ is corresponding to the detailed task unit, $\left(\frac{\frac{T_{1,1a}[(M_2, M_3)_1 \text{ and } (T_6, T_7)_1]}{40}}{\frac{T_{1,1b}[(M_4)_1 \text{ and } (T_6, T_7)_1]}{30}}\right)^1$, which means that operation $O_{1,1}$ can be processed with one of the two task options, $T_{1,1a}$ and $T_{1,1b}$. The $T_{1,1a}$ and $T_{1,1b}$ here indicate that we can choose one of the options "$a$" or "$b$" for the operation $O_{1,1}$ as the first operation (task) to produce part #1. The task $T_{1,1a}$ has its detail resource information, $\frac{T_{1,1a}[(M_2, M_3)_1 \text{ and } (T_6, T_7)_1]}{40}$. It means that for the task option $T_{1,1a}$, it requires one of the machines among $(M_2, M_3)$ and one of the tools among $(T_6, T_7)$. And the duration of task option $T_{1,1a}$ is 40 time units.

Each operation with its resource selection needs a certain period of time to process; we can simplify the problem by fitting the processing time of an operation into a discrete number of time slots. For example, if an operation $o_{m,n}$ requires 35 time units to finish, and we define each time slot (1TS) stands for 10 time units. Therefore, the operation $o_{m,n}$ needs 4 time slots (4TS) to process. Based on this assumption, we can translate Figure 9 to Figure 10 with the simplified processing time (duration) information.

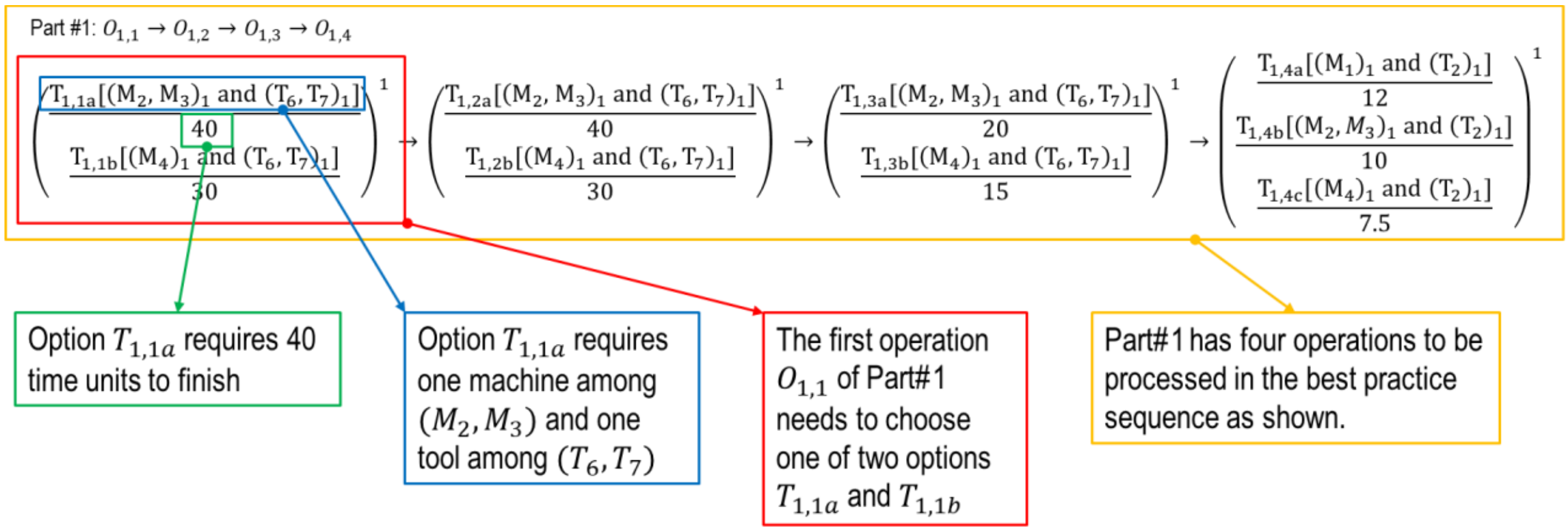

Figure 8. Interpretation for Operation Data Preparation

Part #1: $O_{1,1} \rightarrow O_{1,2} \rightarrow O_{1,3} \rightarrow O_{1,4}$

$$\begin{pmatrix} \frac{T_{1,1a}[(M_2,\ M_3)_1 \text{ and } (T_6, T_7)_1]}{40} \\ \frac{T_{1,1b}[(M_4)_1 \text{ and } (T_6, T_7)_1]}{30} \end{pmatrix}^1 \rightarrow \begin{pmatrix} \frac{T_{1,2a}[(M_2,\ M_3)_1 \text{ and } (T_6, T_7)_1]}{40} \\ \frac{T_{1,2b}[(M_4)_1 \text{ and } (T_6, T_7)_1]}{30} \end{pmatrix}^1 \rightarrow \begin{pmatrix} \frac{T_{1,3a}[(M_2,\ M_3)_1 \text{ and } (T_6, T_7)_1]}{20} \\ \frac{T_{1,3b}[(M_4)_1 \text{ and } (T_6, T_7)_1]}{15} \end{pmatrix}^1$$

$$\rightarrow \begin{pmatrix} \frac{T_{1,4a}[(M_1)_1 \text{ and } (T_2)_1]}{12} \\ \frac{T_{1,4b}[(M_2, M_3)_1 \text{ and } (T_2)_1]}{10} \\ \frac{T_{1,4c}[(M_4)_1 \text{ and } (T_2)_1]}{7.5} \end{pmatrix}^1$$

Part #2: $O_{2,1} \rightarrow O_{2,2} \rightarrow O_{2,3}$

$$\begin{pmatrix} \frac{T_{2,1a}[(M_1)_1 \text{ and } (T_1)_1]}{12} \\ \frac{T_{2,1b}[(M_2, M_3)_1 \text{ and } (T_1)_1]}{10} \\ \frac{T_{2,1c}[(M_4)_1 \text{ and } (T_1)_1]}{7.5} \end{pmatrix}^1 \rightarrow \begin{pmatrix} \frac{T_{2,2a}[(M_2, M_3)_1 \text{ and } (T_{12})_1]}{20} \\ \frac{T_{2,2b}[(M_4)_1 \text{ and } (T_{12})_1]}{15} \end{pmatrix}^1 \rightarrow \begin{pmatrix} \frac{T_{2,3a}[(M_2, M_3)_1 \text{ and } (T_5, T_6, T_{11})_1]}{18} \\ \frac{T_{2,3b}[(M_4)_1 \text{ and } (T_5, T_6, T_{11})_1]}{13.5} \end{pmatrix}^1$$

Part #3: $O_{3,3} \rightarrow O_{3,1} \rightarrow O_{3,2}$

$$\begin{pmatrix} \frac{T_{3,1a}[(M_2, M_3)_1 \text{ and } (T_7, T_8)_1]}{15} \\ \frac{T_{3,1b}[(M_4)_1 \text{ and } (T_7, T_8)_1]}{11.25} \end{pmatrix}^1 \rightarrow \begin{pmatrix} \frac{T_{3,2a}[(M_2, M_3)_1 \text{ and } (T_7, T_8)_1]}{20} \\ \frac{T_{3,2b}[(M_4)_1 \text{ and } (T_7, T_8)_1]}{15} \end{pmatrix}^1 \rightarrow \begin{pmatrix} \frac{T_{3,3a}[(M_2, M_3)_1 \text{ and } (T_7, T_8)_1]}{20} \\ \frac{T_{3,3b}[(M_4)_1 \text{ and } (T_7, T_8)_1]}{15} \end{pmatrix}^1$$

Part #4: $O_{4,2} \rightarrow O_{4,4} \rightarrow O_{4,1} \rightarrow O_{4,3}$

$$\begin{pmatrix} \frac{T_{4,1a}[(M_2)_1 \text{ and } (T_9, T_{10})_1]}{21} \\ \frac{T_{4,1b}[(M_3)_1 \text{ and } (T_9, T_{10})_1]}{18} \end{pmatrix}^1 \rightarrow \begin{pmatrix} \frac{T_{4,2a}[(M_2)_1 \text{ and } (T_1, T_3)_1]}{27} \\ \frac{T_{4,2b}[(M_3)_1 \text{ and } (T_1, T_3)_1]}{25} \end{pmatrix}^1 \rightarrow \begin{pmatrix} \frac{T_{4,3a}[(M_2)_1 \text{ and } (T_6, T_9)_1]}{12} \\ \frac{T_{4,3b}[(M_3)_1 \text{ and } (T_6, T_9)_1]}{15} \end{pmatrix}^1 \rightarrow$$

$$\begin{pmatrix} \frac{T_{4,4a}[(M_2)_1 \text{ and } (T_3)_1]}{18} \\ \frac{T_{4,4b}[(M_3)_1 \text{ and } (T_3)_1]}{25} \end{pmatrix}^1$$

Figure 9. Reformatted Parts Information

Since we want to use the node in the conflicting graph to represent a task with its resource instance, while choosing the best qualified MIS of nodes, tasks with different durations may cause unbalanced conflicting constraints. Because a long duration task only causes one conflicting count with another conflicting task. In order to capture all the possible constraints, as well as simplify the weight factor calculation and fulfill different weights factor assignment strategies, we want to ensure every node in the conflicting graph stands for one task with one combination instance of its required resources for one time slot. Based on the task information in Figure 10, we break down all tasks into single time slots. We name a task that is broken down in such a way as a Unit Task. For example, $\frac{T_{1,1a-1}[(M_2, M_3)_1 \text{ and } (T_6, T_7)_1]}{1TS}$ is a Unit Task, it can be marked as $T_{1,1a-1}$, which means that it is the first Unit Task of option “a” in part #1 operations. According to the details of $T_{1,1a-1}$, it requires one of the machines among $(M_2,\ M_3)$ and one of the tools among $(T_6, T_7)$. Based on the information from Figure 10, the transformed tasks information in Unit Tasks is shown in Figure 11. The information in Figure 11 can be formulated into a dictionary for the implementation of the proposed approach. The format is shown in Figure 12; there are 47 Unit Tasks after breaking up. In the next step, we discuss how we can generate the nodes and edges for generating the conflicting graph for the PPS problem.

Part #1: $O_{1,1} \rightarrow O_{1,2} \rightarrow O_{1,3} \rightarrow O_{1,4}$

$$\begin{pmatrix} \frac{T_{1,1a}[(M_2,\ M_3)_1 \text{ and } (T_6, T_7)_1]}{4TS} \\ \frac{T_{1,1b}[(M_4)_1 \text{ and } (T_6, T_7)_1]}{3TS} \end{pmatrix}^1 \rightarrow \begin{pmatrix} \frac{T_{1,2a}[(M_2,\ M_3)_1 \text{ and } (T_6, T_7)_1]}{4TS} \\ \frac{T_{1,2b}[(M_4)_1 \text{ and } (T_6, T_7)_1]}{3TS} \end{pmatrix}^1$$

$$\rightarrow \left(\frac{T_{1,3a}[(M_2,\ M_3, M_4)_1 \text{ and } (T_6, T_7)_1]}{2TS}\right)^1 \rightarrow \begin{pmatrix} \frac{T_{1,4a}[(M_1)_1 \text{ and } (T_2)_1]}{2TS} \\ \frac{T_{1,4b}[(M_2,\ M_3, M_4)_1 \text{ and } (T_2)_1]}{1TS} \end{pmatrix}^1$$

Part #2: $O_{2,1} \rightarrow O_{2,2} \rightarrow O_{2,3}$

$$\begin{pmatrix} \frac{T_{2,1b}[(M_1)_1 \text{ and } (T_1)_1]}{2TS} \\ \frac{T_{2,1a}[(M_2,\ M_3, M_4)_1 \text{ and } (T_1)_1]}{1TS} \end{pmatrix}^1 \rightarrow \left(\frac{T_{2,2a}[(M_2,\ M_3, M_4)_1 \text{ and } (T_{12})_1]}{2TS}\right)^1$$

$$\rightarrow \left(\frac{T_{2,3a}[(M_2,\ M_3, M_4)_1 \text{ and } (T_6, T_7, T_{11})_1]}{2TS}\right)^1$$

Part #3: $O_{3,3} \rightarrow O_{3,1} \rightarrow O_{3,2}$

$$\left(\frac{T_{3,1a}[(M_2,\ M_3, M_4)_1 \text{ and } (T_7, T_8)_1]}{2TS}\right)^1 \rightarrow \left(\frac{T_{3,2a}[(M_2,\ M_3, M_4)_1 \text{ and } (T_7, T_8)_1]}{2TS}\right)^1$$

$$\rightarrow \left(\frac{T_{3,3a}[(M_2,\ M_3, M_4)_1 \text{ and } (T_7, T_8)_1]}{2TS}\right)^1$$

Part #4: $O_{4,2} \rightarrow O_{4,4} \rightarrow O_{4,1} \rightarrow O_{4,3}$

$$\begin{pmatrix} \frac{T_{4,1a}[(M_2)_1 \text{ and } (T_9, T_{10})_1]}{3TS} \\ \frac{T_{4,1b}[(M_3)_1 \text{ and } (T_9, T_{10})_1]}{2TS} \end{pmatrix}^1 \rightarrow \left(\frac{T_{4,2a}[(M_2, M_3)_1 \text{ and } (T_1, T_3)_1]}{3TS}\right)^1 \rightarrow \left(\frac{T_{4,3a}[(M_2, M_3)_1 \text{ and } (T_6, T_9)_1]}{2TS}\right)^1$$

$$\rightarrow \begin{pmatrix} \frac{T_{4,4a}[(M_2)_1 \text{ and } (T_3)_1]}{2TS} \\ \frac{T_{4,4b}[(M_3)_1 \text{ and } (T_3)_1]}{3TS} \end{pmatrix}^1$$

Figure 10. Tasks Information with Simplified Duration Information

Part #1: $O_{1,1} \rightarrow O_{1,2} \rightarrow O_{1,3} \rightarrow O_{1,4}$

$$\left(\begin{array}{c} \frac{T_{1,1a-1}[(M_2,\ M_3)_1 \text{ and } (T_6, T_7)_1]}{1TS} \rightarrow \frac{T_{1,1a-2}[(M_2,\ M_3)_1 \text{ and } (T_6, T_7)_1]}{1TS} \rightarrow \frac{T_{1,1a-3}[(M_2,\ M_3)_1 \text{ and } (T_6, T_7)_1]}{1TS} \rightarrow \frac{T_{1,1a-4}[(M_2,\ M_3)_1 \text{ and } (T_6, T_7)_1]}{1TS} \\ \frac{T_{1,1b-1}[(M_4)_1 \text{ and } (T_6, T_7)_1]}{1TS} \rightarrow \frac{T_{1,1b-2}[(M_4)_1 \text{ and } (T_6, T_7)_1]}{1TS} \rightarrow \frac{T_{1,1b-3}[(M_4)_1 \text{ and } (T_6, T_7)_1]}{1TS} \end{array}\right)^1$$

$$\rightarrow \left(\begin{array}{c} \frac{T_{1,2a-1}[(M_2,\ M_3)_1 \text{ and } (T_6, T_7)_1]}{1TS} \rightarrow \frac{T_{1,2a-2}[(M_2,\ M_3)_1 \text{ and } (T_6, T_7)_1]}{1TS} \rightarrow \frac{T_{1,2a-3}[(M_2,\ M_3)_1 \text{ and } (T_6, T_7)_1]}{1TS} \rightarrow \frac{T_{1,2a-4}[(M_2,\ M_3)_1 \text{ and } (T_6, T_7)_1]}{1TS} \\ \frac{T_{1,2b-1}[(M_4)_1 \text{ and } (T_6, T_7)_1]}{1TS} \rightarrow \frac{T_{1,2b-2}[(M_4)_1 \text{ and } (T_6, T_7)_1]}{1TS} \rightarrow \frac{T_{1,2b-3}[(M_4)_1 \text{ and } (T_6, T_7)_1]}{1TS} \end{array}\right)^1$$

$$\rightarrow \left(\frac{T_{1,3a-1}[(M_2,\ M_3, M_4)_1 \text{ and } (T_6, T_7)_1]}{1TS} \rightarrow \frac{T_{1,3a-2}[(M_2,\ M_3, M_4)_1 \text{ and } (T_6, T_7)_1]}{1TS}\right)^1 \rightarrow \left(\begin{array}{c} \frac{T_{1,4a-1}[(M_1)_1 \text{ and } (T_2)_1]}{1TS} \rightarrow \frac{T_{1,4a-2}[(M_1)_1 \text{ and } (T_2)_1]}{1TS} \\ \frac{T_{1,4b}[(M_2,\ M_3, M_4)_1 \text{ and } (T_2)_1]}{1TS} \end{array}\right)^1$$

Part #2: $O_{2,1} \rightarrow O_{2,2} \rightarrow O_{2,3}$

$$\left(\begin{array}{c} \frac{T_{2,1b-1}[(M_1)_1 \text{ and } (T_1)_1]}{1TS} \rightarrow \frac{T_{2,1b-2}[(M_1)_1 \text{ and } (T_1)_1]}{1TS} \\ \frac{T_{2,1a}[(M_2,\ M_3, M_4)_1 \text{ and } (T_1)_1]}{1TS} \end{array}\right)^1 \rightarrow \left(\frac{T_{2,2a-1}[(M_2,\ M_3, M_4)_1 \text{ and } (T_{12})_1]}{1TS} \rightarrow \frac{T_{2,2a-2}[(M_2,\ M_3, M_4)_1 \text{ and } (T_{12})_1]}{1TS}\right)^1$$

$$\rightarrow \left(\frac{T_{2,3a-1}[(M_2,\ M_3, M_4)_1 \text{ and } (T_5, T_6, T_{11})_1]}{1TS} \rightarrow \frac{T_{2,3a-2}[(M_2,\ M_3, M_4)_1 \text{ and } (T_6, T_7, T_{11})_1]}{1TS}\right)^1$$

Part #3: $O_{3,3} \rightarrow O_{3,1} \rightarrow O_{3,2}$

$$\left(\frac{T_{3,1a-1}[(M_2,\ M_3, M_4)_1 \text{ and } (T_7, T_8)_1]}{1TS} \rightarrow \frac{T_{3,1a-2}[(M_2,\ M_3, M_4)_1 \text{ and } (T_7, T_8)_1]}{1TS}\right)^1$$

$$\rightarrow \left(\frac{T_{3,2a-1}[(M_2,\ M_3, M_4)_1 \text{ and } (T_7, T_8)_1]}{1TS} \rightarrow \frac{T_{3,2a-2}[(M_2,\ M_3, M_4)_1 \text{ and } (T_7, T_8)_1]}{1TS}\right)^1$$

$$\rightarrow \left(\frac{T_{3,3a-1}[(M_2,\ M_3, M_4)_1 \text{ and } (T_7, T_8)_1]}{1TS} \rightarrow \frac{T_{3,3a-2}[(M_2,\ M_3, M_4)_1 \text{ and } (T_7, T_8)_1]}{1TS}\right)^1$$

Part #4: $O_{4,2} \rightarrow O_{4,4} \rightarrow O_{4,1} \rightarrow O_{4,3}$

$$\left(\begin{array}{c} \frac{T_{4,1a-1}[(M_2)_1 \text{ and } (T_9, T_{10})_1]}{1TS} \rightarrow \frac{T_{4,1a-2}[(M_2)_1 \text{ and } (T_9, T_{10})_1]}{1TS} \rightarrow \frac{T_{4,1a-3}[(M_2)_1 \text{ and } (T_9, T_{10})_1]}{1TS} \\ \frac{T_{4,1b-1}[(M_3)_1 \text{ and } (T_9, T_{10})_1]}{1TS} \rightarrow \frac{T_{4,1b-2}[(M_3)_1 \text{ and } (T_9, T_{10})_1]}{1TS} \end{array}\right)^1$$

$$\rightarrow \left(\frac{T_{4,2a-1}[(M_2, M_3)_1 \text{ and } (T_1, T_3)_1]}{1TS} \rightarrow \frac{T_{4,2a-2}[(M_2, M_3)_1 \text{ and } (T_1, T_3)_1]}{1TS} \rightarrow \frac{T_{4,2a-3}[(M_2, M_3)_1 \text{ and } (T_1, T_3)_1]}{1TS}\right)^1$$

$$\rightarrow \left(\frac{T_{4,3a-1}[(M_2, M_3)_1 \text{ and } (T_6, T_9)_1]}{1TS} \rightarrow \frac{T_{4,3a-2}[(M_2, M_3)_1 \text{ and } (T_6, T_9)_1]}{1TS}\right)^1$$

$$\rightarrow \left(\begin{array}{c} \frac{T_{4,4a-1}[(M_2)_1 \text{ and } (T_3)_1]}{1TS} \rightarrow \frac{T_{4,4a} - 2[(M_2)_1 \text{ and } (T_3)_1]}{1TS} \\ \frac{T_{4,4b-1}[(M_3)_1 \text{ and } (T_3)_1]}{1TS} \rightarrow \frac{T_{4,4b-2}[(M_3)_1 \text{ and } (T_3)_1]}{1TS} \rightarrow \frac{T_{4,4b-3}[(M_3)_1 \text{ and } (T_3)_1]}{1TS} \end{array}\right)^1$$

Figure 11. Transformed Tasks Information in Unit Tasks

```
{
'a01':    [[[['J1'],['11']],['1'],['a']],['2'], [[['1'],['M2','M3']],[['1'],['T6','T7']],2]],
'a012':   [[[['J1'],['11']],['2'],['a']],['2'], [[['1'],['M2','M3']],[['1'],['T6','T7']],2]],
'a013':   [[[['J1'],['11']],['3'],['a']],['2'], [[['1'],['M2','M3']],[['1'],['T6','T7']],2]],
'a014':   [[[['J1'],['11']],['4'],['a']],['2'], [[['1'],['M2','M3']],[['1'],['T6','T7']],2]],
'2a01':    [[[['J1'],['11']],['1'],['b']],['2'], [[['1'],['M4']],[['1'],['T6','T7']],2]],
'2a012':   [[[['J1'],['11']],['2'],['b']],['2'], [[['1'],['M4']],[['1'],['T6','T7']],2]],
'2a013':   [[[['J1'],['11']],['3'],['b']],['2'], [[['1'],['M4']],[['1'],['T6','T7']],2]],

'a02':    [[[['J1'],['12']],['1'],['a']],['2'], [[['1'],['M2','M3']],[['1'],['T6','T7']],2]],
'a022':   [[[['J1'],['12']],['2'],['a']],['2'], [[['1'],['M2','M3']],[['1'],['T6','T7']],2]],
'a023':   [[[['J1'],['12']],['3'],['a']],['2'], [[['1'],['M2','M3']],[['1'],['T6','T7']],2]],
'a024':   [[[['J1'],['12']],['4'],['a']],['2'], [[['1'],['M2','M3']],[['1'],['T6','T7']],2]],
'2a02':    [[[['J1'],['12']],['1'],['b']],['2'], [[['1'],['M4']],[['1'],['T6','T7']],2]],
'2a022':   [[[['J1'],['12']],['2'],['b']],['2'], [[['1'],['M4']],[['1'],['T6','T7']],2]],
'2a023':   [[[['J1'],['12']],['3'],['b']],['2'], [[['1'],['M4']],[['1'],['T6','T7']],2]],

'a03':    [[[['J1'],['13']],['1'],['a']],['1'], [[['1'],['M2','M3','M4']],[['1'],['T6','T7']],2]],
'a032':   [[[['J1'],['13']],['2'],['a']],['1'], [[['1'],['M2','M3','M4']],[['1'],['T6','T7']],2]],

'a04':    [[[['J1'],['14']],['1'],['a']],['2'], [[['1'],['M1']],[['1'],['T2']],2]],
'a042':   [[[['J1'],['14']],['2'],['a']],['2'], [[['1'],['M1']],[['1'],['T2']],2]],
'2a04':    [[[['J1'],['14']],['1'],['b']],['2'], [[['1'],['M2','M3','M4']],[['1'],['T2']],2]],

'b01':    [[[['J2'],['21']],['1'],['a']],['2'], [[['1'],['M2','M3','M4']],[['1'],['T1']],2]],
'2b01':    [[[['J2'],['21']],['1'],['b']],['2'], [[['1'],['M1']],[['1'],['T1']],2]],
'2b012':   [[[['J2'],['21']],['2'],['b']],['2'], [[['1'],['M1']],[['1'],['T1']],2]],

'b02':    [[[['J2'],['22']],['1'],['a']],['1'], [[['1'],['M2','M3','M4']],[['1'],['T12']],2]],
'b022':   [[[['J2'],['22']],['2'],['a']],['1'], [[['1'],['M2','M3','M4']],[['1'],['T12']],2]],

'b03':    [[[['J2'],['23']],['1'],['a']],['1'], [[['1'],['M2','M3','M4']],[['1'],['T6','T7','T11']],2]],
'b032':   [[[['J2'],['23']],['2'],['a']],['1'], [[['1'],['M2','M3','M4']],[['1'],['T6','T7','T11']],2]],

'c01':    [[[['J3'],['31']],['1'],['a']],['1'], [[['1'],['M2','M3','M4']],[['1'],['T7','T8']],2]],
'c012':   [[[['J3'],['31']],['2'],['a']],['1'], [[['1'],['M2','M3','M4']],[['1'],['T7','T8']],2]],

'c02':    [[[['J3'],['32']],['1'],['a']],['1'], [[['1'],['M2','M3','M4']],[['1'],['T7','T8']],2]],
'c022':   [[[['J3'],['32']],['2'],['a']],['1'], [[['1'],['M2','M3','M4']],[['1'],['T7','T8']],2]],

'c03':    [[[['J3'],['33']],['1'],['a']],['1'], [[['1'],['M2','M3','M4']],[['1'],['T7','T8']],2]],
'c032':   [[[['J3'],['33']],['2'],['a']],['1'], [[['1'],['M2','M3','M4']],[['1'],['T7','T8']],2]],

'd01':    [[[['J4'],['41']],['1'],['a']],['2'], [[['1'],['M2']],[['1'],['T9','T10']],2]],
'd012':   [[[['J4'],['41']],['2'],['a']],['2'], [[['1'],['M2']],[['1'],['T9','T10']],2]],
'd013':   [[[['J4'],['41']],['3'],['a']],['2'], [[['1'],['M2']],[['1'],['T9','T10']],2]],
'2d01':    [[[['J4'],['41']],['1'],['b']],['2'], [[['1'],['M3']],[['1'],['T9','T10']],2]],
'2d012':   [[[['J4'],['41']],['2'],['b']],['2'], [[['1'],['M3']],[['1'],['T9','T10']],2]],

'd02':    [[[['J4'],['42']],['1'],['a']],['1'], [[['1'],['M2','M3']],[['1'],['T1','T3']],2]],
'd022':   [[[['J4'],['42']],['2'],['a']],['1'], [[['1'],['M2','M3']],[['1'],['T1','T3']],2]],
'd023':   [[[['J4'],['42']],['3'],['a']],['1'], [[['1'],['M2','M3']],[['1'],['T1','T3']],2]],

'd03':    [[[['J4'],['43']],['1'],['a']],['1'], [[['1'],['M2','M3']],[['1'],['T6','T9']],2]],
'd032':   [[[['J4'],['43']],['2'],['a']],['1'], [[['1'],['M2','M3']],[['1'],['T6','T9']],2]],

'd04':    [[[['J4'],['44']],['1'],['b']],['2'], [[['1'],['M3']],[['1'],['T3']],2]],
'd042':   [[[['J4'],['44']],['2'],['b']],['2'], [[['1'],['M3']],[['1'],['T3']],2]],
'd043':   [[[['J4'],['44']],['3'],['b']],['2'], [[['1'],['M3']],[['1'],['T3']],2]],
'2d04':    [[[['J4'],['44']],['1'],['a']],['2'], [[['1'],['M2']],[['1'],['T3']],2]],
'2d042':   [[[['J4'],['44']],['2'],['a']],['2'], [[['1'],['M2']],[['1'],['T3']],2]],
}
```

Figure 12. Scheduling Problem Input Format

Step 2. Generating Nodes and Edges of the Conflicting Graph

A conflicting graph consists of two essentials, the nodes and edges. A node is representing one possible resource combination instance of a Unit Task. And the edges are representing the resource constraints of the instances of the Unit Tasks.

Step 2.1 Generating the Nodes

In order to explain how to generate nodes for the conflicting graph, let us take a Unit Task example from Figure 11, $\mathrm{T}_{2,1b-1}$, which is the first Unit Task in option "b" of the first operation in part #2 production processes. Based on the details, $\frac{\mathrm{T}_{2,1b-1}[(\mathrm{M}_1)_1 \text{ and } (\mathrm{T}_1)_1]}{1TS}$, of this Unit Task, it can be represented by one node, because it only has one possible resource instance, machine $M_1$ and tool $T_1$. On the same idea, all the nodes stand for all the possible resource instance of all the Unit Tasks can be generated for the conflicting graph. The node details of the example problem are shown in the first two columns in Figure 15.

Step 2.2 Generating the Edges

We developed the following four rules for generating edges in the conflicting graph.

(1) For any two nodes from the same Unit Task, they are connected by an edge. It implies the constraint that for each Unit Task, it can only be scheduled once.

(2) For any two nodes from the same operation, if they belong to different task options, they are connected by an edge. It implies the constraint that for each operation, we can only schedule it with only one task option.

(3) For any two nodes from the same operation and the same task option, but different Unit Task, if their resources are not the same, they are connected by an edge. It implies the constraint that once an operation is started, the resources have been selected cannot be changed until it is finished.

(4) For the nodes from different parts, if any of their resources is the same, they are connected by an edge. It implies the resource constraints that one resource can be occupied by only one operation during the same time period.

Besides the rules mentioned above, note that there are no edges between the nodes of two different operations for the same part because they cannot be scheduled in the same time slot, and the selection has no effect on each other. This situation is ensured by the weight assignment strategies, which are discussed in detail in the following sections.

To better illustrate the rules for generating the edges of the conflicting graph, let us take the two operations $O_{2,1} \rightarrow O_{2,2}$ ($T_{2,1} \rightarrow T_{2,2}$) of Part #2 from the example problem plus a given operation $O_{i,1}$ ($T_{i,1}$) of Part #$i$, tasks details are shown as below:

$$\begin{pmatrix} \frac{\mathrm{T}_{2,1a}[(\mathrm{M}_2,\ \mathrm{M}_3, \mathrm{M}_4)_1 \text{ and } (\mathrm{T}_1)_1]}{1TS} \\ \frac{\mathrm{T}_{2,1b-1}[(\mathrm{M}_1)_1 \text{ and } (\mathrm{T}_1)_1]}{1TS} \rightarrow \frac{\mathrm{T}_{2,1b-2}[(\mathrm{M}_1)_1 \text{ and } (\mathrm{T}_1)_1]}{1TS} \end{pmatrix}^1 \rightarrow \left( \frac{\mathrm{T}_{2,2a-1}[(\mathrm{M}_2,\ \mathrm{M}_3, \mathrm{M}_4)_1 \text{ and } (\mathrm{T}_{12})_1]}{1TS} \rightarrow \frac{\mathrm{T}_{2,2a-2}[(\mathrm{M}_2,\ \mathrm{M}_3, \mathrm{M}_4)_1 \text{ and } (\mathrm{T}_{12})_1]}{1TS} \right)^1$$

And

$$\left(\frac{\mathrm{T}_{i,1\mathrm{a}-1}[(\mathrm{M}_4)_1 \text{ and } (\mathrm{T}_6)_1]}{1TS}\right)^1$$

A conflict graph can be constructed, as shown in Figure 13. The colors differentiate the Unit Tasks, and the numbers on edges indicate the rule used while generating the edges.

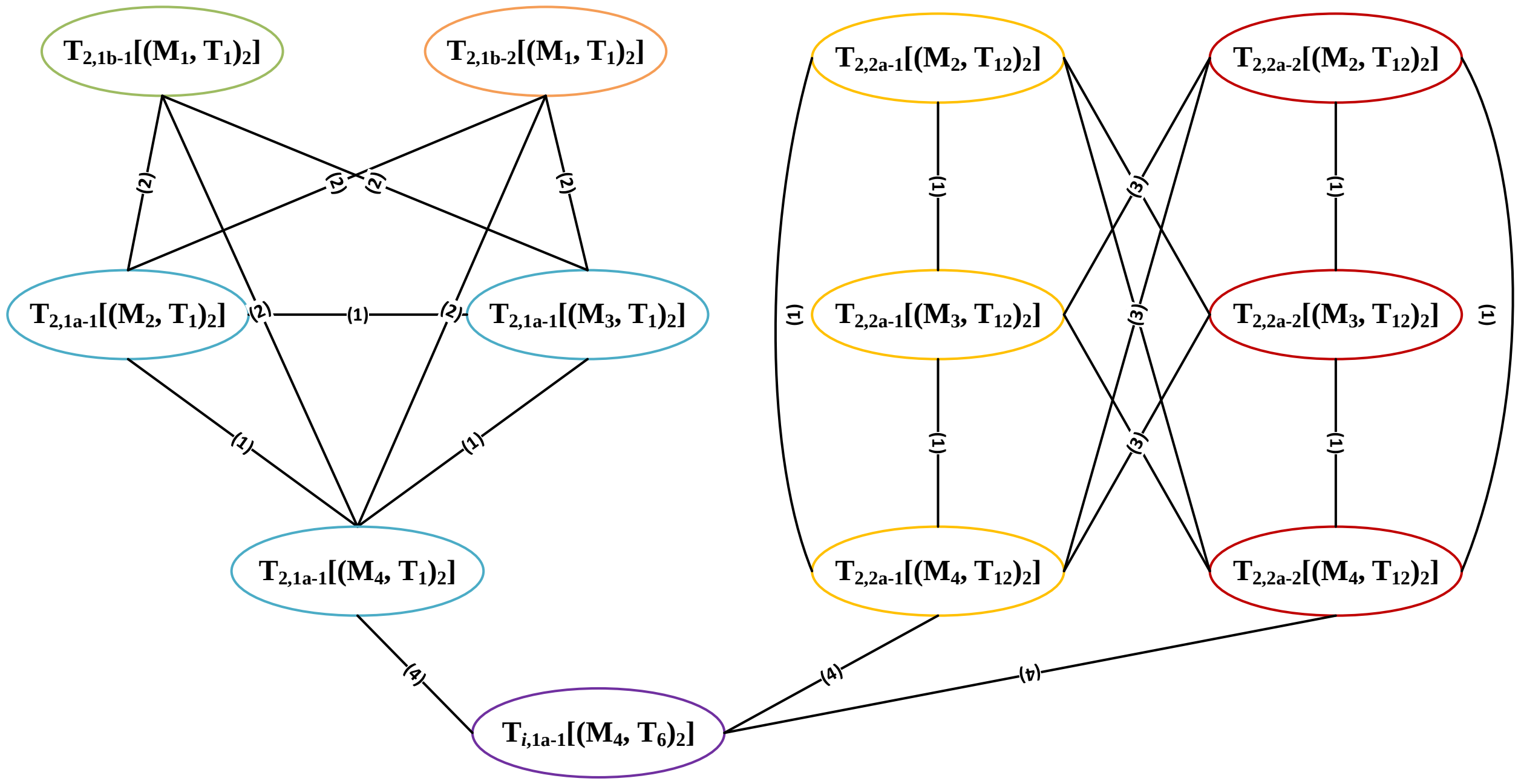

Figure 13. The Conflict Graph of operations $O_{2,1} \rightarrow O_{2,2}$ of Part #2 and the operation $O_{i,1}$ ($T_{i,1}$) of Part #$i$

With the same idea, a conflicting graph for all four parts in the example problem is constructed as Figure 14. The graph has 161 nodes and 4718 edges. The node labels and the connection details of the conflicting graph are shown in Figure 15. For example, the node '0' represents $\mathrm{T}_{1,1\mathrm{a}-1}(M_2, T_6)$, which is one of the resource selections of the Unit Task, $\frac{\mathrm{T}_{1,1\mathrm{a}-1}[(\mathrm{M}_2, \mathrm{M}_3)_1 \text{ and } (\mathrm{T}_6, \mathrm{T}_7)_1]}{1TS}$. Note that the color clusters in Figure 15 are for differentiating different operations. With the conflicting graph ready, in the next section, we explain how we generate weights for Unit Tasks and how we assign weight factors to nodes so that the MWIS algorithms can be configured to schedule the nodes to achieve the objective of minimizing the makespan for the PPS problem.

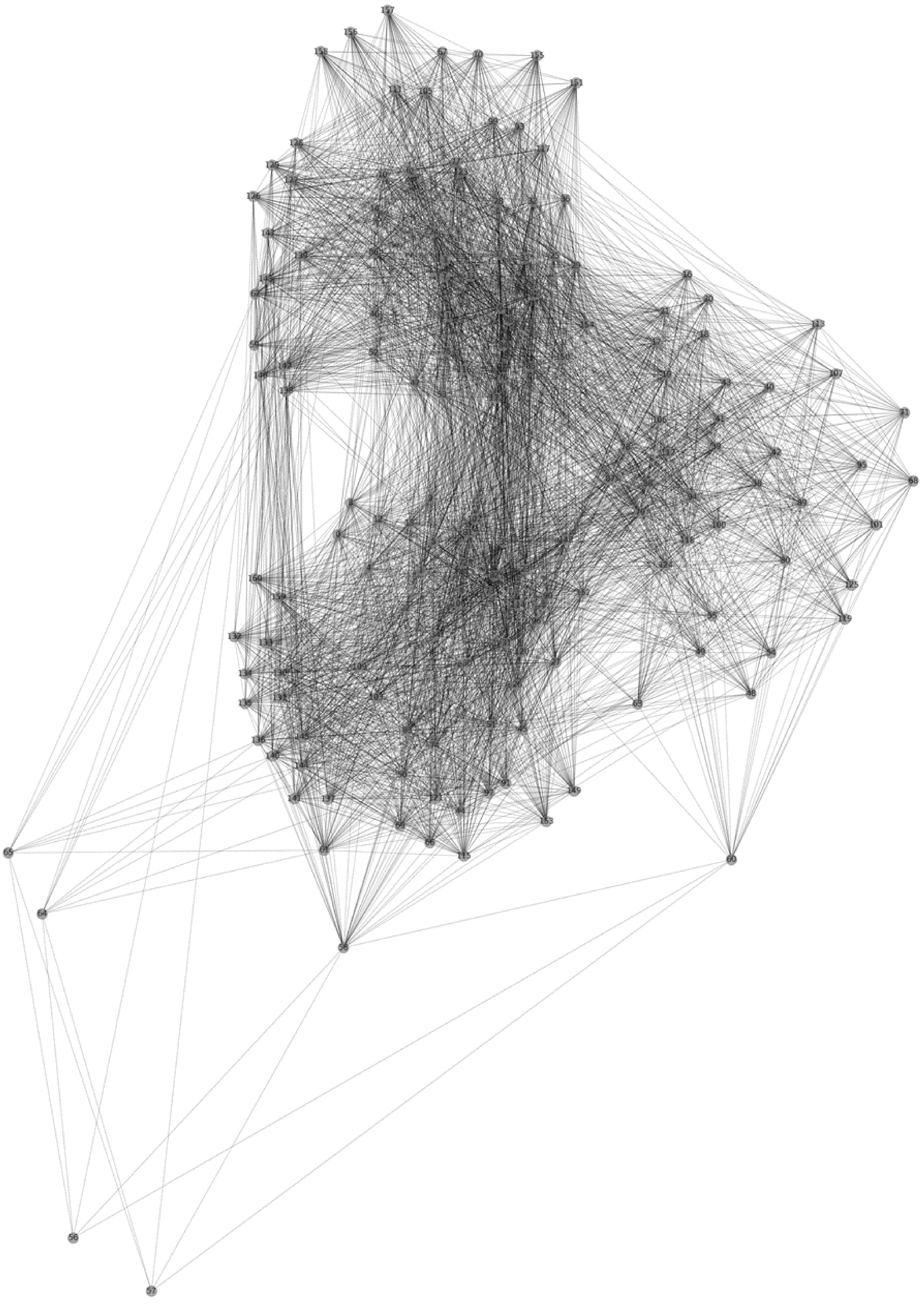

Figure 14. Conflict Graph for the Example Problem

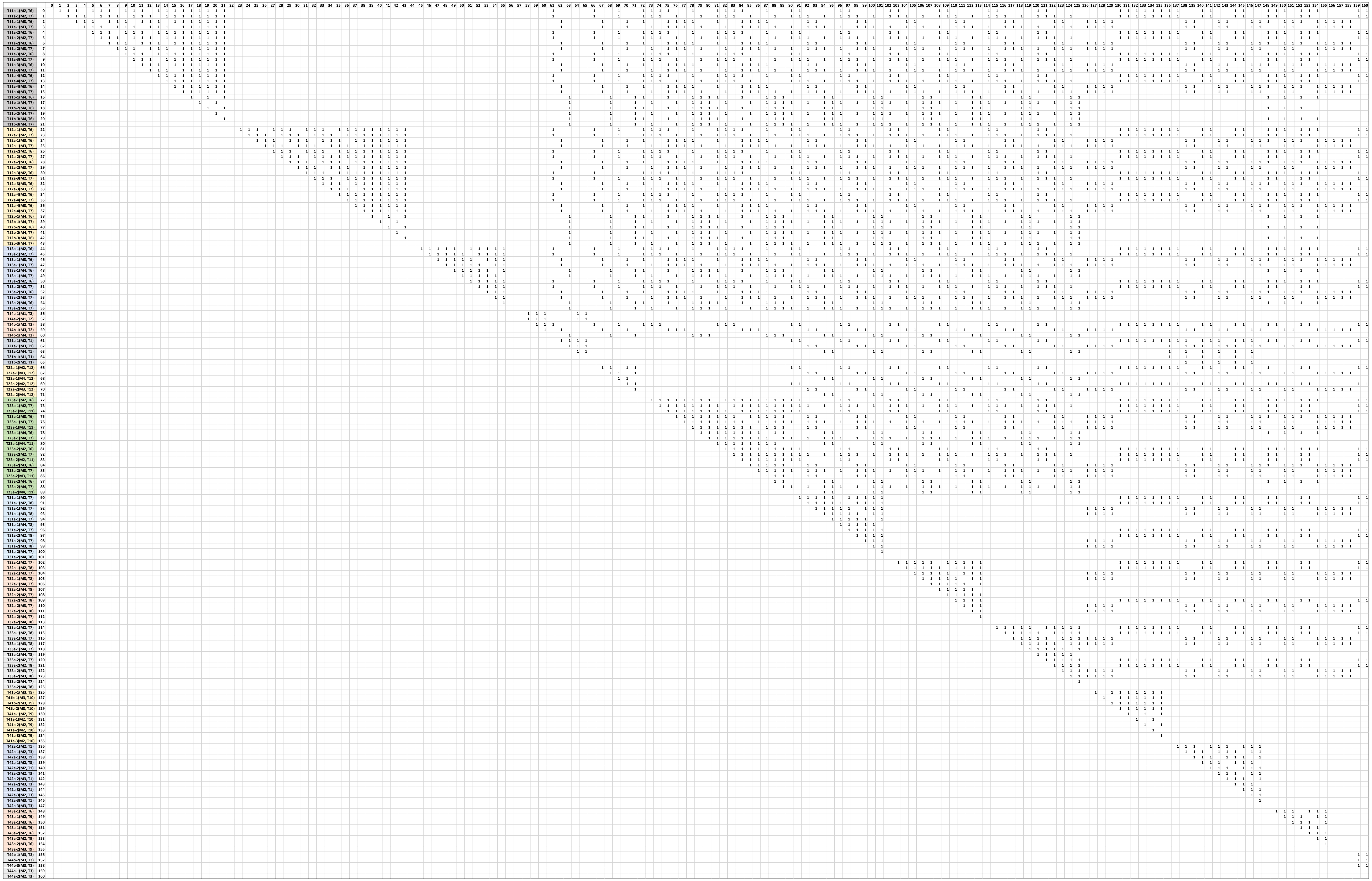

Figure 15. Graph Connection Details for the Example Problem

# 5 Weight Factors Calculation and the Configurations of MWIS Algorithms

With the problem formulated as a conflicting graph, our goal is to find the nodes to schedule for each time slot towards the objective of minimizing the total number of required time slots to finish all the operations. The weight factors assigned to nodes in the conflicting graph are used as the guidance for task and resource selections towards the optimal solution of the PPS problem.

From Figure 10, only the node from the first Unit Task of each option of each part can be scheduled for the current time slot. We name such a Unit Task as a Unit Task Candidate, and the nodes from Unit Task Candidates as Candidate Nodes. The simple idea is that we want to schedule as much as possible Unit Task Candidates at each time slot, and we want to ensure that these scheduled Unit Task Candidates have the most constraints for the rest of Unit Tasks. Because once a Unit Task is scheduled for the current time slot, it is removed from the graph of the following procedures. By doing so, we can remove as many as possible Candidate Nodes at each time slot, and if we can ensure that by removing those nodes, we can remove the most constraints for the remaining Unit Tasks. By discharging the most constraints at each time slot, we have more freedom to schedule more Unit Task Candidates in the following time slots. In this sense, we can achieve the optimal or near-optimal result of the PPS problem. In order to execute this idea, we developed heuristics to generate weights and configure these heuristics with MWIS algorithms proposed in our work (Sun et al., preprint). We are focusing on the weight calculation and the configurations with MWIS algorithms for the PPS problem in Section 5.

## 5.1 *The Weights Calculation*

From the edges generating rules, Candidate Nodes, which are not compatible due to constraints, are connected. In other words, they are not independent. By applying the MWIS algorithms, we can find the most weighted set of independent candidate nodes, which can be scheduled for the current time slot. We assume that the nodes belong to the same Unit Task should have the same weights. Then, the weight of a Unit Task can be determined based on the conflicting condition of this Unit Task among all Unit Tasks remaining. We can calculate the weights for all Unit Tasks remaining and configure the weight factors for Unit Tasks Candidates with different MWIS algorithms.

We define two types of weight to describe the conflicting condition of a Unit Task.

1. The Unit Task connection weight,

$$W_{connection}(T_{p,t}, T_{p',t'})$$

2. The Unit Task length weight,

$$W_{length}(T_{p,t})$$

Where the two Unit Tasks, $T_{p,t}$ and $T_{p',t'}$, belong to two different parts, $P_p$ and $P_{p'}$. $p \neq p'$, $t \in [1, t_{p_max}]$, $t' \in [1, t'_{p'_max}]$, $p$ & $p' \in [1, p_{max}]$, where $t_{p_max}$ is the last Unit Task (the task with the greatest index) in part $P_p$ and $t'_{p'_max}$ is the last Unit Task in part $P_{p'}$. $p_{max}$ is the index of the last part (the part with the greatest index).

The Unit Task connection weight, $W_{connection}(T_{p,t}, T_{p',t'})$, is based on the connection rate between two Unit Tasks $T_{p,t}$ and $T_{p',t'}$ from different parts $P_p$ and $P_{p'}$. Being inspired by the graph density definition. Let $N(T_{p,t})$ be a set of $n(T_{p,t})$ number of nodes from the Unit Task $T_{p,t}$, and $e(T_{p,t}, T_{p',t'})$ is the number of edges between set $N(T_{p,t})$ and set $N(T_{p\prime,t\prime})$. Then, we have the Unit Task connection weights:

$$W_{connection}\left(T_{p,t}, T_{p',t'}\right) = \frac{e(T_{p,t}, T_{p',t'})}{n(T_{p,t}) * n(T_{p\prime,t\prime})}$$

The Unit Task length weight, $W_{length}(T_{p,t})$, is the length weight coefficient, $LW_c$, multiply by the number, $r\left(T_{p,t}\right)$, of remaining time slots needed to finish part $P_p$. Where we have:

$$W_{length}\left(T_{p,t}\right) = LW_c * r\left(T_{p,t}\right)$$

Note that the length weight coefficient, $LW_c$, is used to describe the level priority given to a Unit Task based on the number of time slots remaining for finishing the part individually. Based on our testing, we define three levels of length weight coefficient, median, high and low, as $LW_c^M$, $LW_c^H$ and $LW_c^L$ respectively. And they are defined as follows:

- Let $LW_c^M = 1$, to keep the length weight coefficient in the same scale as the Unit Task connection weight. In this case, the resource constraints and sequencing constraints are considered as equal while selecting nodes.
- Let $LW_c^H = p_{max} + \sum_{p=1}^{p=p_{max}} r\left(T_{p,1}\right)$, which is the total number of time slots of all the parts, to ensure the parts need more remaining time slots are given priority.
- $LW_c^L = 0.01$, to keep the length weight coefficient a minimum effect on node selection. In this case, the resource constraints are more emphasized compare to the sequencing constraints while selecting nodes.

The total weight, $W_{total}\left(T_{p,t}\right)$, of the nodes of a Unit Task, $T_{p,t}$, is the sum of the Unit Task connection weight between itself and all other Unit Tasks of different parts, plus the Unit Task length weight. Formally,

$$W_{total}\left(T_{p,t}\right) = W_{length}\left(T_{p,t}\right) + \sum_{p'=1,t'=1}^{p'=p_{max},t'=t_{p_max}} W_{connection}(T_{p,t}, T_{p',t'})$$

Note that the total weight, $W_{total}\left(T_{p,t}\right)$, as the initial weight value, its purpose is to describe the conflicts that a Unit Task can possibly cause in a PPS problem. The final weight factors need to be configured with the MWIS algorithms for solving the PPS problem. An instance of the weights of the example problem can be calculated as Table 3 below. Each column describes the part ID, operation ID, Unit Tasks, nodes, and the value of the initial weights, respectively. Note that in Table 3, we choose to use the high length weight coefficient, $LW_c^H = 32$.

**Table 3. Unit Tasks and Nodes**

| Part-ID | Op-ID | Unit Tasks | Nodes | Initial Weights |
|---|---|---|---|---|
| **Part 1** | $O_{1,1}$ | $T_{1,1a\text{-}1}$ | 0, 1, 2, 3 | 157.097 |
| | | $T_{1,1a\text{-}2}$ | 4, 5, 6, 7 | 141.097 |
| | | $T_{1,1a\text{-}3}$ | 8, 9, 10, 11 | 125.097 |
| | | $T_{1,1a\text{-}4}$ | 12, 13, 14, 15 | 117.097 |
| | | $T_{1,1b-1}$ | 16, 17 | 154.722 |
| | | $T_{1,1b-2}$ | 18, 19 | 138.722 |
| | | $T_{1,1b-3}$ | 20, 21 | 122.722 |
| | $O_{1,2}$ | $T_{1,2a-1}$ | 22, 23, 24, 25 | 101.097 |
| | | $T_{1,2a-2}$ | 26, 27, 28, 29 | 85.097 |
| | | $T_{1,2a-3}$ | 30, 31, 32, 33 | 69.097 |
| | | $T_{1,2a-4}$ | 34, 35, 36, 37 | 61.097 |
| | | $T_{1,2b-1}$ | 38, 39 | 98.722 |
| | | $T_{1,2b-2}$ | 40, 41 | 82.722 |
| | | $T_{1,2b-3}$ | 42, 43 | 66.722 |
| | $O_{1,3}$ | $T_{1,3a-1}$ | 44, 45, 46, 47, 48, 49 | 88.611 |
| | | $T_{1,3a-2}$ | 50, 51, 52, 53, 54, 55 | 56.611 |
| | $O_{1,4}$ | $T_{1,4a-1}$ | 56 | 8.5 |
| | | $T_{1,4a-2}$ | 57 | 0.5 |
| | | $T_{1,4b-1}$ | 58, 59, 60 | 11.417 |
| **Part 2** | $O_{2,1}$ | $T_{2,1a-1}$ | 61, 62, 63 | 76.750 |
| | | $T_{2,1b-1}$ | 64 | 73.25 |
| | | $T_{2,1b-2}$ | 65 | 65.25 |
| | $O_{2,2}$ | $T_{2,2a-1}$ | 66, 67, 68 | 104.499 |
| | | $T_{2,2a-2}$ | 69, 70, 71 | 72.499 |
| | $O_{2,3}$ | $T_{2,3a-1}$ | 72, 73, 74, 75, 76, 77, 78, 79, 80 | 43.389 |
| | | $T_{2,3a-2}$ | 81, 82, 83, 84, 85, 86, 87, 88, 89 | 11.389 |
| **Part 3** | $O_{3,3}$ | $T_{3,1a-1}$ | 90, 91, 92, 93, 94, 95 | 169.722 |
| | | $T_{3,1a-2}$ | 96, 97, 98, 99, 100, 101 | 137.722 |
| | $O_{3,1}$ | $T_{3,2a-1}$ | 102, 103, 104, 105, 106, 107 | 105.722 |
| | | $T_{3,2a-2}$ | 108, 109, 110, 111, 112, 113 | 73.722 |
| | $O_{3,2}$ | $T_{3,3a-1}$ | 114, 115, 116, 117, 118, 119 | 41.722 |
| | | $T_{3,3a-2}$ | 120, 121, 122, 123, 124, 125 | 9.722 |
| **Part 4** | $O_{4,2}$ | $T_{4,1b-1}$ | 126, 127 | 147.167 |
| | | $T_{4,1b-2}$ | 128, 129 | 131.167 |
| | | $T_{4,1a-1}$ | 130, 131 | 147.167 |
| | | $T_{4,1a-2}$ | 132, 133 | 131.167 |
| | | $T_{4,1a-3}$ | 134, 135 | 123.167 |
| | $O_{4,4}$ | $T_{4,2a-1}$ | 136, 137, 138, 139 | 215.0 |
| | | $T_{4,2a-2}$ | 140, 141, 142, 143 | 183.0 |
| | | $T_{4,2a-3}$ | 144, 145, 146, 147 | 151.0 |
| | $O_{4,1}$ | $T_{4,3a-1}$ | 148, 149, 150, 151 | 120.139 |
| | | $T_{4,3a-2}$ | 152, 153, 154, 155 | 88.139 |
| | $O_{4,3}$ | $T_{4,4b-1}$ | 156 | 27.167 |
| | | $T_{4,4b-2}$ | 157 | 11.167 |
| | | $T_{4,4b-3}$ | 158 | 3.167 |
| | | $T_{4,4a-1}$ | 159 | 27.168 |
| | | $T_{4,4a-2}$ | 160 | 11.167 |

### 5.2 *Weight Factor Arrangements with MWIS Algorithms*

We have calculated the weight factors for the Unit Tasks, and now we explain how to finalize the weight factors with the MWIS algorithms. We developed three weight factor arrangements for the MWIS-based algorithms and seven weight factor arrangements for the AMISL-based algorithms. The weight factor arrangements, together with the MWIS algorithms, make twenty-eight different heuristics configurations for solving the PPS problem.

Before we start to talk about the weight factor arrangements, let us first recall the eight MWIS algorithms from our work (Sun et al., preprint). These algorithms are:

- Algorithm A1 MWIS: the proposed exact MWIS algorithm.
- Algorithm A2 AMISL: the proposed exact AMISL-based MWIS algorithm.
- Algorithm A3 GWMIN: the GWMIN approximation algorithm from literature.
- Algorithm A4 MWIS_CS_GWMIN: it is an algorithm composed of Algorithm A1 and Algorithm A3. This algorithm computes Compare Sets based on the whole induced subgraph at each level using Algorithm A3 GWMIN.
- Algorithm A5 MWIS_SubCS_GWMIN: it is an algorithm composed of Algorithm A1 and Algorithm A3. This algorithm computes Compare Sets based on the induced CSSs, excluding the current removed node, using Algorithm A3 GWMIN.
- Algorithm A6 GWMIN2: the GWMIN2 approximation algorithm from literature.
- Algorithm A7 MWIS_CS_GWMIN2: it is an algorithm composed of Algorithm A1 and Algorithm A6. This algorithm computes Compare Sets based on the whole induced subgraph at each level using Algorithm A6 GWMIN2.
- Algorithm A8 MWIS_SubCS_GWMIN2: it is an algorithm composed of Algorithm A1 and Algorithm A6. This algorithm computes Compare Sets based on the induced CSSs, excluding the current removed node, using Algorithm A6 GWMIN2.

The algorithms list above except Algorithm A2 AMISL are MWIS-based algorithms; they require the weights of all nodes to be positive ($\geq 0$) to make valid comparations in steps so that the final MWIS can be calculated. In this case, the flexibility of weight arrangements is limited, but this is easy to apply approximation strategies to reduce the complexity to speed up the computation. However, Algorithm A2 AMISL first look for all the **M**aximal **I**ndependent **S**ets (MIS), then get the set with the maximum total weight. In this case, the negative and zero weights are allowed. But Algorithm A2 AMISL may have an unreasonable complexity when there is a large number of large size MISs. Algorithm A2 AMISL is also hard to applied approximation strategies. The details of the three weight factor arrangements for the MWIS based algorithms and the seven weight factor arrangements for the AMISL based MWIS algorithms are discussed below. The idea is that while searching for the nodes for the current time slot, the Unit Tasks that can only be scheduled a good number of time slots later may have limited impact. Based on this idea, the wright factor arrangements are created by only checking different limited numbers of steps ahead and aiming to find the best set of the nodes for the current time slot to achieve the objective of minimizing the makespan.

(1) The weight factor arrangements for MWIS based algorithms

For the MWIS based algorithms, we assign weight factors to the Candidate Nodes of Unit Task Candidates according to the three arrangements described below. Then, a small positive value (for instance, 0.0000001) is assigned to the non-candidate nodes. With the weight factors ready, we can apply one of the seven MWIS-based algorithms to find the set of Candidate Nodes with the maximum total weight with the maximum number of nodes. For this setup, the Candidate Nodes associated with the most uncommon resources for the non-candidate nodes are scheduled for the current time slot. So that there are fewer conflicts for the following time slots if the operations scheduled for the current time slot must be continued for more time slots. The Unit Tasks Candidates with the associated resources represented by the set of nodes are scheduled for the current time slot. The three weight factor arrangements are as follows:

(1.1) MWIS A1: MWIS Weights 1

In each time slot, for each $p \in [1, p_{max}]$, assign weight factors to the Unit Task Candidates, $T^C_{p,t_{min}}$, which is the first Unit Task, $T_{p,t_{min}}$, that can be scheduled for part $P_p$. The value of weight factors of the candidate nodes in $T^C_{p,t_{min}}$ is the weight of $T_{p,t_{min}}$, as the equation below,

$$W_{MWIS_A1_condidate}\left(T^C_{p,t_{min}}\right) = W_{total}\left(T_{p,t_{min}}\right)$$

(1.2) MWIS A2: MWIS Weights 2

In each time slot, for each $p \in [1, p_{max}]$, assign weight factors to the Unit Task Candidates, $T^C_{p,t_{min}}$. The value of weight factors of the Candidate Nodes in $T^C_{p,t_{min}}$ is the sum of the weights of $T_{p,t_{min}}$ and $T_{p,(\ t_{min}+1)}$, where $T_{p,(\ t_{min}+1)}$ is the following Unit Task of $T_{p,t_{min}}$, as the equation below,

$$W_{MWIS_A2_condidate}\left(T^C_{p,t_{min}}\right) = W_{total}\left(T_{p,t_{min}}\right) + W_{total}\left(T_{p,t_{min}+1}\right)$$

(1.3) MWIS A3: MWIS Weights 3

In each time slot, for each $p \in [1, p_{max}]$, assign weight factors to the Unit Task Candidates, $T^C_{p,t_{min}}$. The value of weight factors of the Candidate Nodes in $T^C_{p,t_{min}}$ is the sum of the weights of $T_{p,t_{min}}$, $T_{p,(\ t_{min}+1)}$ and $T_{p,(\ t_{min}+2)}$, where $T_{p,(\ t_{min}+2)}$ is the following Unit Task of $T_{p,(\ t_{min}+1)}$, as the equation below,

$$W_{MWIS_A3_condidate}\left(T^C_{p,t_{min}}\right) = W_{total}\left(T_{p,t_{min}}\right) + W_{total}\left(T_{p,t_{min}+1}\right) + W_{total}\left(T_{p,t_{min}+2}\right)$$

(2) The weight factor arrangements for Algorithm A2 AMISL

For the AMISL based algorithms, we assign weight factors to the nodes indicated by the seven different weight factor arrangements described below. Then, a small negative value (for instance, -0.0000001) is assigned to the unaddressed nodes. With the weight factors ready, applied Algorithm A2 AMISL to find the set of Candidate Nodes with the maximum total weight with the minimum number of nodes. For this setup, the Candidate Nodes associated with the most common resources for the unaddressed nodes are scheduled for the current time slot, so that the most constraints are removed for the following time slots by scheduling such a set of Candidate Nodes. The Unit Tasks Candidates with the associated resources represented by the set of nodes are scheduled for the current time slot. The seven weight factor arrangements are as follows:

(2.1) AMISL A1: AMISL Weights 1

In each time slot, for each $p \in [1, p_{max}]$, assign weight factors to the Unit Task Candidates, $T^C_{p,t_{min}}$. The value of weight factors of the Candidate Nodes in $T^C_{p,t_{min}}$ is the weight of $T_{p,t_{min}}$, as the equation below,

$$W_{AMISL_A1_condidate}\left(T^C_{p,t_{min}}\right) = W_{total}\left(T_{p,t_{min}}\right)$$

(2.2) AMISL A2: AMISL Weights 2 Aggregation

In each time slot, for each $p \in [1, p_{max}]$, assign weight factors to the Unit Task Candidates, $T^C_{p,t_{min}}$. The value of weight factors of the Candidate Nodes in $T^C_{p,t_{min}}$ is the sum of the weights of $T_{p,t_{min}}$ and $T_{p,(\ t_{min}+1)}$, where $T_{p,(\ t_{min}+1)}$ is the following Unit Task of $T_{p,t_{min}}$, as the equation below,

$$W_{AMISL_A2_condidate}\left(T^C_{p,t_{min}}\right) = W_{total}\left(T_{p,t_{min}}\right) + W_{total}\left(T_{p,t_{min}+1}\right)$$

(2.3) AMISL A3: AMISL Weights 2 Aggregation + Non-aggregation

In each time slot, for each $p \in [1, p_{max}]$, assign weight factors to the Unit Task Candidates, $T^C_{p,t_{min}}$, and the following Unit Task, $T_{p,(\ t_{min}+1)}$. The value of weight factors of the Candidate Nodes in $T^C_{p,t_{min}}$ is the sum of the weights of $T_{p,t_{min}}$ and $T_{j,(\ t_{min}+1)}$, as the equation below,

$$W_{AMISL_A3_condidate}\left(T^C_{p,t_{min}}\right) = W_{total}\left(T_{p,t_{min}}\right) + W_{total}\left(T_{p,t_{min}+1}\right)$$

The value of weight factors of nodes in $T_{p,t_{min}+1}$ is the weight of $T_{p,t_{min}+1}$, as the equation below,

$$W_{AMISL_A3_following}\left(T_{p,t_{min}+1}\right) = W_{total}\left(T_{p,t_{min}+1}\right)$$

(2.4) AMISL A4: AMISL Weights 2 Non-aggregation

In each time slot, for each $p \in [1, p_{max}]$, assign weight factors to the Unit Task Candidates, $T^C_{p,t_{min}}$, and the following Unit Task, $T_{p,(\ t_{min}+1)}$. The value of weight factors of the Candidate Nodes in $T^C_{p,t_{min}}$ is the weight of $T_{p,t_{min}}$, as the equation below,

$$W_{AMISL_A4_condidate}\left(T^C_{p,t_{min}}\right) = W_{total}\left(T_{p,t_{min}}\right)$$

The value of weight factors of nodes in $T_{p,t_{min}+1}$ is the weight of $T_{p,t_{min}+1}$, as the equation below,

$$W_{AMISL_A4_following}\left(T_{p,t_{min}+1}\right) = W_{total}\left(T_{p,t_{min}+1}\right)$$

(2.5) AMISL A5: AMISL Weights 3 Aggregation

In each time slot, for each $p \in [1, p_{max}]$, assign weight factors to the Unit Task Candidates, $T^C_{p,t_{min}}$. The value of weight factors of the Candidate Nodes in $T^C_{p,t_{min}}$ is the sum of the weights of $T_{p,t_{min}}$, $T_{p,(\ t_{min}+1)}$ and $T_{p,(\ t_{min}+2)}$, as the equation below,

$$W_{AMISL_A5_condidate}\left(T^C_{p,t_{min}}\right) = W_{total}\left(T_{p,t_{min}}\right) + W_{total}\left(T_{p,t_{min}+1}\right) + W_{total}\left(T_{p,t_{min}+2}\right)$$

(2.6) AMISL A6: AMISL Weights 3 Aggregation + Non-aggregation

In each time slot, for each $p \in [1, p_{max}]$, assign weight factors to the Unit Task Candidates, $T^C_{p,t_{min}}$, and the two following Unit Tasks, $T_{p,(\ t_{min}+1)}$ and $T_{p,(\ t_{min}+2)}$. The value of weight factors of the Candidate Nodes in $T^C_{p,t_{min}}$ is the sum of the weights of $T_{p,t_{min}}$, $T_{p,(\ t_{min}+1)}$ and $T_{p,(\ t_{min}+2)}$, as the equation below,

$$W_{MWIS_A6_condidate}\left(T^C_{p,t_{min}}\right) = W_{total}\left(T_{p,t_{min}}\right) + W_{total}\left(T_{p,t_{min}+1}\right) + W_{total}\left(T_{p,t_{min}+2}\right)$$

The value of weight factors of nodes in $T_{p,t_{min}+1}$ is the weight of $T_{p,t_{min}+1}$, as the equation below,

$$W_{AMISL_A6_following}\left(T_{p,t_{min}+1}\right) = W_{total}\left(T_{p,t_{min}+1}\right)$$

The value of weight factors of nodes in $T_{p,t_{min}+2}$ is the weight of $T_{p,t_{min}+2}$, as the equation below,

$$W_{AMISL_A6_following_two}(T_{p,t_{min}+2}) = W_{total}(T_{p,t_{min}+2})$$

(2.7) AMISL A7: AMISL Weights 3 Non-aggregation

In each time slot, for each $p \in [1, p_{max}]$, assign weight factors to the Unit Task Candidates, $T^C_{p,t_{min}}$, and the two following Unit Tasks, $T_{p,(\ t_{min}+1)}$ and $T_{p,(\ t_{min}+2)}$. The value of weight factors of the Candidate Nodes in $T^C_{p,t_{min}}$ is the weights of $T_{p,t_{min}}$, as the equation below,

$$W_{MWIS_A7_condidate}(T^C_{p,t_{min}}) = W_{total}(T_{p,t_{min}})$$

The value of weight factors of nodes in $T_{p,t_{min}+1}$ is the weight of $T_{p,t_{min}+1}$, as the equation below,

$$W_{AMISL_A7_following}(T_{p,t_{min}+1}) = W_{total}(T_{p,t_{min}+1})$$

The value of weight factors of nodes in $T_{p,t_{min}+2}$ is the weight of $T_{p,t_{min}+2}$, as the equation below,

$$W_{AMISL_A7_following_two}(T_{p,t_{min}+2}) = W_{total}(T_{p,t_{min}+2})$$

## 5.3 *Heuristics Configurations*

The eight MWIS algorithms, together with the ten weight arrangements, can be configured into twenty-eight heuristics configurations for solving the PPS problem. The heuristics configurations are shown in Table 4. Each column describes the heuristics configuration ID, algorithm ID, weight arrangement strategies, whether it is an MWIS-based algorithm and whether it is an approximation algorithm, respectively.

**Table 4. Heuristics Configurations**

| **Heuristics-ID** | **Algorithm-ID** | **Weight arrangement strategies** | **MWIS based?** | **Appr?** |
|---|---|---|---|---|
| **1** | A1 MWIS | MWIS A1: MWIS Weights 1 | Yes | No |
| **2** | A1 MWIS | MWIS A2: MWIS Weights 2 | Yes | No |
| **3** | A1 MWIS | MWIS A3: MWIS Weights 3 | Yes | No |
| **4** | A2 AMISL | AMISL A1: AMISL Weights 1 | No | No |
| **5** | A2 AMISL | AMISL A2: AMISL Weights 2 Agg | No | No |
| **6** | A2 AMISL | AMISL A3: AMISL Weights 2 Agg + Nagg | No | No |
| **7** | A2 AMISL | AMISL A4: AMISL Weights 2 Nagg | No | No |
| **8** | A2 AMISL | AMISL A5: AMISL Weights 3 Agg | No | No |
| **9** | A2 AMISL | AMISL A6: AMISL Weights 3 Agg + Nagg | No | No |
| **10** | A2 AMISL | AMISL A7: AMISL Weights 3 Nagg | No | No |
| **11** | A3 GWMIN | MWIS A1: MWIS Weights 1 | Yes | Yes |
| **12** | A4 MWIS_CS_GWMIN | MWIS A1: MWIS Weights 1 | Yes | Yes |
| **13** | A5 MWIS_SubCS_GWMIN | MWIS A1: MWIS Weights 1 | Yes | Yes |
| **14** | A3 GWMIN | MWIS A2: MWIS Weights 2 | Yes | Yes |
| **15** | A4 MWIS_CS_GWMIN | MWIS A2: MWIS Weights 2 | Yes | Yes |
| **16** | A5 MWIS_SubCS_GWMIN | MWIS A2: MWIS Weights 2 | Yes | Yes |
| **17** | A3 GWMIN | MWIS A3: MWIS Weights 3 | Yes | Yes |
| **18** | A4 MWIS_CS_GWMIN | MWIS A3: MWIS Weights 3 | Yes | Yes |
| **19** | A5 MWIS_SubCS_GWMIN | MWIS A3: MWIS Weights 3 | Yes | Yes |
| **20** | A6 GWMIN2 | MWIS A1: MWIS Weights 1 | Yes | Yes |
| **21** | A7 MWIS_CS_GWMIN2 | MWIS A1: MWIS Weights 1 | Yes | Yes |
| **22** | A8 MWIS_SubCS_GWMIN2 | MWIS A1: MWIS Weights 1 | Yes | Yes |
| **23** | A6 GWMIN2 | MWIS A2: MWIS Weights 2 | Yes | Yes |
| **24** | A7 MWIS_CS_GWMIN2 | MWIS A2: MWIS Weights 2 | Yes | Yes |
| **25** | A8 MWIS_SubCS_GWMIN2 | MWIS A2: MWIS Weights 2 | Yes | Yes |
| **26** | A6 GWMIN2 | MWIS A3: MWIS Weights 3 | Yes | Yes |
| **27** | A7 MWIS_CS_GWMIN2 | MWIS A3: MWIS Weights 3 | Yes | Yes |
| **28** | A8 MWIS_SubCS_GWMIN2 | MWIS A3: MWIS Weights 3 | Yes | Yes |

# 6 Solving the Example Problem via the Proposed Approach

In this section, we summarize the proposed method for solving the PPS problem with the example PPS problem, as shown in Figure 2 and Table 1. The major steps of the proposed approach are described below:

Step #1: Prepare the input information.

In step one, we reformat the operation information with the best practice operation sequence and simplify the problem by breaking down the processing time into time slots. The operation information of the four parts in the example problem is reformatted as Figure 9 based on operation sequencing constraints shown as the top part of Figure 2. Figure 10 can be transformed based on Figure 9 by breaking down the operations into Unit Tasks. Here, the processing time for each Unit Task is one time slot, which stands for 10 time units.

Step #2: Generate the conflicting graph for the PPS problem.

In step two, the nodes for the conflict graph is generated based on the different possible resource selections for each Unit Task. And the edges of the conflicting graph are generated based on the constraints. Figure 14 is the conflicting graph for the example problem, which has 4718 edges and 161 nodes, and Figure 15 shows the details of the edges.

Step #3: Based on the selected heuristics configuration, arrange weight factors and compute the MWIS.

In step three, we select Heuristics #13, which assigns weight factors as MWIS A1: MWIS Weights 1 and uses Algorithm A5 MWIS_SubCS_GWMIN to compute the MWIS for the nodes to schedule for the current time slot. Note that we choose to use the high length weight coefficient, $LW_c^H$, which $LW_c^H = 32$ for the example problem. The final weight factors at the first time slot for the Unit Task Candidates and Candidate Nodes of the example problem are shown in Table 5. Each column describes the part ID, operation ID, Unit Tasks, nodes of the Unit Tasks and the Final weight factors, respectively. The MWIS found by Heuristics #13 is the node set, ['0', '4', '8', '12', '22', '26', '30', '34', '44', '50', '139', '143', '147', '126', '128', '151', '155', '156', '157', '158', '58', '95', '101', '107', '113', '119', '125', '64', '65']. It means that the Unit Task Candidates with their resources, $\mathrm{T}_{1,1a-1}[(\mathrm{M}_2)_1 \text{ and } (\mathrm{T}_6)_1]$, $\mathrm{T}_{2,1b-1}[(\mathrm{M}_1)_1 \text{ and } (\mathrm{T}_1)_1]$, $\mathrm{T}_{3,1a-1}[(\mathrm{M}_4)_1 \text{ and } (\mathrm{T}_8)_1]$ and $\mathrm{T}_{4,1b-1}[(\mathrm{M}_3)_1 \text{ and } (\mathrm{T}_9)_1]$, are scheduled for the current time slot.

Step #4: Update the remaining Unit Tasks and the conflicting graph

In step four, remove the Unit Tasks that have been scheduled and remove the Unit Tasks that cannot be scheduled because of the constraints that no changing resources is allowed before an operation is finished. Then, update the conflicting graph and the weight factors. Figure 16 is the updated task information for the remaining Unit Tasks. And the updated remaining conflicting graph, the node labels, and edge connection details for the following time slot are shown as Figure 17 and Figure 18, respectively.

**Table 5. Final Weight Factors Unit Task Candidates and Candidate Nodes via Heuristics #13 on the Example Problem**

| Part-ID | Op-ID | Unit Tasks | Nodes | Final Weights |
|---|---|---|---|---|
| **Part 1** | $O_{1,1}$ | $T_{1,1a-1}$ | 0, 1, 2, 3 | 157.097 |
| | | $T_{1,1a-2}$ | 4, 5, 6, 7 | 141.097 |
| | | $T_{1,1a-3}$ | 8, 9, 10, 11 | 125.097 |
| | | $T_{1,1a-4}$ | 12, 13, 14, 15 | 117.097 |
| | | $T_{1,1b-1}$ | 16, 17 | 154.722 |
| | | $T_{1,1b-2}$ | 18, 19 | 138.722 |
| | | $T_{1,1b-3}$ | 20, 21 | 122.722 |
| **Part 2** | $O_{2,1}$ | $T_{2,1a-1}$ | 61, 62, 63 | 76.750 |
| | | $T_{2,1b-1}$ | 64 | 73.25 |
| | | $T_{2,1b-2}$ | 65 | 65.25 |
| **Part 3** | $O_{3,3}$ | $T_{3,1a-1}$ | 90, 91, 92, 93, 94, 95 | 169.722 |
| | | $T_{3,1a-2}$ | 96, 97, 98, 99, 100, 101 | 137.722 |
| **Part 4** | $O_{4,2}$ | $T_{4,1b-1}$ | 126, 127 | 147.167 |
| | | $T_{4,1b-2}$ | 128, 129 | 131.167 |
| | | $T_{4,1a-1}$ | 130, 131 | 147.167 |
| | | $T_{4,1a-2}$ | 132, 133 | 131.167 |
| | | $T_{4,1a-3}$ | 134, 135 | 123.167 |

Step #5: Checking the ending condition

In step five, we need to make a judgment. If there is at least one remaining Unit Task, go to step #3. If there is no remaining Unit Task, the PPS problem computation is finished, and the output schedule is the combination of Unit Task Candidates and Candidate Nodes found at each time slot.

The results of the example problem with Heuristics #13 is illustrated in Figure 19. Our approach can get to a near-optimal solution finishing in 107.5 time unit compare to the optimum solution finishing in 98 time units, which is a 9.69% error. The computation of our approach takes about 20 seconds, which is much faster (seconds versus days) compare to the optimum solutions using integer programming.

Table 6 shows the performance of our approach in terms of accuracy and computation time. Each column describes the heuristics ID, the minimum makespan in a number of time slots, average clock time in 3-run, whether it is an approximation algorithm, error in a number of time slots, and error rate, respectively. The accuracy is the error rate by comparing the result of my approach with the optimal solution, and computation time is the clock time taken to finish the computation. For the example problem, we can get a near-optimal in seconds with a minimum 10% error but much faster. In the following section, we test our approach on a real-world example from the literature, and further test cases are designed to exam the accuracy, robustness, and scalability of our approach.

Job #1: $O_{11} \rightarrow O_{12} \rightarrow O_{13} \rightarrow O_{14}$

$$\left(\frac{\mathrm{T}_{11a-2}[(\mathrm{M}_2,\ \mathrm{M}_3)_1 \text{ and } (\mathrm{T}_6,\mathrm{T}_7)_1]}{1TS} \rightarrow \frac{\mathrm{T}_{11a-3}[(\mathrm{M}_2,\ \mathrm{M}_3)_1 \text{ and } (\mathrm{T}_6,\mathrm{T}_7)_1]}{1TS} \rightarrow \frac{\mathrm{T}_{11a-4}[(\mathrm{M}_2,\ \mathrm{M}_3)_1 \text{ and } (\mathrm{T}_6,\mathrm{T}_7)_1]}{1TS}\right)^1$$

$$\rightarrow \left(\begin{matrix} \frac{\mathrm{T}_{12a-1}[(\mathrm{M}_2,\ \mathrm{M}_3)_1 \text{ and } (\mathrm{T}_6,\mathrm{T}_7)_1]}{1TS} \rightarrow \frac{\mathrm{T}_{12a-2}[(\mathrm{M}_2,\ \mathrm{M}_3)_1 \text{ and } (\mathrm{T}_6,\mathrm{T}_7)_1]}{1TS} \rightarrow \frac{\mathrm{T}_{12a-3}[(\mathrm{M}_2,\ \mathrm{M}_3)_1 \text{ and } (\mathrm{T}_6,\mathrm{T}_7)_1]}{1TS} \rightarrow \frac{\mathrm{T}_{12a-4}[(\mathrm{M}_2,\ \mathrm{M}_3)_1 \text{ and } (\mathrm{T}_6,\mathrm{T}_7)_1]}{1TS} \\ \frac{\mathrm{T}_{12b-1}[(\mathrm{M}_4)_1 \text{ and } (\mathrm{T}_6,\mathrm{T}_7)_1]}{1TS} \rightarrow \frac{\mathrm{T}_{12b-2}[(\mathrm{M}_4)_1 \text{ and } (\mathrm{T}_6,\mathrm{T}_7)_1]}{1TS} \rightarrow \frac{\mathrm{T}_{12b-3}[(\mathrm{M}_4)_1 \text{ and } (\mathrm{T}_6,\mathrm{T}_7)_1]}{1TS} \end{matrix}\right)^1$$

$$\rightarrow \left(\frac{\mathrm{T}_{13a-1}[(\mathrm{M}_2,\ \mathrm{M}_3,\mathrm{M}_4)_1 \text{ and } (\mathrm{T}_6,\mathrm{T}_7)_1]}{1TS} \rightarrow \frac{\mathrm{T}_{13a-2}[(\mathrm{M}_2,\ \mathrm{M}_3,\mathrm{M}_4)_1 \text{ and } (\mathrm{T}_6,\mathrm{T}_7)_1]}{1TS}\right)^1 \rightarrow \left(\begin{matrix} \frac{\mathrm{T}_{14a-1}[(\mathrm{M}_1)_1 \text{ and } (\mathrm{T}_2)_1]}{1TS} \rightarrow \frac{\mathrm{T}_{14a-2}[(\mathrm{M}_1)_1 \text{ and } (\mathrm{T}_2)_1]}{1TS} \\ \frac{\mathrm{T}_{14b}[(\mathrm{M}_2,\ \mathrm{M}_3,\mathrm{M}_4)_1 \text{ and } (\mathrm{T}_2)_1]}{1TS} \end{matrix}\right)^1$$

Job #2: $O_{21} \rightarrow O_{22} \rightarrow O_{23}$

$$(\frac{\mathrm{T}_{21b-2}[(\mathrm{M}_1)_1 \text{ and } (\mathrm{T}_1)_1]}{1TS})^1 \rightarrow \left(\frac{\mathrm{T}_{22a-1}[(\mathrm{M}_2,\ \mathrm{M}_3,\mathrm{M}_4)_1 \text{ and } (\mathrm{T}_{12})_1]}{1TS} \rightarrow \frac{\mathrm{T}_{22a-2}[(\mathrm{M}_2,\ \mathrm{M}_3,\mathrm{M}_4)_1 \text{ and } (\mathrm{T}_{12})_1]}{1TS}\right)^1$$

$$\rightarrow \left(\frac{\mathrm{T}_{23a-1}[(\mathrm{M}_2,\ \mathrm{M}_3,\mathrm{M}_4)_1 \text{ and } (\mathrm{T}_5,\mathrm{T}_6,\mathrm{T}_{11})_1]}{1TS} \rightarrow \frac{\mathrm{T}_{23a-2}[(\mathrm{M}_2,\ \mathrm{M}_3,\mathrm{M}_4)_1 \text{ and } (\mathrm{T}_6,\mathrm{T}_7,\mathrm{T}_{11})_1]}{1TS}\right)^1$$

Job #3: $O_{33} \rightarrow O_{31} \rightarrow O_{32}$

$$\left(\frac{\mathrm{T}_{31a-2}[(\mathrm{M}_2,\ \mathrm{M}_3,\mathrm{M}_4)_1 \text{ and } (\mathrm{T}_7,\mathrm{T}_8)_1]}{1TS}\right)^1 \rightarrow \left(\frac{\mathrm{T}_{32a-1}[(\mathrm{M}_2,\ \mathrm{M}_3,\mathrm{M}_4)_1 \text{ and } (\mathrm{T}_7,\mathrm{T}_8)_1]}{1TS} \rightarrow \frac{\mathrm{T}_{32a-2}[(\mathrm{M}_2,\ \mathrm{M}_3,\mathrm{M}_4)_1 \text{ and } (\mathrm{T}_7,\mathrm{T}_8)_1]}{1TS}\right)^1$$

$$\rightarrow \left(\frac{\mathrm{T}_{33a-1}[(\mathrm{M}_2,\ \mathrm{M}_3,\mathrm{M}_4)_1 \text{ and } (\mathrm{T}_7,\mathrm{T}_8)_1]}{1TS} \rightarrow \frac{\mathrm{T}_{33a-2}[(\mathrm{M}_2,\ \mathrm{M}_3,\mathrm{M}_4)_1 \text{ and } (\mathrm{T}_7,\mathrm{T}_8)_1]}{1TS}\right)^1$$

Job #4: $O_{42} \rightarrow O_{44} \rightarrow O_{41} \rightarrow O_{43}$

$$\left(\frac{\mathrm{T}_{41b-2}[(\mathrm{M}_3)_1 \text{ and } (\mathrm{T}_9,\mathrm{T}_{10})_1]}{1TS}\right)^1 \rightarrow \left(\frac{\mathrm{T}_{42a-1}[(\mathrm{M}_2,\mathrm{M}_3)_1 \text{ and } (\mathrm{T}_1,\mathrm{T}_3)_1]}{1TS} \rightarrow \frac{\mathrm{T}_{42a-2}[(\mathrm{M}_2,\mathrm{M}_3)_1 \text{ and } (\mathrm{T}_1,\mathrm{T}_3)_1]}{1TS} \rightarrow \frac{\mathrm{T}_{42a-3}[(\mathrm{M}_2,\mathrm{M}_3)_1 \text{ and } (\mathrm{T}_1,\mathrm{T}_3)_1]}{1TS}\right)^1$$

$$\rightarrow \left(\frac{\mathrm{T}_{43a-1}[(\mathrm{M}_2,\mathrm{M}_3)_1 \text{ and } (\mathrm{T}_6,\mathrm{T}_9)_1]}{1TS} \rightarrow \frac{\mathrm{T}_{43a-2}[(\mathrm{M}_2,\mathrm{M}_3)_1 \text{ and } (\mathrm{T}_6,\mathrm{T}_9)_1]}{1TS}\right)^1$$

$$\rightarrow \left(\begin{matrix} \frac{\mathrm{T}_{44a-1}[(\mathrm{M}_2)_1 \text{ and } (\mathrm{T}_3)_1]}{1TS} \rightarrow \frac{\mathrm{T}_{44a} - 2[(\mathrm{M}_2)_1 \text{ and } (\mathrm{T}_3)_1]}{1TS} \\ \frac{\mathrm{T}_{44b-1}[(\mathrm{M}_3)_1 \text{ and } (\mathrm{T}_3)_1]}{1TS} \rightarrow \frac{\mathrm{T}_{44b-2}[(\mathrm{M}_3)_1 \text{ and } (\mathrm{T}_3)_1]}{1TS} \rightarrow \frac{\mathrm{T}_{44b-3}[(\mathrm{M}_3)_1 \text{ and } (\mathrm{T}_3)_1]}{1TS} \end{matrix}\right)^1$$

Figure 16. Updated Remaining Tasks Information for the Following Time Slot

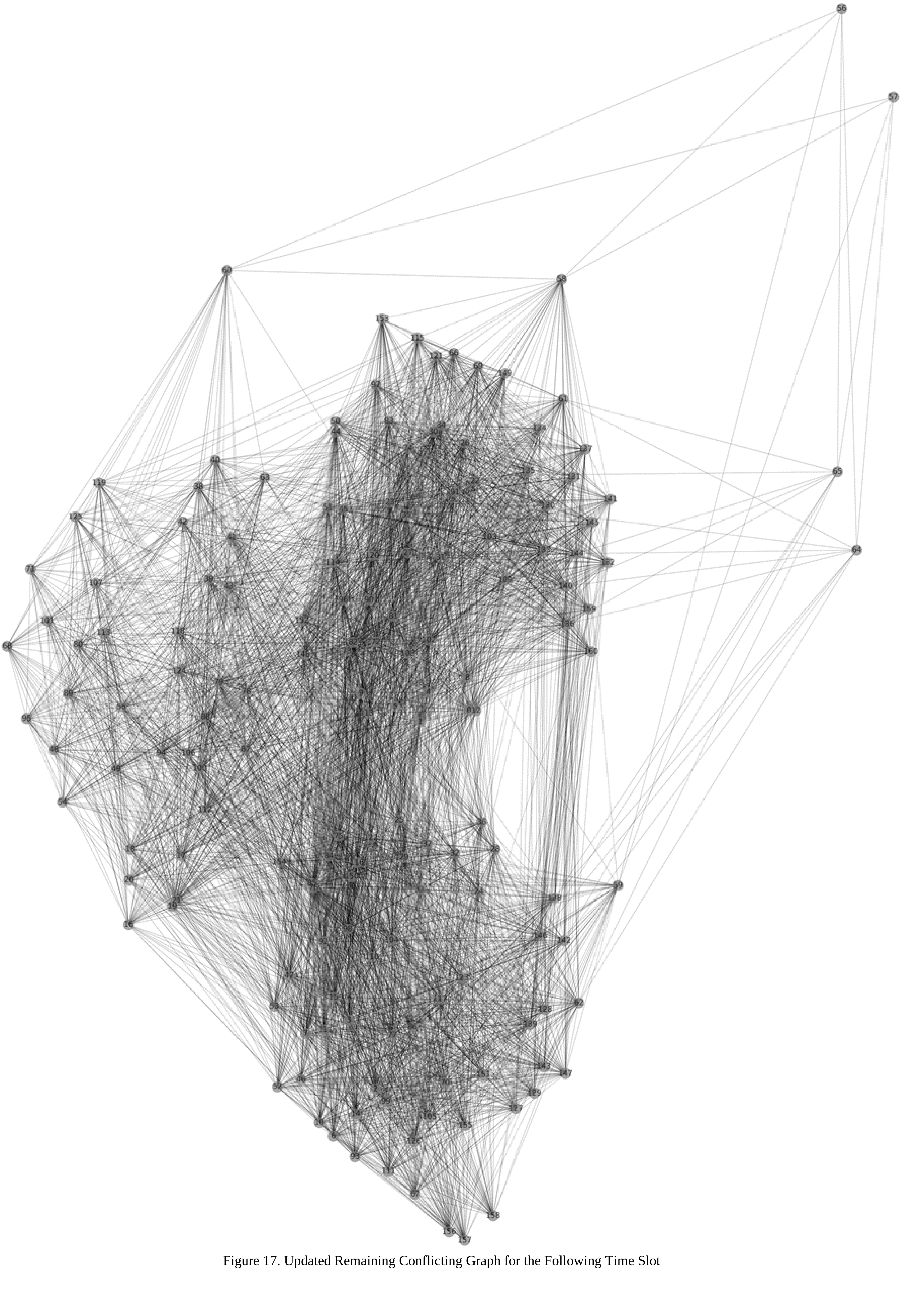

Figure 17. Updated Remaining Conflicting Graph for the Following Time Slot

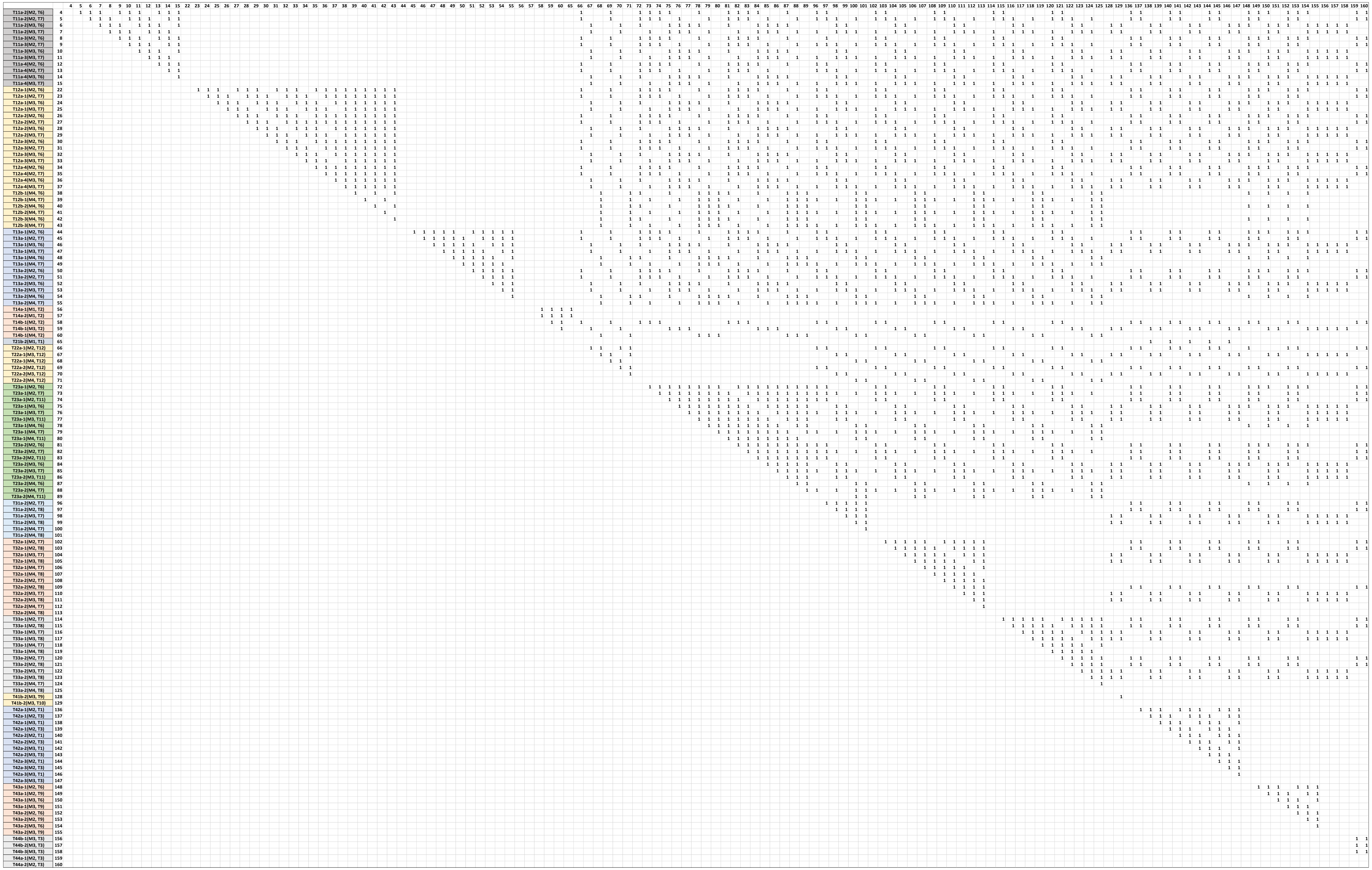

Figure 18. Updated Remaining Edge Connection Details for the Following Time Slot

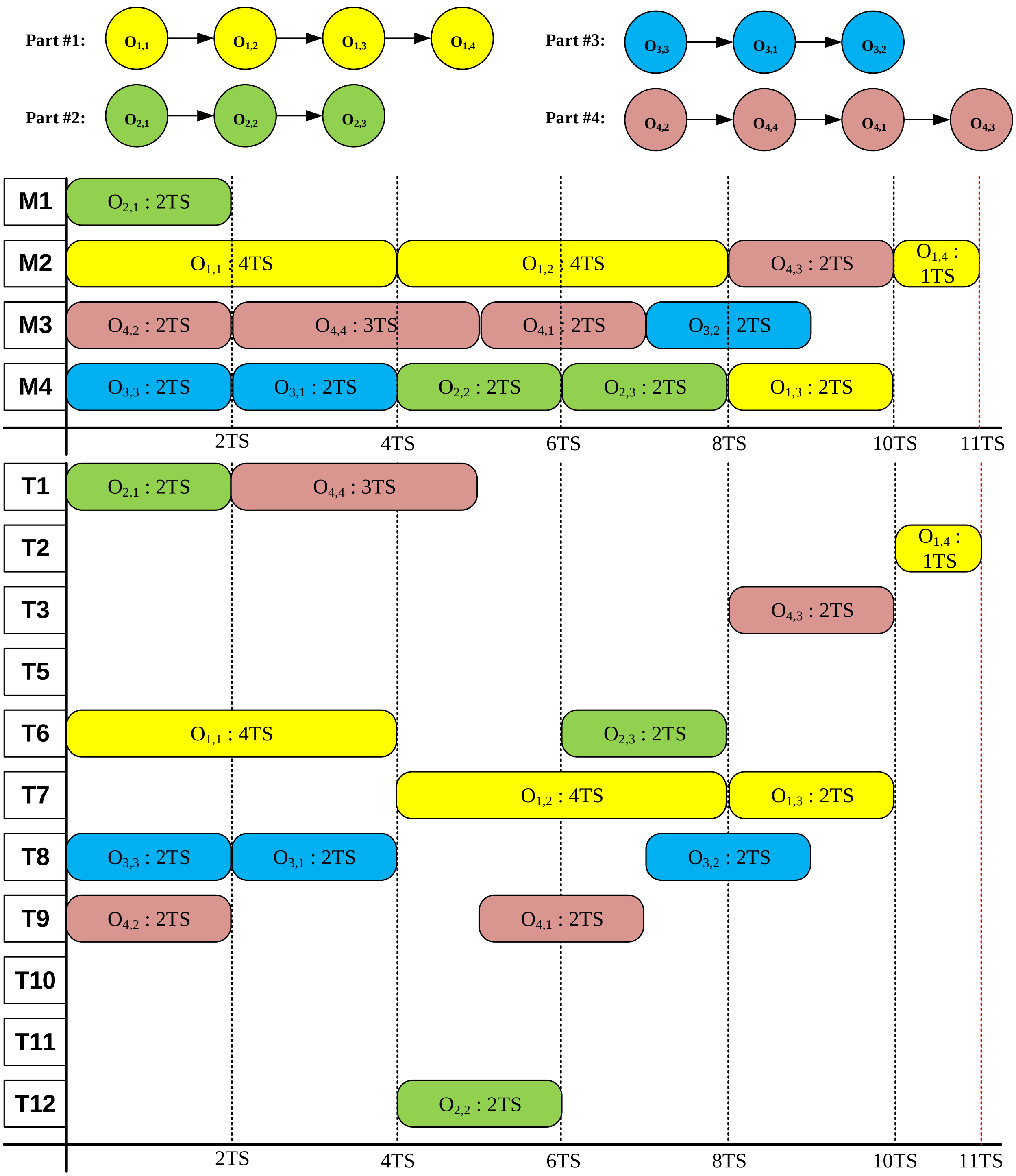

Figure 19. Schedule Created with Heuristics #13

**Table 6. Outputs of the Heuristics Configurations on the Example PPS Problem**

| Methods | Minimum makespan (in time slots) | Clock time, 3-run average | Is approx? | Error | Error rate |
|---|---|---|---|---|---|
| **Heuristics#1** | 11 | 11768.55 | No | 1 | 10% |
| **Heuristics#2** | 11 | 14392.62 | No | 1 | 10% |
| **Heuristics#3** | 11 | 13370.07 | No | 1 | 10% |
| **Heuristics#4** | 11 | 13344.73 | No | 1 | 10% |
| **Heuristics#5** | 11 | 10904.36 | No | 1 | 10% |
| **Heuristics#6** | 11 | 12042.11 | No | 1 | 10% |
| **Heuristics#7** | 14 | 11942.88 | No | 4 | 40% |
| **Heuristics#8** | 11 | 10178.87 | No | 1 | 10% |
| **Heuristics#9** | 11 | 10496.09 | No | 1 | 10% |
| **Heuristics#10** | 11 | 10833.84 | No | 1 | 10% |
| **Heuristics#11** | 12 | 8.75 | Yes | 2 | 20% |
| **Heuristics#12** | 11 | 38.73 | Yes | 1 | 10% |
| **Heuristics#13** | 11 | 26.02 | Yes | 1 | 10% |
| **Heuristics#14** | 11 | 9.23 | Yes | 1 | 10% |
| **Heuristics#15** | 11 | 39.22 | Yes | 1 | 10% |
| **Heuristics#16** | 11 | 28.09 | Yes | 1 | 10% |
| **Heuristics#17** | 11 | 9.16 | Yes | 1 | 10% |
| **Heuristics#18** | 11 | 37.66 | Yes | 1 | 10% |
| **Heuristics#19** | 11 | 26.74 | Yes | 1 | 10% |
| **Heuristics#20** | 14 | 9.99 | Yes | 4 | 40% |
| **Heuristics#21** | 11 | 43.34 | Yes | 1 | 10% |
| **Heuristics#22** | 11 | 27.02 | Yes | 1 | 10% |
| **Heuristics#23** | 14 | 9.78 | Yes | 4 | 40% |
| **Heuristics#24** | 11 | 41.62 | Yes | 1 | 10% |
| **Heuristics#25** | 11 | 27.59 | Yes | 1 | 10% |
| **Heuristics#26** | 14 | 9.91 | Yes | 4 | 40% |
| **Heuristics#27** | 11 | 40.42 | Yes | 1 | 10% |
| **Heuristics#28** | 11 | 28.64 | Yes | 1 | 10% |

# 7 A Real-world Example Using the Proposed Approach

Based on the case study from Zhang et al.'s work (Zhang et al., 2014) and combined with the details from Zhang et al.'s references (Chu & Gadh, 1996; Zhang et al., 2003; Li et al., 2005; Li & McMahon, 2007), we constructed a real-world PPS problem to verify our approach. The resources, machines, and cutting tools of a specific job shop are defined in Table 7. The four parts of the problem are shown in Figure 20. The relevant technical specifications of the four parts are defined in Tables 8 to 11.

**Table 7. The Resource of a Job Shop – Machines and Tools**

| Types | No. |
|---|---|
| **Machines** | |
| **Drilling press** | $M_1$ |
| **Three-axis vertical milling machine I** | $M_2$ |
| **Three-axis vertical milling machine II** | $M_3$ |
| **CNC three-axis vertical milling machine** | $M_4$ |
| **Boring machine** | $M_5$ |

| Cutting tools | |
|---|---|
| **Drill 1** | $T_1$ |
| **Drill 2** | $T_2$ |
| **Drill 3** | $T_3$ |
| **Drill 4** | $T_4$ |
| **Tapping tool** | $T_5$ |
| **Mill 1** | $T_6$ |
| **Mill 2** | $T_7$ |
| **Mill 3** | $T_8$ |
| **Reaming tool** | $T_9$ |
| **Boring tool** | $T_{10}$ |
| **Slot cutter** | $T_{11}$ |
| **Chamfer tool** | $T_{12}$ |

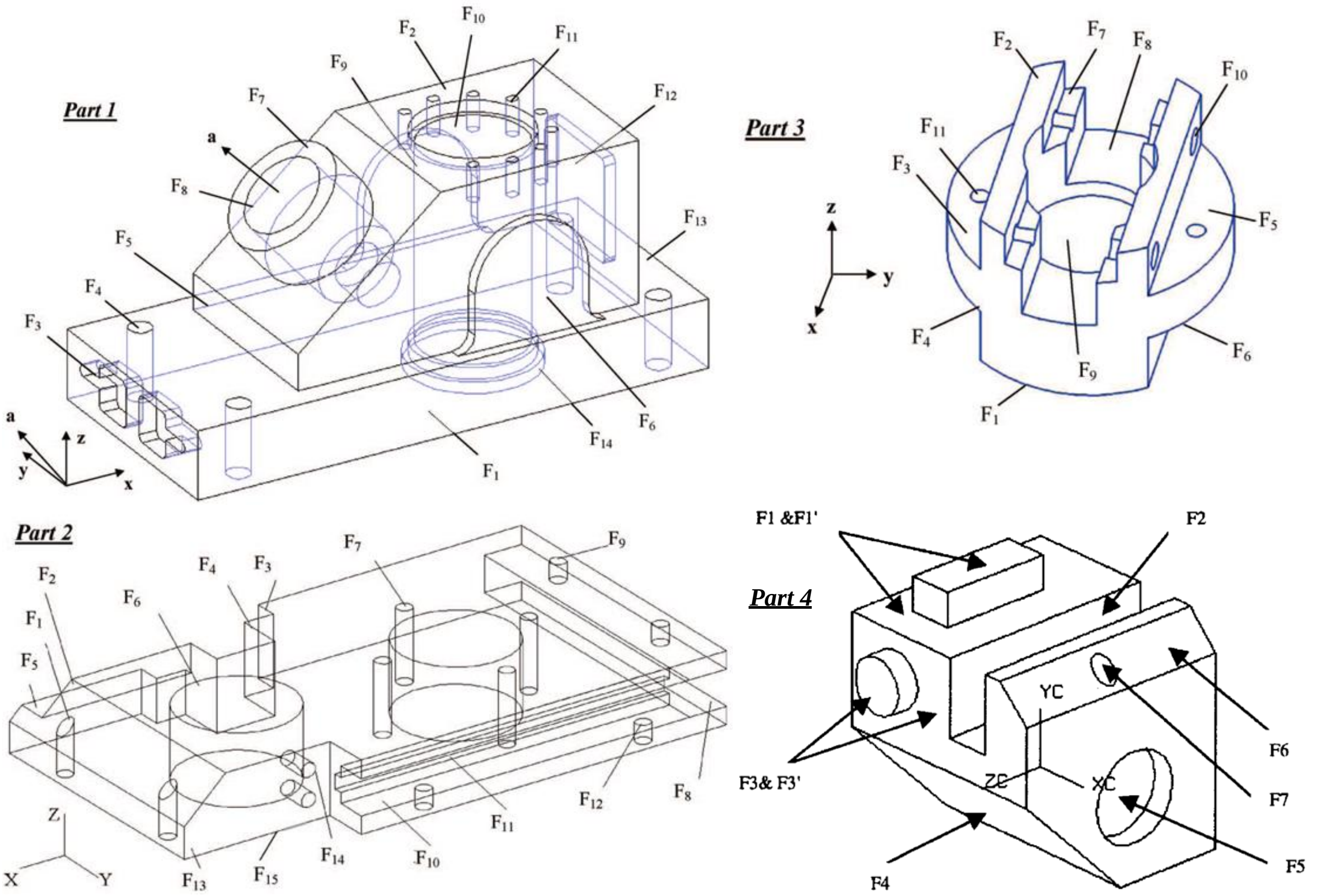

Figure 20. The description of 4 parts of PPS

**Table 8. The Technical Specifications for Part #1**

| Features | Operations | Machine Candidates | Tool Candidates | Machining time for each candidate machine (s) |
|---|---|---|---|---|
| **F1** | Milling (Oper1) | $M_2$, $M_3$, $M_4$ | $T_6$, $T_7$, $T_8$ | 40, 40, 30 |
| **F2** | Milling (Oper2) | $M_2$, $M_3$, $M_4$ | $T_6$, $T_7$, $T_8$ | 40, 40, 30 |
| **F3** | Milling (Oper3) | $M_2$, $M_3$, $M_4$ | $T_6$, $T_7$, $T_8$ | 20, 20, 15 |
| **F4** | Drilling (Oper4) | $M_1$, $M_2$, $M_3$, $M_4$ | $T_2$ | 12, 10, 10, 7.5 |
| **F5** | Milling (Oper5) | $M_2$, $M_3$, $M_4$ | $T_6$, $T_7$ | 35, 35, 26.25 |
| **F6** | Milling (Oper6) | $M_2$, $M_3$, $M_4$ | $T_7$, $T_8$ | 15, 15, 11.25 |
| **F7** | Milling (Oper7) | $M_2$, $M_3$, $M_4$ | $T_7$, $T_8$ | 30, 30, 22.5 |
| **F8** | Milling (Oper8) | $M_1$, $M_2$, $M_3$, $M_4$ | $T_2$, $T_3$, $T_4$ | 21.6, 18, 18, 13.5 |
| | Reaming (Oper9) | $M_2$, $M_3$, $M_4$ | $T_9$ | 10, 10, 7.5 |
| | Boring (Oper10) | $M_2$, $M_3$, $M_4$, $M_5$ | $T_{10}$ | 10, 10, 7.5, 12 |
| **F9** | Milling (Oper11) | $M_2$, $M_3$, $M_4$ | $T_7$, $T_8$ | 15, 15, 11.25 |
| **F10** | Drilling (Oper12) | $M_1$, $M_2$, $M_3$, $M_4$ | $T_2$, $T_3$, $T_4$ | 48, 40, 40, 30 |
| | Reaming (Oper13) | $M_2$, $M_3$, $M_4$ | $T_9$ | 25, 25, 18.75 |
| | Boring (Oper14) | $M_2$, $M_3$, $M_4$, $M_5$ | $T_{10}$ | 25, 25, 18.75, 30 |
| **F11** | Milling (Oper15) | $M_1$, $M_2$, $M_3$, $M_4$ | $T_1$ | 26.4, 22, 22, 16.5 |
| | Tapping (Oper16) | $M_2$, $M_3$, $M_4$ | $T_5$ | 20, 20, 15 |
| **F12** | Milling (Oper17) | $M_2$, $M_3$, $M_4$ | $T_7$, $T_8$ | 16, 16, 12 |
| **F13** | Milling (Oper18) | $M_2$, $M_3$, $M_4$ | $T_6$, $T_7$ | 35, 35, 26.25 |
| **F14** | Reaming (Oper19) | $M_2$, $M_3$, $M_4$ | $T_9$ | 12, 12, 9 |
| | Boring (Oper20) | $M_2$, $M_3$, $M_4$, $M_5$ | $T_{10}$ | 12, 12, 9, 14.4 |

**Table 9. The Technical Specifications for Part #2**

| Features | Operations | Machine Candidates | Tool Candidates | Machining time for each candidate machine (s) |
|---|---|---|---|---|
| **F1** | Drilling (Oper1) | $M_1$, $M_2$, $M_3$, $M_4$ | $T_1$ | 12, 10, 10, 7.5 |
| **F2** | Milling (Oper2) | $M_2$, $M_3$, $M_4$ | $T_{12}$ | 20, 20, 15 |
| **F3** | Milling (Oper3) | $M_2$, $M_3$, $M_4$ | $T_5$, $T_6$, $T_{11}$ | 18, 18, 13.5 |
| **F4** | Milling (Oper4) | $M_2$, $M_3$, $M_4$ | $T_6$, $T_7$, $T_8$ | 16, 16, 12 |
| **F5** | Milling (Oper5) | $M_2$, $M_3$, $M_4$ | $T_6$, $T_7$, $T_8$ | 15, 15, 11.25 |
| **F6** | Drilling (Oper6) | $M_1$, $M_2$, $M_3$, $M_4$ | $T_2$ | 30, 25, 25, 18.75 |
| | Reaming (Oper7) | $M_2$, $M_3$, $M_4$ | $T_9$ | 25, 25, 18.75 |
| **F7** | Drilling (Oper8) | $M_1$, $M_2$, $M_3$, $M_4$ | $T_1$ | 14.4, 12, 12, 9 |
| **F8** | Milling (Oper9) | $M_2$, $M_3$, $M_4$ | $T_6$, $T_7$, $T_8$ | 15, 15, 11.25 |
| **F9** | Drilling (Oper10) | $M_1$, $M_2$, $M_3$, $M_4$ | $T_1$ | 9.6, 8, 8, 6 |
| **F10** | Milling (Oper11) | $M_2$, $M_3$, $M_4$ | $T_6$, $T_7$, $T_8$ | 10, 10, 7.5 |
| **F11** | Milling (Oper12) | $M_2$, $M_3$, $M_4$ | $T_6$, $T_7$, $T_8$ | 10, 10, 7.5 |
| **F12** | Drilling (Oper13) | $M_1$, $M_2$, $M_3$, $M_4$ | $T_1$ | 9.6, 8, 8, 6 |
| **F13** | Milling (Oper14) | $M_2$, $M_3$, $M_4$ | $T_6$, $T_7$, $T_8$ | 16, 16, 12 |
| **F14** | Drilling (Oper15) | $M_1$, $M_2$, $M_3$, $M_4$ | $T_1$ | 9.6, 8, 8, 6 |
| **F15** | Milling (Oper16) | $M_1$, $M_2$, $M_3$, $M_4$ | $T_6$, $T_7$, $T_8$ | 36, 30, 30, 22.5 |

**Table 10. The Technical Specifications for Part #3**

| Features | Operations | Machine Candidates | Tool Candidates | Machining time for each candidate machine (s) |
|---|---|---|---|---|
| **F1** | Milling (Oper1) | $M_2$, $M_3$, $M_4$ | $T_6$, $T_7$, $T_8$ | 20, 15, 20 |
| **F2** | Milling (Oper2) | $M_2$, $M_3$, $M_4$ | $T_6$, $T_7$, $T_8$ | 20, 15, 20 |
| **F3** | Milling (Oper3) | $M_2$, $M_3$, $M_4$ | $T_6$, $T_7$, $T_8$ | 15, 15, 11.25 |
| **F4** | Milling (Oper4) | $M_1$, $M_2$, $M_3$, $M_4$ | $T_2$ | 15, 15, 11.25, 18 |
| **F5** | Milling (Oper5) | $M_2$, $M_3$, $M_4$ | $T_6$, $T_7$, $T_8$ | 15, 15, 11.25 |
| **F6** | Milling (Oper6) | $M_2$, $M_3$, $M_4$ | $T_7$, $T_8$ | 15, 15, 11.25 |
| **F7** | Milling (Oper7) | $M_2$, $M_3$, $M_4$ | $T_7$, $T_8$, $T_{11}$ | 15, 15, 11.25 |
| **F8** | Milling (Oper8) | $M_2$, $M_3$, $M_4$ | $T_6$, $T_7$, $T_8$, $T_{11}$ | 25, 25, 18.75 |
| **F9** | Drilling (Oper9) | $M_1$, $M_2$, $M_3$, $M_4$ | $T_2$, $T_3$, $T_4$ | 30, 25, 25, 18.75 |
| | Reaming (Oper10) | $M_2$, $M_3$, $M_4$ | $T_9$ | 20, 20, 15 |
| | Boring (Oper11) | $M_2$, $M_3$, $M_4$, $M_5$ | $T_{10}$ | 20, 20, 15, 24 |
| **F10** | Drilling (Oper12) | $M_1$, $M_2$, $M_3$, $M_4$ | $T_1$ | 9.6, 8, 8, 6 |
| | Tapping (Oper13) | $M_2$, $M_3$, $M_4$ | $T_5$ | 8, 8, 6 |
| **F11** | Drilling (Oper14) | $M_1$, $M_2$, $M_3$, $M_4$ | $T_9$ | 6, 5, 5, 3.75 |

**Table 11. The Technical Specifications for Part #4**

| Features | Operations | Machine Candidates | Tool Candidates | Machining time for each candidate machine (s) |
|---|---|---|---|---|
| **F1** | Milling (Oper1) | $M_2$, $M_4$ | $T_6$, $T_9$ | 12 |
| **F2** | Milling (Oper2) | $M_2$, $M_4$ | $T_9$, $T_{10}$ | 21 |
| **F3** | Milling (Oper3) | $M_2$, $M_4$ | $T_9$ | 18 |
| **F4** | Milling (Oper4) | $M_2$, $M_4$ | $T_1$, $T_9$ | 27 |
| **F5** | Drilling (Oper5) | $M_1$, $M_2$, $M_4$ | $T_2$ | 20 |
| **F6** | Milling (Oper6) | $M_2$, $M_4$ | $T_1$, $T_9$ | 18 |
| **F7** | Drilling (Oper7) | $M_1$, $M_2$, $M_4$ | $T_2$ | 20 |

We define each time slot representing 15 time units. The top segment of Figure 22 illustrates the best practice operation sequence of the four parts. All the operations are then transformed into the input format. Figure 21 shows the transformed operations of Part #1. There are 119 Unit Tasks for all the four parts. We can generate the conflicting graph, which has 47525 edges and 580 nodes. For a problem in such a size, the heuristics configurations with faster approximation-based algorithms are preferred.

Using the Heuristics #19, Algorithm A5 MWIS_SubCS_GWMIN and MWIS A3: MWIS Weights 3, and using the median length weight factor, $LW_c^M = 1$. The schedule with the resource allocations generated is shown in Figure 22. Table 12 shows the outputs of Heuristics #11 to Heuristics #28 on the real-world PPS problem. Among all the Heuristics tested, the Heuristics #19 achieved the optimum solution with 31 time slots. The results with an error rate of less than 5% take 7000~11000 seconds of clock time for finishing the computation.

In Zhang et al.'s work (Zhang et al., 2014), they assume that tools are always available without causing any constraints. This assumption is based on the understanding that the machining tools are mostly available, but the machines are more critical resources in a flexible job shop. By removing the tools from the constraints, we formulate a lite version of the real-world PPS problem. The edge number is reduced to 8771, and the node number is reduced to 292 in the conflicting graph. Table 13 shows the outputs of Heuristics #11 to Heuristics #28 on this simplified real-world PPS problem. Note that the optimum solution for this instance is also 31 time slots; it is calculated by manipulating the IP model in a trial and error fashion. The results with an error rate of less than 5% take less than 700 seconds clock time for finishing the computation. Although our approach almost doubles the computation time compare to Zhang et al.'s work, the runtime is still acceptable. We can say that our approach has acceptable practicability and feasibility on real-world PPS problem. To further justify this conclusion, the following section discusses the details regarding the scalability and accuracy of the proposed approach.

**Table 12. Outputs of Heuristics on Real-world PPS Problem**

| Methods | Length Weight | Minimum makespan (in time slots) | Clock time 3-run average (s) | Error | Error rate |
|---|---|---|---|---|---|
| **Heuristics#11** | LW=86 | 37 | 2533.453125 | 6 | 19.35% |
| **Heuristics#12** | LW=86 | 33 | 9576.869792 | 2 | 6.45% |
| **Heuristics#13** | LW=1 | 32 | 7474.770833 | 1 | 3.23% |
| **Heuristics#14** | LW=86 | 33 | 2542.958333 | 2 | 6.45% |
| **Heuristics#15** | LW=86 | 32 | 8680.5 | 1 | 3.23% |
| **Heuristics#16** | LW=1 | 32 | 7131.817708 | 1 | 3.23% |
| **Heuristics#17** | LW=1 | 33 | 2498.739583 | 2 | 6.45% |
| **Heuristics#18** | LW=1 | 33 | 9011.416667 | 2 | 6.45% |
| **Heuristics#19** | LW=1 | 31 | 6256.010417 | 0 | 0.00% |
| **Heuristics#20** | LW=1 | 53 | 6122.572917 | 22 | 70.97% |
| **Heuristics#21** | LW=1 | 36 | 23818.79167 | 5 | 16.13% |
| **Heuristics#22** | LW=86 | 32 | 10684.86458 | 1 | 3.23% |
| **Heuristics#23** | LW=0.001 | 52 | 4757.463542 | 21 | 67.74% |
| **Heuristics#24** | LW=1 | 33 | 21079.84375 | 2 | 6.45% |
| **Heuristics#25** | LW=86 | 35 | 9119.651042 | 4 | 12.90% |
| **Heuristics#26** | LW=1 | 58 | 4884.520833 | 27 | 87.10% |
| **Heuristics#27** | LW=0.001 | 35 | 19428.39583 | 4 | 12.90% |
| **Heuristics#28** | LW=86 | 32 | 8462.151042 | 1 | 3.23% |

```
{
'a01':   [[[['J1'],['11']],['1'],['a']],['2'], [[['1'],['M2','M3']],[['1'],['T6','T7','T8']],2]],
'a012':   [[[['J1'],['11']],['2'],['a']],['2'], [[['1'],['M2','M3']],[['1'],['T6','T7','T8']],2]],
'a013':   [[[['J1'],['11']],['3'],['a']],['2'], [[['1'],['M2','M3']],[['1'],['T6','T7','T8']],2]],
'2a01':   [[[['J1'],['11']],['1'],['b']],['2'], [[['1'],['M4']],[['1'],['T6','T7','T8']],2]],
'2a012':   [[[['J1'],['11']],['2'],['b']],['2'], [[['1'],['M4']],[['1'],['T6','T7','T8']],2]],
'a02':   [[[['J1'],['12']],['1'],['a']],['2'], [[['1'],['M2','M3']],[['1'],['T6','T7','T8']],2]],
'a022':   [[[['J1'],['12']],['2'],['a']],['2'], [[['1'],['M2','M3']],[['1'],['T6','T7','T8']],2]],
'a023':   [[[['J1'],['12']],['3'],['a']],['2'], [[['1'],['M2','M3']],[['1'],['T6','T7','T8']],2]],
'2a02':   [[[['J1'],['12']],['1'],['b']],['2'], [[['1'],['M4']],[['1'],['T6','T7','T8']],2]],
'2a022':   [[[['J1'],['12']],['2'],['b']],['2'], [[['1'],['M4']],[['1'],['T6','T7','T8']],2]],
'a03':   [[[['J1'],['13']],['1'],['a']],['2'], [[['1'],['M2','M3']],[['1'],['T6','T7','T8']],2]],
'a032':   [[[['J1'],['13']],['2'],['a']],['2'], [[['1'],['M2','M3']],[['1'],['T6','T7','T8']],2]],
'2a03':   [[[['J1'],['13']],['1'],['b']],['2'], [[['1'],['M4']],[['1'],['T6','T7','T8']],2]],
'a05':   [[[['J1'],['14']],['1'],['a']],['2'], [[['1'],['M2','M3']],[['1'],['T6','T7']],2]],
'a052':   [[[['J1'],['14']],['2'],['a']],['2'], [[['1'],['M2','M3']],[['1'],['T6','T7']],2]],
'a053':   [[[['J1'],['14']],['3'],['a']],['2'], [[['1'],['M2','M3']],[['1'],['T6','T7']],2]],
'2a05':   [[[['J1'],['14']],['1'],['b']],['2'], [[['1'],['M4']],[['1'],['T6','T7']],2]],
'2a052':   [[[['J1'],['14']],['2'],['b']],['2'], [[['1'],['M4']],[['1'],['T6','T7']],2]],
'a06':   [[[['J1'],['15']],['1'],['a']],['1'], [[['1'],['M2','M3','M4']],[['1'],['T7','T8']],2]],
'a11':   [[[['J1'],['16']],['1'],['a']],['1'], [[['1'],['M2','M3','M4']],[['1'],['T7','T8']],2]],
'a18':   [[[['J1'],['17']],['1'],['a']],['2'], [[['1'],['M2','M3']],[['1'],['T6','T7']],2]],
'a182':   [[[['J1'],['17']],['2'],['a']],['2'], [[['1'],['M2','M3']],[['1'],['T6','T7']],2]],
'a183':   [[[['J1'],['17']],['3'],['a']],['2'], [[['1'],['M2','M3']],[['1'],['T6','T7']],2]],
'2a18':   [[[['J1'],['17']],['1'],['b']],['2'], [[['1'],['M4']],[['1'],['T6','T7']],2]],
'2a182':   [[[['J1'],['17']],['2'],['b']],['2'], [[['1'],['M4']],[['1'],['T6','T7']],2]],
'a04':   [[[['J1'],['18']],['1'],['a']],['1'], [[['1'],['M1','M2','M3','M4']],[['1'],['T2']],2]],
'a07':   [[[['J1'],['19']],['1'],['a']],['1'], [[['1'],['M2','M3','M4']],[['1'],['T7','T8']],2]],
'a072':   [[[['J1'],['19']],['2'],['a']],['1'], [[['1'],['M2','M3','M4']],[['1'],['T7','T8']],2]],
'a12':   [[[['J1'],['110']],['1'],['a']],['2'], [[['1'],['M1','M2','M3']],[['1'],['T2','T3','T4']],2]],
'a121':   [[[['J1'],['110']],['2'],['a']],['2'], [[['1'],['M1','M2','M3']],[['1'],['T2','T3','T4']],2]],
'a122':   [[[['J1'],['110']],['3'],['a']],['2'], [[['1'],['M1','M2','M3']],[['1'],['T2','T3','T4']],2]],
'2a12':   [[[['J1'],['110']],['1'],['b']],['2'], [[['1'],['M4']],[['1'],['T2','T3','T4']],2]],
'2a122':   [[[['J1'],['110']],['2'],['b']],['2'], [[['1'],['M4']],[['1'],['T2','T3','T4']],2]],
'a13':   [[[['J1'],['111']],['1'],['a']],['1'], [[['1'],['M2','M3','M4']],[['1'],['T9']],2]],
'a132':   [[[['J1'],['111']],['2'],['a']],['1'], [[['1'],['M2','M3','M4']],[['1'],['T9']],2]],
'a14':   [[[['J1'],['112']],['1'],['a']],['1'], [[['1'],['M2','M3','M4','M5']],[['1'],['T10']],2]],
'a142':   [[[['J1'],['112']],['2'],['a']],['1'], [[['1'],['M2','M3','M4','M5']],[['1'],['T10']],2]],
'a15':   [[[['J1'],['113']],['1'],['a']],['1'], [[['1'],['M1','M2','M3','M4']],[['1'],['T1']],2]],
'a152':   [[[['J1'],['113']],['2'],['a']],['1'], [[['1'],['M1','M2','M3','M4']],[['1'],['T1']],2]],
'a16':   [[[['J1'],['114']],['1'],['a']],['2'], [[['1'],['M2','M3']],[['1'],['T5']],2]],
'a162':   [[[['J1'],['114']],['2'],['a']],['2'], [[['1'],['M2','M3']],[['1'],['T5']],2]],
'2a16':   [[[['J1'],['114']],['1'],['b']],['2'], [[['1'],['M4']],[['1'],['T5']],2]],
'a17':   [[[['J1'],['115']],['1'],['a']],['2'], [[['1'],['M2','M3']],[['1'],['T7','T8']],2]],
'a172':   [[[['J1'],['115']],['2'],['a']],['2'], [[['1'],['M2','M3']],[['1'],['T7','T8']],2]],
'2a17':   [[[['J1'],['115']],['1'],['b']],['2'], [[['1'],['M4']],[['1'],['T7','T8']],2]],
'a08':   [[[['J1'],['116']],['1'],['a']],['2'], [[['1'],['M1','M2','M3']],[['1'],['T2','T3','T4']],2]],
'a082':   [[[['J1'],['116']],['2'],['a']],['2'], [[['1'],['M1','M2','M3']],[['1'],['T2','T3','T4']],2]],
'2a08':   [[[['J1'],['116']],['1'],['b']],['2'], [[['1'],['M4']],[['1'],['T2','T3','T4']],2]],
'a09':   [[[['J1'],['117']],['1'],['a']],['1'], [[['1'],['M2','M3','M4']],[['1'],['T9']],2]],
'a10':   [[[['J1'],['118']],['1'],['a']],['1'], [[['1'],['M2','M3','M4','M5']],[['1'],['T10']],2]],
'a19':   [[[['J1'],['119']],['1'],['a']],['1'], [[['1'],['M2','M3','M4']],[['1'],['T9']],2]],
'a20':   [[[['J1'],['120']],['1'],['a']],['1'], [[['1'],['M2','M3','M4','M5']],[['1'],['T10']],2]],
```

Figure 21. Transformed Operations of Part #1

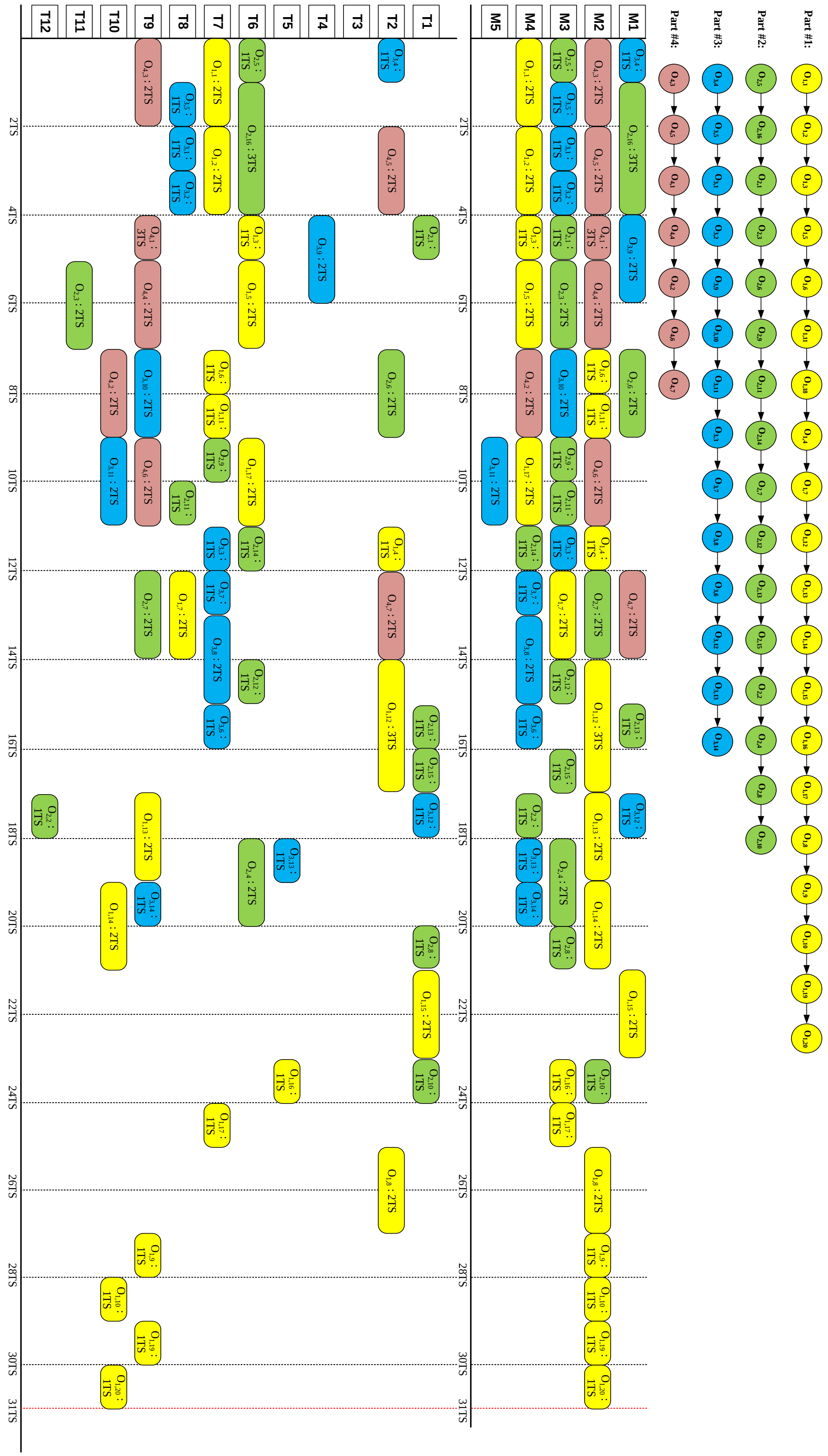

Figure 22. Schedule Created with Heuristics #19

**Table 13. Outputs of Heuristics on Real-world PPS Problem without Tool Constraints**

| Methods | Length Weight | Minimum makespan (in time slots) | Clock time 3-run average (s) | Error | Error rate |
|---|---|---|---|---|---|
| **Heuristics#11** | LW=86 | 37 | 147.2188 | 6 | 19.35% |
| **Heuristics#12** | LW=1 | 35 | 704.8594 | 4 | 12.90% |
| **Heuristics#13** | LW=86 | 33 | 521.4063 | 2 | 6.45% |
| **Heuristics#14** | LW=1 | 33 | 134.8698 | 2 | 6.45% |
| **Heuristics#15** | LW=1 | 32 | 671.2188 | 1 | 3.23% |
| **Heuristics#16** | LW=1 | 35 | 467.6979 | 4 | 12.90% |
| **Heuristics#17** | LW=1 | 34 | 124.8281 | 3 | 9.68% |
| **Heuristics#18** | LW=1 | 32 | 686.7813 | 1 | 3.23% |
| **Heuristics#19** | LW=1 | 34 | 449.1823 | 3 | 9.68% |
| **Heuristics#20** | LW=1 | 43 | 287.8906 | 12 | 38.71% |
| **Heuristics#21** | LW=0.001 | 35 | 1629.25 | 4 | 12.90% |
| **Heuristics#22** | LW=86 | 32 | 580.8958 | 1 | 3.23% |
| **Heuristics#23** | LW=1 | 43 | 293.9688 | 12 | 38.71% |
| **Heuristics#24** | LW=0.001 | 35 | 1671.74 | 4 | 12.90% |
| **Heuristics#25** | LW=1 | 31 | 591.6198 | 0 | 0.00% |
| **Heuristics#26** | LW=1 | 43 | 283.3698 | 12 | 38.71% |
| **Heuristics#27** | LW=1 | 36 | 1806.609 | 5 | 16.13% |
| **Heuristics#28** | LW=1 | 33 | 598.1563 | 2 | 6.45% |

# 8 Results and Discussions on Test Instances

We create nineteen test instances based on the structure of the real-world PPS example problem with randomized sequencing constraints and resource combinations. Since our approach returns feasible results on all the test instances and the real-world example, we assume that our approach has a satisfactory robustness on similar types of the PPS problem. Then, the discussion is focusing on the scalability and accuracy. The scalability analysis shows how the proposed approach behaves on different size and variance of the inputs. It can be evaluated based on the computation time versus the different input sizes, node numbers, and edge numbers of the different conflicting graphs. The accuracy refers to how likely the proposed approach can get to the optimum solution. It can be measured by the average and maximum error rate of all the test instances.

## *8.1 Scalability*

The essential understanding of our approach to PPS problems, MWIS algorithms are the determinant of the computation speed of different heuristics configurations. For those heuristics configurations based on the same MWIS algorithm, the ones with more complex weight factor calculations are slower. But this difference is minimal.

Figures 23 and 24 show how the computation time is changing with node number and edge number on Heuristics #1~10, respectively. The Heuristics #1~10, which are based on the two exact MWIS algorithms, Algorithm A1 MWIS and Algorithm A2 AMISL, are much slower than all other heuristics configurations. The computation time could be hours when there are about 140 nodes and 4000 edges, which could be much smaller than a typical PPS problem. Although the worst case of the two algorithms can be exponentially slow, the PPS problem considered here may not always be the worst case. As shown in Figure 23 and Figure 24, the Heuristics #1~10 match higher-order (order 4 or higher) polynomial trendlines, but they are faster than the exponential trendline.

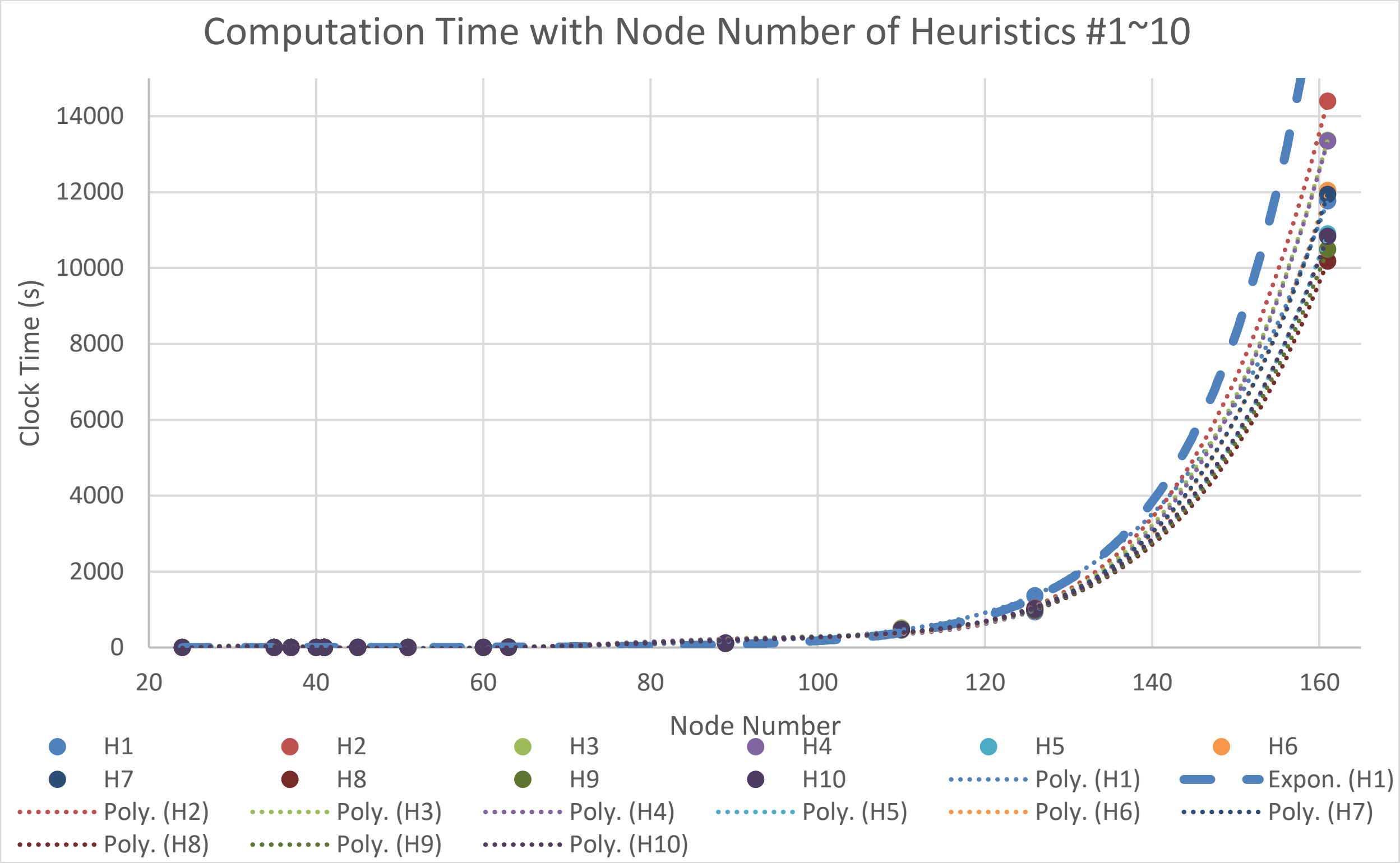

Figure 23. Computation Time with Node Number of Heuristics #1~10

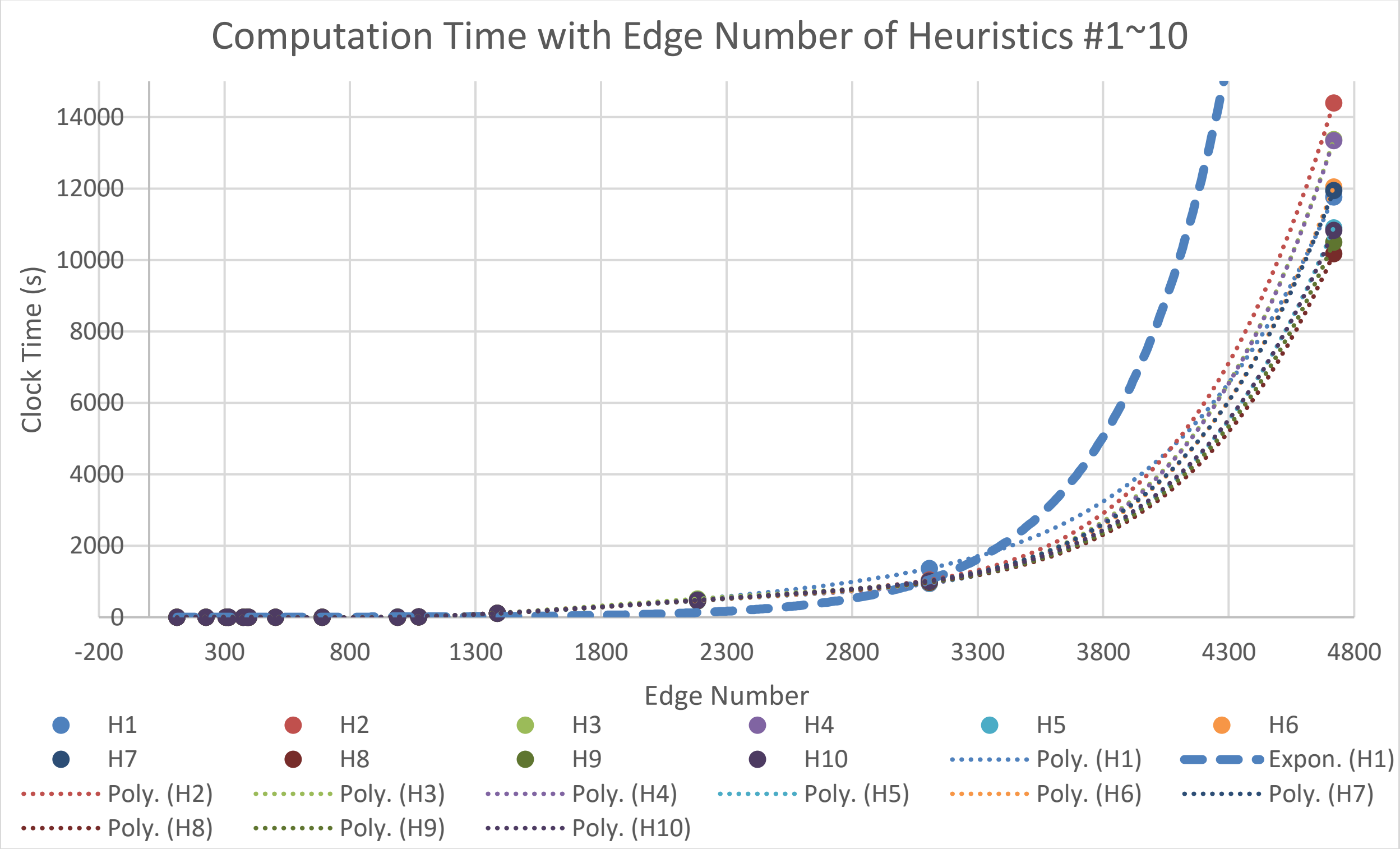

Figure 24. Computation Time with Edge Number of Heuristics #1~10

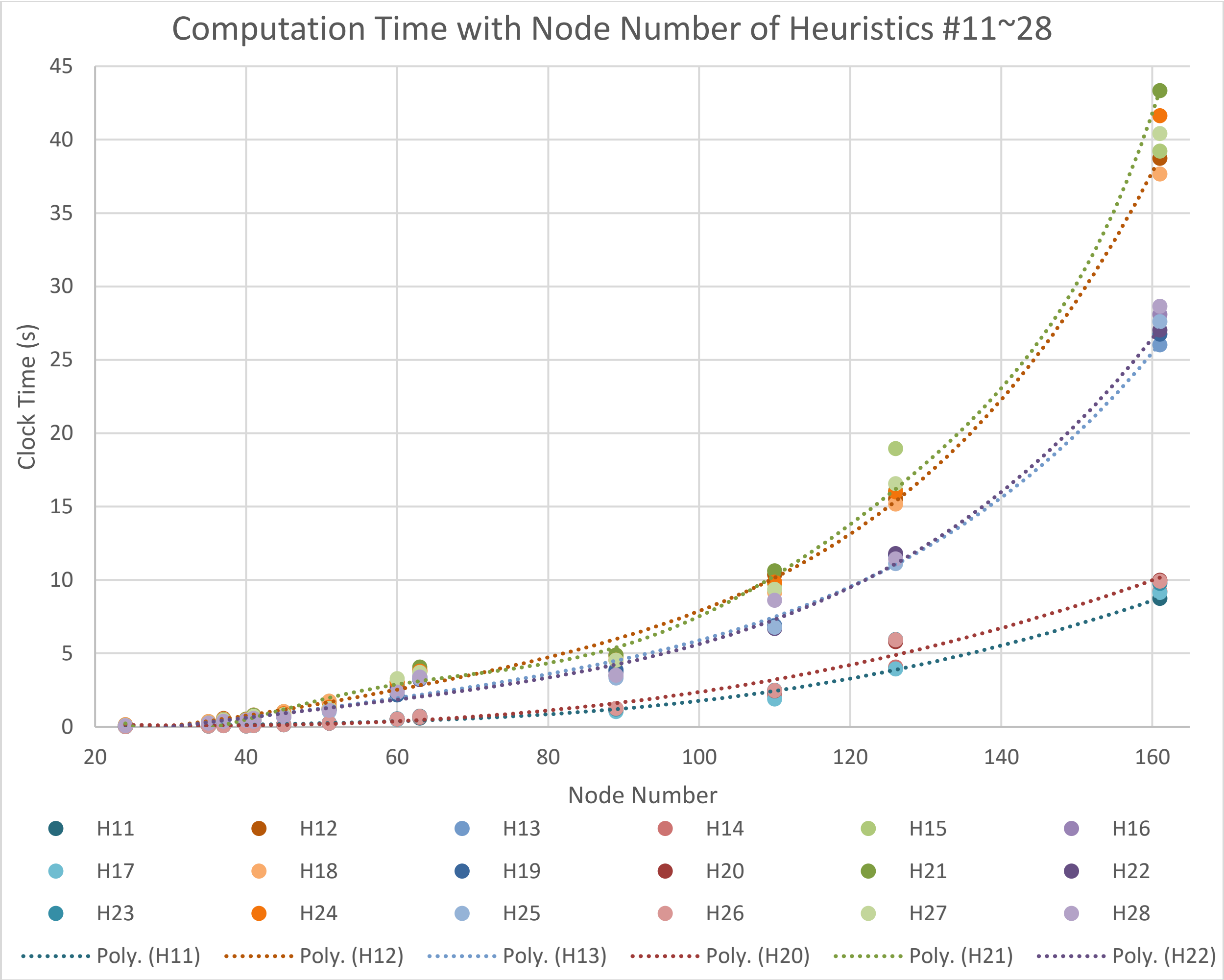

Figure 25. Computation Time with Node Number of Heuristics #11~28

For Heuristics #11~28, how the computation time is changing with node number and edge number is represented in Figure 25 and Figure 26, respectively. The Heuristics #11~28 are based on the approximation MWIS algorithms, GWMIN, GWMIN2, and their combinations. Heuristics #11, Heuristics #14, Heuristics #17, Heuristics #20, Heuristics #23, and Heuristics #26, which utilizing the Algorithm A3 GWMIN and Algorithm A6 GWMIN2 are the fastest. Heuristics #13, Heuristics #16, Heuristics #19, Heuristics #22, Heuristics #25 and Heuristics #28, which utilizing Algorithm A5 MWIS_SubCS_GWMIN and Algorithm A8 MWIS_SubCS_GWMIN2 are following. Heuristics #12, Heuristics #15, Heuristics #18, Heuristics #21, Heuristics #24 and Heuristics #27, which utilizing Algorithm A4 MWIS_CS_GWMIN and Algorithm A7 MWIS_CS_GWMIN2 are the slowest. The computational speed of these Heuristics follows the similar trendlines of the approximation MWIS algorithms, as discussed in our work (Sun et al., preprint). And Heuristics based on approximation MWIS algorithms are much feasible in the sense of computation time.

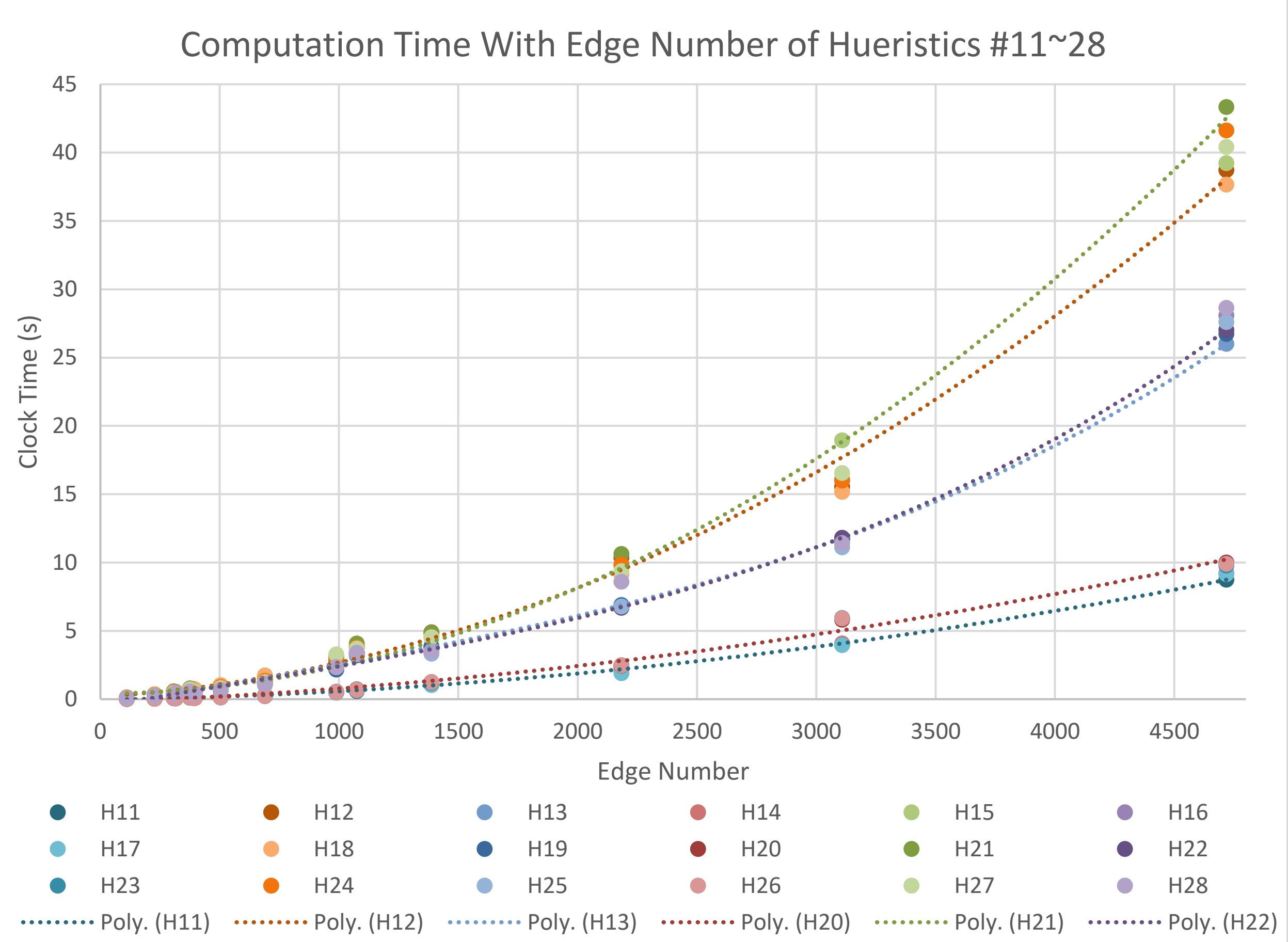

Figure 26. Computation Time with Edge Number of Heuristics #11~28

## 8.2 Accuracy

Figure 27 shows the average and maximum error rate for all heuristics configurations. The detailed information of the Heuristics configurations is as Table 4. The detailed accuracy summary of all the heuristics configurations on all tests is in Appendix I.

Assume $TS_{optimum}$ is the minimum number of time slots need for the PPS problem on the test instance, and $TS$ is the number of time slots found by our approach. The error rate is calculated using the function below.

$$Error\ Rate = \frac{TS - TS_{optimum}}{TS_{optimum}} \times 100\%$$

Note that the $TS_{optimum}$ is calculated based on the IP model with manipulating inputs to get the optimum result with reasonable computation time, and the error rate of each heuristics configuration is calculated based on the best accuracy among the three different length weight factors.

Let the threshold for heuristics configuration selection be the average error of less than 7% and the maximum error of less than 20%. For Heuristics #1-10 with the exact MWIS algorithms, from the best to the worst, Heuristics #2, Heuristics #8, Heuristics #5, Heuristics #3 and Heuristics #4 are satisfactory. For Heuristics #11-28 with the approximation MWIS algorithms, from the best to the worst, Heuristics #16, Heuristics #19, Heuristics #28, Heuristics #25, Heuristics #15, Heuristics #18, Heuristics #14, and Heuristics #17 are satisfactory.

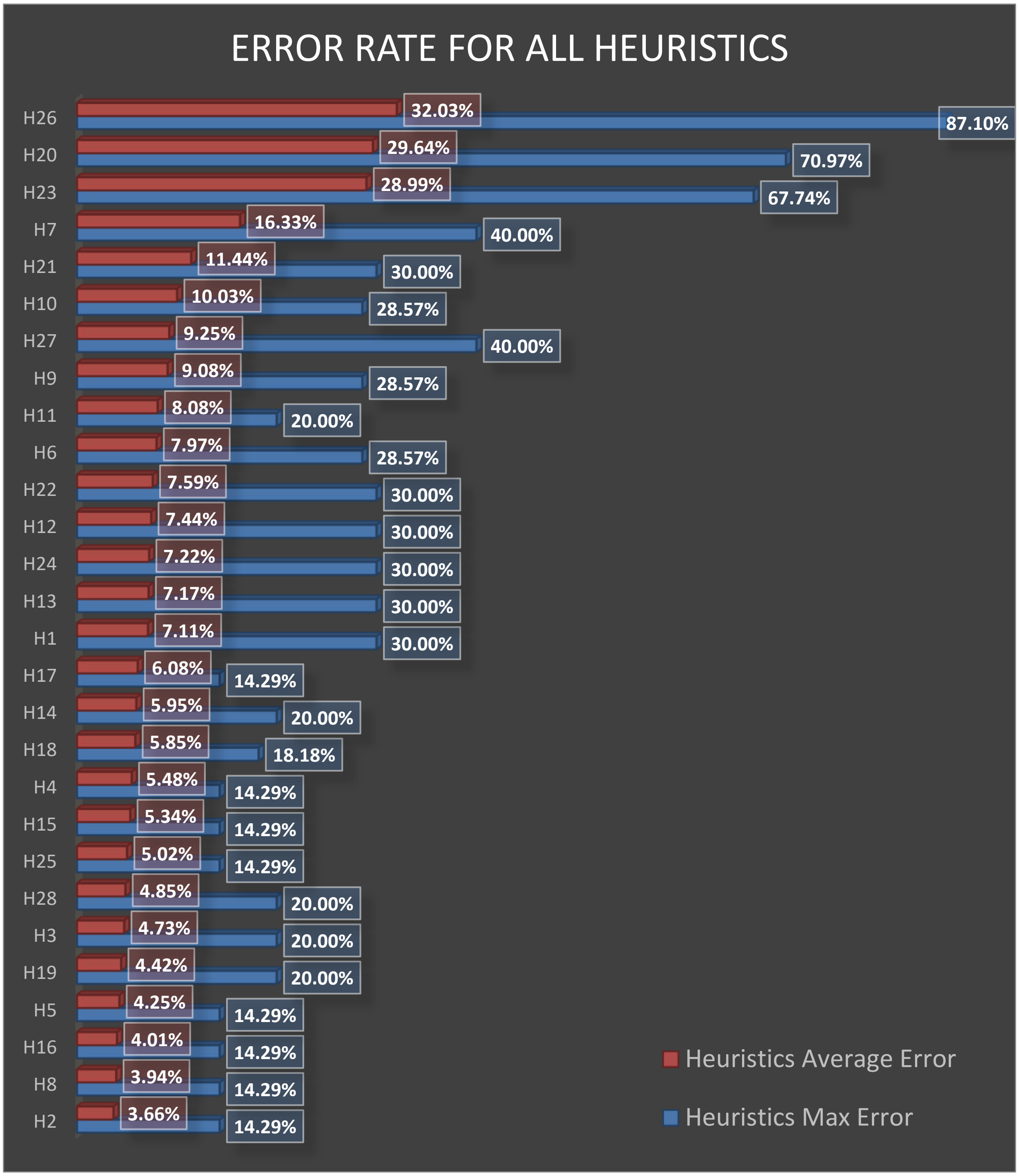

Figure 27. The Average and Maximum Error Rate for All Heuristics Configurations

Based on the computational experiments in our work (Sun et al., preprint), the general accuracy of the MWIS algorithms can be listed below from the best accuracy to the worst:

- Algorithm A1 MWIS

- Algorithm A2 AMISL (same as Algorithm MWIS)
- Algorithm A5 MWIS_SubCS_GWMIN
- Algorithm A8 MWIS_SubCS_GWMIN2
- Algorithm A4 MWIS_CS_GWMIN
- Algorithm A3 GWMIN
- Algorithm A7 MWIS_CS_GWMIN2
- Algorithm A6 GWMIN2

Compare with the results shown in Figure 27, with the same weight factors assignment, a more accurate MWIS algorithm leads to a better accuracy output of the PPS problem. None of the satisfactory heuristics is using the least accurate MWIS algorithms, Algorithm A7 MWIS_CS_GWMIN2 and A6 GWMIN2. In other words, while using the proposed approach for the PPS problem, a relatively accurate MWIS algorithm is required. This is the evidence of the necessity of the better accuracy MWIS algorithms proposed in our work (Sun et al., preprint).

$$\text{Job \#1: } \frac{T_{11}[(R_1,R_2,R_3)_2]}{2} \rightarrow \frac{T_{12}[(R_1,R_2,R_3)_2]}{1} \rightarrow \frac{T_{13}[(R_1,R_2,R_3)_1]}{2} \rightarrow \frac{T_{14}[(R_4)_1]}{3}$$

$$\text{Job \#2: } \frac{T_{21}[(R_1,R_2,R_3)_1]}{1} \rightarrow \frac{T_{22}[(R_1)_1 \text{and} (R_2,R_3)_1]}{2} \rightarrow \frac{T_{23}[(R_4)_1]}{3}$$

$$\text{Job \#3: } \frac{T_{31}[(R_1,R_2)_1]}{1} \rightarrow \begin{pmatrix} \frac{T_{32a}[(R_1,R_2)_2]}{1} \\ \frac{T_{32b}[(R_3)_1]}{2} \end{pmatrix}^{1} \rightarrow \frac{T_{33}[(R_3,R_4)_1]}{3}$$

$$\text{Job \#4: } \frac{T_{41}[(R_1,R_2,R_3)_2]}{2}$$

$$\text{Job \#5: } \frac{T_{51}[(R_1,R_2,R_3)_2]}{2}$$

Figure 28. Details of Test Instances T24

The above-mentioned heuristics configurations may not able to reach the optimum results on some of the test instances. These bad instances are T6, T11, T12, T13, T14, T17, T18, T19, T20, and T24. Figure 29 shows the average and maximum error rate for all heuristics configurations on these bad instances. These instances have concentrated resource requirements. Let us take the instance T24 (Figure 28) as an example. The jobs in the instances have a significant difference in the number of time slots for finishing. Also, the beginning Unit Tasks are concentrated on the resources $R_1, R_2$, and $R_3$, and the ending Unit Tasks are concentrated on the resources $R_4$. Since the MWIS algorithm tries to schedule as many nodes as possible, it may cause the ending Unit Tasks all leftover, but they cannot be processed on parallel machines. We iterate the three levels of length weight coefficient, median, high and low, as $LW_c^M$, $LW_c^H$ and $LW_c^L$, respectively with the proposed heuristics configurations to balance the length of each job and the concentrated resources requirements. So that the maximum error rate of each satisfactory heuristics configuration is not exceeding 20%.

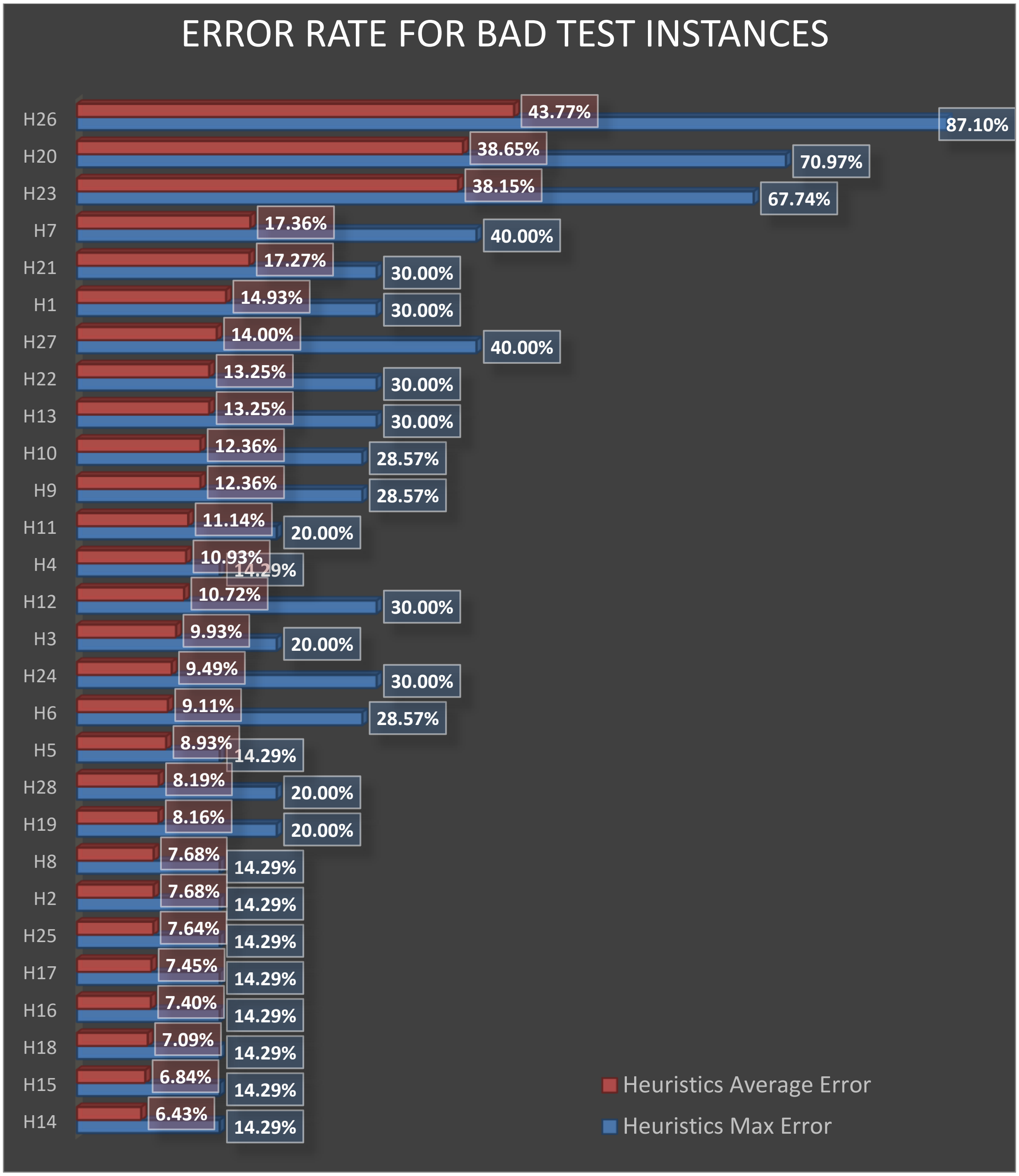

Figure 29. The Average and Maximum Error Rate on Bad Test Instances

Another interesting finding is that the Heuristics #14 and #17, which are using the approximation algorithms GWMIN, perform well on these bad test instances. The hypothesis is that the GWMIN generates the selection of the node with the maximum weight. This may avoid the concentrating resources blocking the optimum results.

Based on the discussions on scalability and accuracy, the better heuristics configurations for the PPS problem are listed as below,

- Heuristics #16, Algorithm MWIS_SubCS_GWMIN, MWIS A2: MWIS Weights 2
- Heuristics #19, Algorithm MWIS_SubCS_GWMIN, MWIS A3: MWIS Weights 3
- Heuristics #28, Algorithm MWIS_SubCS_GWMIN2, MWIS A3: MWIS Weights 3
- Heuristics #25, Algorithm MWIS_SubCS_GWMIN2, MWIS A2: MWIS Weights 2
- Heuristics #15, Algorithm MWIS_CS_GWMIN, MWIS A2: MWIS Weights 2
- Heuristics #18, Algorithm MWIS_CS_GWMIN, MWIS A3: MWIS Weights 3
- Heuristics #14, Algorithm GWMIN, MWIS A2: MWIS Weights 2
- Heuristics #17, Algorithm GWMIN, MWIS A3: MWIS Weights 3

# 9 Conclusions

This paper considers a general type of PPS problem, and proposes a novel graph-based approach for formulating and solving the resource-constrained PPS optimization problem. Unlike the commonly used iteration type of approaches, such as generic algorithms and metaheuristics, or the mixed-integer programming approaches, our approach provides a different angle to address the PPS problem by enabling graph theory concepts and tools. It shows advantages over other approaches, as illustrated in Table 14. The PPS problem is formulated as a conflicting weighted graph, and the two procedures, the resource selection and process scheduling, of the PPS problem, are integrated. This idea extends the universality of the formulation of the graph coloring based scheduling. The new approach requires minimum iteration. And it is guaranteed to return a feasible solution due to the nature of solving the MWIS problem on a conflicting weighted graph. We develop different weight factor calculation strategies and arrangements as the guidance for achieving the optimization objective. With carefully defined weight factors and "good-performance" MWIS algorithms, the new approach has satisfactory accuracy and computational speed. A set of "good-performance" heuristics configurations are found based on test results. Figure 30 is the summary of the performance of the heuristics configurations. All these heuristics configurations considered as satisfactory have the average error of less than 7% and the maximum error of less than 20%.

**Table 14. Comparing the New Approach with Other Methods***

| Measurements | Generic Algorithms | Simulated Anneal | Tabu Search | Mixed-integer Programming | Partial Solutions | Graph Coloring Scheduling |
|---|---|---|---|---|---|---|
| **Accuracy** | = | = | = | - | + | NA |
| **Computational speed** | = | = | = | + | = | NA |
| **Universality** | - | - | = | - | NA | + |
| **Dependence on iterations** | + | + | + | + | + | NA |
| **Feasibility** | + | + | + | = | + | NA |
| **Separated solutions of each time slot** | + | + | + | + | + | NA |

*'+': The new approach is better on the measurement to compare with the other method.
'=': The new approach is similar or potentially better compare with the other method.
'-': The new approach is not as good as the other method.

‘NA’: It is hard to compare the new approach with the other method.

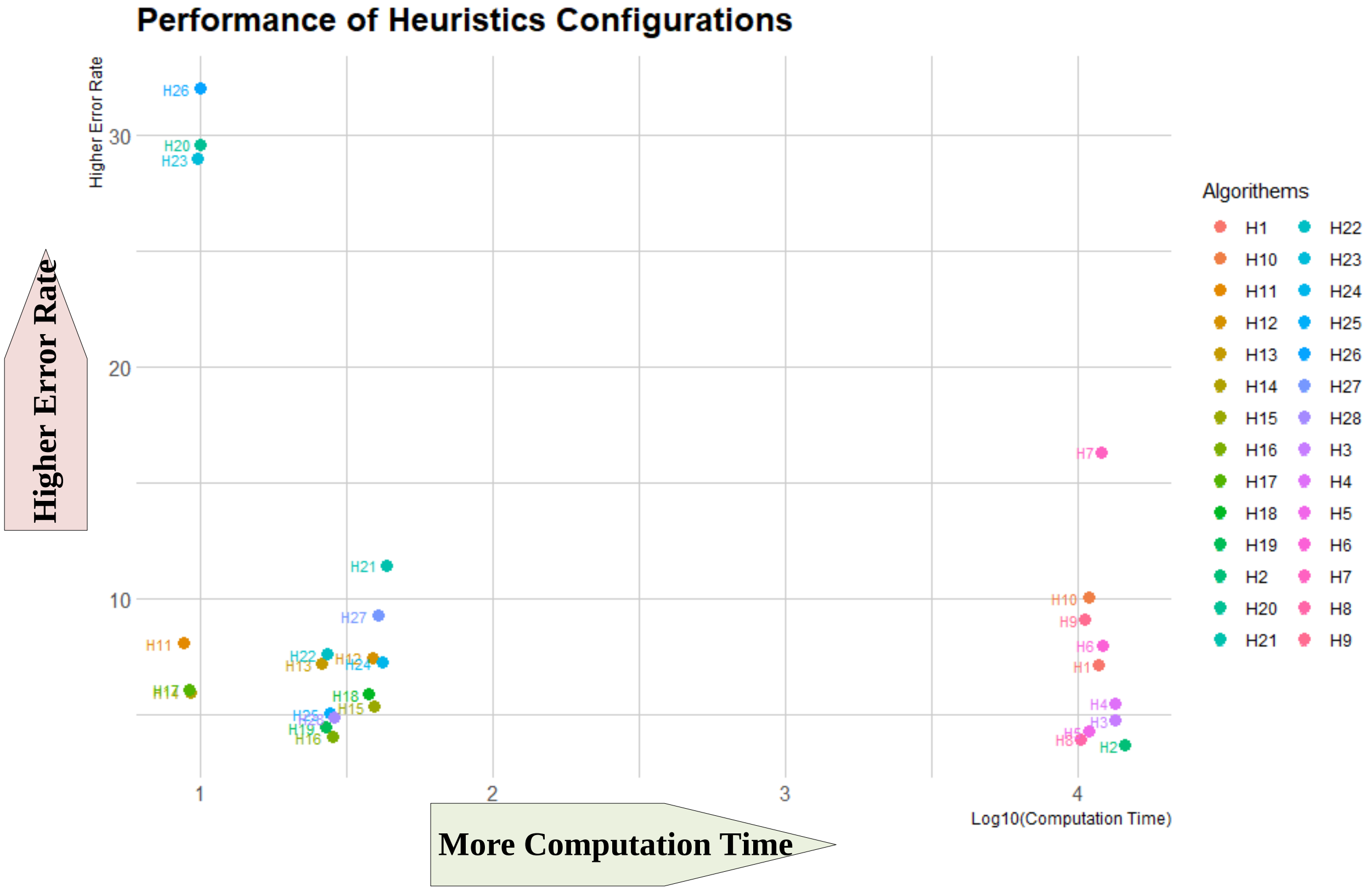

Figure 30. Performance of the Heuristics Configurations

In the future work, we aim to improve the graph-based PPS problem formulation in the following three aspects: (1) To speed up the computation, (2) to improve the accuracy, and (3) to improve the universality.

(1) To speed up the computation:

- For the MWIS algorithms, the NHHA framework (Sun et al., preprint) can be transformed into multi-threading. Each connected subgraph after node removal is independent so that the computation of each connected subgraph can be assigned to different threads.
- For the formulation of the PPS problem, Unit Tasks of the operations that are constrained to be processed in the far future (a good number of time slots later) may have a very limited impact on the scheduling of earlier time slots. While generating the conflicting graph, we may only consider the most recent several Unit Tasks of each part so that the size of the conflicting graph can be reduced.

(2) To improve the accuracy:

- The current weight calculation and weight factor arrangements can be fine-tuned and closely- integrated with the MWIS algorithms based on the part and resource information to achieve better node selections. The examples can be specific heuristics for weight calculation, machine learning methods to optimize the value of the weight factor.

- Stochastic Optimization (SO) methods are optimization methods that generate and use random variables (Spall, 2003). This method can be applied to bring in probabilistic in the schedule generation process when solving the MWIS problem for each time slot. It enables the possibility of iteratively selecting different sets of nodes for each time slot. By applying this method, the trapping of bad node selections may be avoided.

(3) To improve the universality:

- Our approach for the PPS problem can be easily implemented for a dynamic job taking environment by updating the conflicting graph for each time slot. The traditional approach requires taking consideration of known operations and iterates to get an optimum schedule for recent periods, which requires searching in a vast solution space. Unlike iteration-based approaches, the new approach computes the schedule of each time slot separately, which may only require partial operation information of each job. And for each time slot, the new approach tries to utilize the resources as much as possible by solving the MWIS problem.
- Our approach for the PPS problem can be easily implemented with the flexible operation sequencing constraints by updating the conflicting graph for each time slot. In this case, all the Unit Tasks that are not restricted by the sequencing constraints are considered as Unit Task Candidates to be selected by solving the MWIS problem.

The conflicting weighted graph may be extended to a multi-connected graph, directed graph, weighted edges to represent more information for the optimization problem modeling. We wish to improve the approach by, such as enabling the multi-objective optimization, introducing more variables for the details of the PPS problem, introducing probabilistic variables. And further, we are willing to extend this graph-based formulation technique to real-world applications such as logistics and transportation, supply-chain management, product design, development and manufacturing, and analysis and optimization of complex systems.

# Appendix I: The PPS Test Results Summary on Accuracy

| Test Accuracy Summary (different length weight coefficients) | | | | | | |
|---|---|---|---|---|---|---|
| Heuristics | Length Weight | Error Sum | Min Error | Max Error | Standard Deviation | Average Error |
| H1 | LW median | 146.79% | 0.00% | 30.00% | 0.116101709 | 7.73% |
| H1 | LW high | 66.79% | 0.00% | 14.29% | 0.054035859 | 3.52% |
| H1 | LW low | 379.65% | 0.00% | 50.00% | 0.215917236 | 19.98% |
| H2 | LW median | 116.79% | 0.00% | 20.00% | 0.079669269 | 6.15% |
| H2 | LW high | 96.79% | 0.00% | 20.00% | 0.073307066 | 5.09% |
| H2 | LW low | 425.39% | 0.00% | 60.00% | 0.20627864 | 22.39% |
| H3 | LW median | 82.34% | 0.00% | 14.29% | 0.063872671 | 4.33% |
| H3 | LW high | 96.79% | 0.00% | 20.00% | 0.073307066 | 5.09% |
| H3 | LW low | 337.89% | 0.00% | 50.00% | 0.199032284 | 17.78% |
| H4 | LW median | 132.67% | 0.00% | 30.00% | 0.107201134 | 6.98% |
| H4 | LW high | 72.67% | 0.00% | 14.29% | 0.05359337 | 3.82% |
| H4 | LW low | 377.15% | 0.00% | 60.00% | 0.227042888 | 19.85% |
| H5 | LW median | 132.67% | 0.00% | 20.00% | 0.090325798 | 6.98% |
| H5 | LW high | 92.67% | 0.00% | 20.00% | 0.06424635 | 4.88% |
| H5 | LW low | 467.89% | 0.00% | 60.00% | 0.211785288 | 24.63% |
| H6 | LW median | 239.09% | 0.00% | 28.57% | 0.103530837 | 12.58% |
| H6 | LW high | 157.31% | 0.00% | 28.57% | 0.074081227 | 8.28% |
| H6 | LW low | 503.60% | 0.00% | 60.00% | 0.21305678 | 26.51% |
| H7 | LW median | 274.65% | 0.00% | 40.00% | 0.117823431 | 14.46% |
| H7 | LW high | 264.65% | 0.00% | 40.00% | 0.11744188 | 13.93% |
| H7 | LW low | 423.74% | 0.00% | 60.00% | 0.213851589 | 22.30% |
| H8 | LW median | 72.67% | 0.00% | 14.29% | 0.05359337 | 3.82% |
| H8 | LW high | 92.67% | 0.00% | 20.00% | 0.06424635 | 4.88% |
| H8 | LW low | 266.55% | 0.00% | 50.00% | 0.194635336 | 14.03% |
| H9 | LW median | 168.06% | 0.00% | 28.57% | 0.066280593 | 8.85% |
| H9 | LW high | 157.31% | 0.00% | 28.57% | 0.074081227 | 8.28% |
| H9 | LW low | 453.60% | 0.00% | 50.00% | 0.188410498 | 23.87% |
| H10 | LW median | 238.06% | 0.00% | 28.57% | 0.120058681 | 12.53% |
| H10 | LW high | 177.15% | 0.00% | 28.57% | 0.069662802 | 9.32% |
| H10 | LW low | 487.89% | 0.00% | 60.00% | 0.204165627 | 25.68% |
| H11 | LW median | 275.40% | 0.00% | 40.00% | 0.132007015 | 14.49% |
| H11 | LW high | 201.21% | 0.00% | 36.36% | 0.102077836 | 10.59% |
| H11 | LW low | 371.62% | 0.00% | 42.86% | 0.166448838 | 19.56% |
| H12 | LW median | 219.18% | 0.00% | 40.00% | 0.127558102 | 11.54% |
| H12 | LW high | 144.00% | 0.00% | 18.18% | 0.06672139 | 7.58% |
| H12 | LW low | 393.70% | 0.00% | 42.86% | 0.129630104 | 20.72% |
| H13 | LW median | 194.71% | 0.00% | 42.86% | 0.137130947 | 10.25% |
| H13 | LW high | 89.69% | 0.00% | 14.29% | 0.062333895 | 4.72% |
| H13 | LW low | 325.46% | 0.00% | 50.00% | 0.146692817 | 17.13% |
| H14 | LW median | 209.92% | 0.00% | 30.00% | 0.104197662 | 11.05% |
| H14 | LW high | 151.82% | 0.00% | 30.00% | 0.089640345 | 7.99% |
| H14 | LW low | 319.95% | 0.00% | 40.00% | 0.157068297 | 16.84% |
| H15 | LW median | 205.23% | 0.00% | 40.00% | 0.112344169 | 10.80% |
| H15 | LW high | 114.64% | 0.00% | 18.18% | 0.059177995 | 6.03% |
| H15 | LW low | 418.39% | 0.00% | 50.00% | 0.126877852 | 22.02% |
| H16 | LW median | 146.14% | 0.00% | 20.00% | 0.094973115 | 7.69% |
| H16 | LW high | 106.14% | 0.00% | 20.00% | 0.06475161 | 5.59% |
| H16 | LW low | 423.45% | 0.00% | 50.00% | 0.130585361 | 22.29% |
| H17 | LW median | 192.74% | 0.00% | 30.00% | 0.090580368 | 10.14% |
| H17 | LW high | 147.55% | 0.00% | 18.18% | 0.063333126 | 7.77% |
| H17 | LW low | 233.49% | 0.00% | 30.00% | 0.114784161 | 12.29% |
| H18 | LW median | 161.10% | 0.00% | 40.00% | 0.100517064 | 8.48% |
| H18 | LW high | 132.16% | 0.00% | 28.57% | 0.076243832 | 6.96% |
| H18 | LW low | 340.30% | 0.00% | 40.00% | 0.106109163 | 17.91% |
| H19 | LW median | 104.92% | 0.00% | 14.29% | 0.06548711 | 5.52% |
| H19 | LW high | 125.82% | 0.00% | 20.00% | 0.071838329 | 6.62% |
| H19 | LW low | 347.65% | 0.00% | 50.00% | 0.18967755 | 18.30% |
| H20 | LW median | 534.04% | 5.56% | 70.97% | 0.193136428 | 28.11% |
| H20 | LW high | 567.68% | 5.56% | 80.65% | 0.214868683 | 29.88% |
| H20 | LW low | 553.39% | 5.56% | 80.65% | 0.21783028 | 29.13% |
| H21 | LW median | 242.68% | 0.00% | 30.00% | 0.11708745 | 12.77% |
| H21 | LW high | 184.85% | 0.00% | 20.00% | 0.077665972 | 9.73% |
| H21 | LW low | 365.80% | 0.00% | 30.00% | 0.092706741 | 19.25% |
| H22 | LW median | 277.62% | 0.00% | 50.00% | 0.191346926 | 14.61% |
| H22 | LW high | 99.69% | 0.00% | 14.29% | 0.062348484 | 5.25% |
| H22 | LW low | 375.14% | 0.00% | 50.00% | 0.166974759 | 19.74% |
| H23 | LW median | 580.58% | 5.56% | 87.10% | 0.232079338 | 30.56% |
| H23 | LW high | 580.58% | 5.56% | 87.10% | 0.232079338 | 30.56% |
| H23 | LW low | 518.50% | 5.56% | 67.74% | 0.186594926 | 27.29% |
| H24 | LW median | 201.95% | 0.00% | 22.58% | 0.085814202 | 10.63% |
| H24 | LW high | 159.04% | 0.00% | 20.00% | 0.067677392 | 8.37% |
| H24 | LW low | 353.30% | 0.00% | 30.00% | 0.101758463 | 18.59% |
| H25 | LW median | 162.59% | 0.00% | 20.00% | 0.094948821 | 8.56% |
| H25 | LW high | 142.27% | 0.00% | 20.00% | 0.071557047 | 7.49% |
| H25 | LW low | 408.11% | 0.00% | 50.00% | 0.144216269 | 21.48% |
| H26 | LW median | 580.58% | 5.56% | 87.10% | 0.232079338 | 30.56% |
| H26 | LW high | 580.58% | 5.56% | 87.10% | 0.232079338 | 30.56% |
| H26 | LW low | 571.49% | 5.56% | 87.10% | 0.233725333 | 30.08% |
| H27 | LW median | 233.95% | 0.00% | 30.00% | 0.113169757 | 12.31% |
| H27 | LW high | 219.46% | 0.00% | 28.57% | 0.108478598 | 11.55% |
| H27 | LW low | 325.79% | 0.00% | 30.00% | 0.105947041 | 17.15% |
| H28 | LW median | 128.47% | 0.00% | 14.29% | 0.07665969 | 6.76% |
| H28 | LW high | 149.69% | 0.00% | 20.00% | 0.103908679 | 7.88% |
| H28 | LW low | 360.55% | 0.00% | 50.00% | 0.175904418 | 18.98% |